\documentclass[12pt]{article}
\usepackage{tikz}
\usetikzlibrary{matrix,arrows,decorations.pathmorphing}
\usepackage{jheppub}
\usepackage{amsmath,amssymb,euscript,array,mathrsfs}
\usepackage{arydshln}
\usepackage{todonotes}
\usepackage{graphicx}

\newcommand{\startappendix}{
\setcounter{section}{0}
\renewcommand{\thesection}{\Alph{section}}}
\newcommand{\Appendix}[1]{
\refstepcounter{section}
\begin{flushleft}
{\large\bf Appendix \thesection: #1}
\end{flushleft}}

\usepackage{epsfig}

\def\XXint#1#2#3{{\setbox0=\hbox{$#1{#2#3}{\int}$}
     \vcenter{\hbox{$#2#3$}}\kern-.5\wd0}}

\newcommand{\IM}{\operatorname{Im}}

\newcommand{\RE}{\operatorname{Re}}
\def\AMP{{\EuScript A}}

\def\c{\gamma}
\def\d{\delta}

\def\l{\lambda}

\def\p{\pi}

\def\w{\omega}

\def\D{\Delta}

\def\det{{\rm det}}

\def\RR{{\mathfrak R}}

\def\Dbarslash{\,\,{\raise.15ex\hbox{/}\mkern-12mu {\bar D}}}
\def\Dslash{\,\,{\raise.15ex\hbox{/}\mkern-12mu D}}
\def\delslash{\,\,{\raise.15ex\hbox{/}\mkern-9mu \partial}}
\def\delbarslash{\,\,{\raise.15ex\hbox{/}\mkern-9mu {\bar\partial}}}

\def\RR{{\mathfrak R}}

\newcommand{\EQ}[1]{\begin{equation}\begin{split} #1 \end{split}\end{equation}}

\newcommand{\SP}[1]{\begin{equation}\begin{split} #1
\end{split}\end{equation}}

\title{The Effect of Gravitational Tidal Forces on Renormalized Quantum Fields}
\author{Timothy J. Hollowood and Graham M. Shore}
\affiliation{Department of Physics,\\ Swansea University,\\
Swansea,\\ SA2 8PP, UK.}
\emailAdd{t.hollowood@swansea.ac.uk, g.m.shore@swansea.ac.uk}

\abstract{The effect of gravitational tidal forces on renormalized
  quantum fields propagating in curved spacetime is investigated and a
  generalisation of the optical theorem to curved spacetime is
  proved. In the case of QED, the interaction of tidal forces with the
  vacuum polarization cloud of virtual $e^+ e^-$ pairs dressing the
  renormalized photon has been shown to produce several novel
  phenomena. In particular, the photon field amplitude can locally
  increase as well as decrease, corresponding to a negative imaginary
  part of the refractive index, in apparent violation of unitarity and
  the optical theorem. Below threshold decays into $e^+ e^-$ pairs may
  also occur. In this paper, these issues are studied from the point
  of view of a non-equilibrium initial-value problem, with the field
  evolution from an initial null surface being calculated for
  physically distinct initial conditions and for both scalar field
  theories and QED.  It is shown how a generalised version of the
  optical theorem, valid in curved spacetime, allows a local increase
  in amplitude while maintaining consistency with unitarity. The
  picture emerges of the field being dressed and undressed as it
  propagates through curved spacetime, with the local gravitational
  tidal forces determining the degree of dressing and hence the
  amplitude of the renormalized quantum field. These effects are
  illustrated with many examples, including a description of the 
  undressing of a photon in the vicinity of a black hole singularity. }

\notoc
\begin{document}

\maketitle

\section{Introduction}

Investigations of the effect of vacuum polarization on photon
propagation in gravitational backgrounds have revealed many unexpected
features of far wider importance for quantum field theory in curved
spacetime. In particular, the geometry induces a much richer analytic
structure for Green functions which, while preserving causality,
fundamentally changes the assumptions behind established theorems
in S-matrix theory and dispersion relations in flat space 
quantum field theory \cite{Hollowood:2007kt,Hollowood:2007ku,
Hollowood:2008kq,Hollowood:2009qz,Hollowood:2010bd}. 
A variety of new phenomena associated with the lack of translation
invariance also imply that many standard results based around
unitarity and the optical theorem must be reassessed in curved
spacetime.

In QED, gravitational tidal forces act on the virtual cloud
of electron-positron pairs which dress the photon with the result that
photons do not simply propagate along classical null
geodesics. Rather, the spacetime acts as an optical medium with a
non-trivial refractive index $n(\w)$ \cite{Shore:1995fz,Shore:2003jx,
Shore:2003zc,Shore:2007um}. 
Moreover, as originally found by
Drummond and Hathrell \cite{Drummond:1979pp}, the low-frequency phase
velocity arising from this effect may be super-luminal, which standard
dispersion relations would imply is incompatible with causality.

In a series of papers \cite{Hollowood:2007kt,Hollowood:2007ku,
Hollowood:2008kq,Hollowood:2009qz,Hollowood:2010bd},
we have developed the theory of photon propagation in QED in curved 
spacetime and calculated the full frequency dependence of the
refractive index in terms of a geometric quantity, the Van
Vleck-Morette determinant, which characterises the null geodesic
congruence around the classical trajectory. The important role of the
Penrose limit \cite{Penrose,Blau:2004yi,Blau:2006ar}
in identifying the salient features of the background
geometry was identified and exploited to develop a rich phenomenology
covering photon propagation in a wide class of spacetimes. In
particular, it was shown how the curved spacetime geometry induces a
novel analytic structure for the Green functions and refractive index
which modifies fundamental properties of QFT and S-matrix
theory such as hermitian analyticity and the Kramers-Kronig dispersion
relation \cite{Hollowood:2008kq}. 
This improved understanding of analyticity in curved
spacetime allows a reconciliation of the apparent paradox between
low-frequency super-luminal motion and causality.

While this work resolved the problem of how causality is realised in
QFT in curved spacetime, it led to a further apparent paradox, this
time with unitarity. It was found that in certain backgrounds,
including those associated with black holes, the 
imaginary part of the refractive index can be negative
\cite{Hollowood:2008kq,Hollowood:2009qz}. This is in
marked contrast to a conventional optical medium where $\IM 
n(\w)$ is always positive, corresponding to scattering of photons from 
the beam and a reduction in the field amplitude. A negative $\IM 
n(\w)$ would imply gain, with energy pumped in from an external
source. In quantum field theory in flat space, the optical theorem 
relates $\IM  n(\w)$ to the rate of production of real $e^+ e^-$
pairs and therefore vanishes. Even if we considered an off-shell ``photon" which is
above the pair production threshold, $\IM  n(\w)$ would still be
manifestly positive. In contrast, in curved spacetime we find examples
where $\IM  n(\w)$ can be negative, or non-vanishing and positive 
below the $e^+ e^-$ threshold. Clearly, understanding these phenomena 
and reconciling them with unitarity requires a careful reformulation
of the optical theorem in curved spacetime. 

For instance, for QED in the weak curvature expansion we find that the
refractive index matrix takes the form
\EQ{
n_{ij}(u;\w)=\delta_{ij}-&\frac{3\alpha}{180\pi m^2}\big(13R_{uu}(u)\delta_{ij}-4
R_{iuju}(u)\big)\\ &-\frac{i\alpha \omega}{1260\pi m^4}\big(25\dot R_{uu}(u)\delta_{ij}-6
\dot R_{iuju}(u)\big)+\cdots\ ,
\label{zaz}
}
where $u$ is the affine parameter along the photon ray and $i,j=1,2$
are the two transverse space-like directions along which the
polarization vector points. $R_{iuju}$ are components of the Riemann
tensor and $R_{uu}=R^i{}_{uiu}$.   
The first correction in \eqref{zaz} is the original Drummond-Hathrell result
\cite{Drummond:1979pp}, which can lead to either a sub- or super-luminal low-frequency phase
velocity. The second term, however, gives a non-vanishing contribution
to $\IM n(\omega)$
and depending on whether the derivatives $\dot R_{iuju}(u)$ are
positive or negative along 
the photon trajectory can lead to an amplification or attenuation 
of the amplitude.

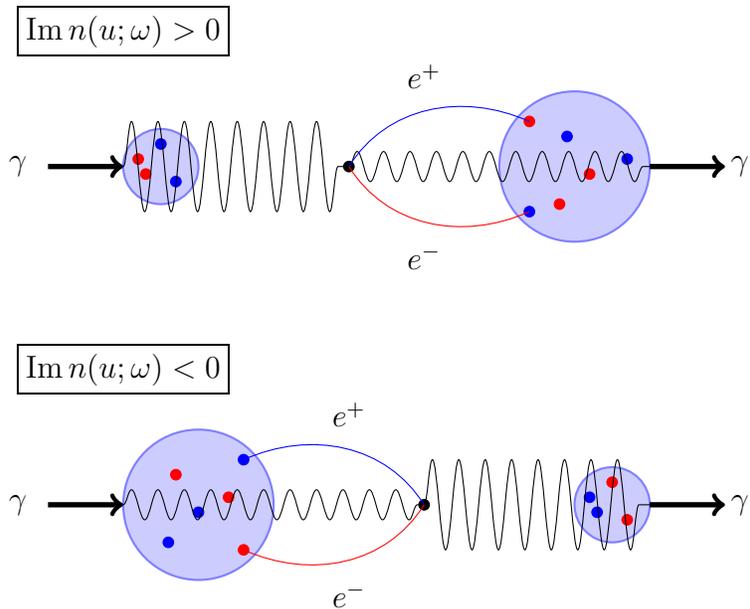
\begin{figure}[t]
\begin{center}
\begin{tikzpicture}[scale=1] 
\node at (-1,1.8) (i5) {$\boxed{\IM n(u;\omega)<0}$};
\draw [draw=blue!50,fill=blue!20,thick] (0,0) circle (1cm);
\draw [draw=blue!50,fill=blue!20,thick] (5.5,0) circle (0.5cm);
\draw [->,line width=2pt] (-2,0) -- (-1,0);
\draw [->,line width=2pt] (6,0) -- (7,0);
\filldraw[black] (3,0) circle (2pt);
\filldraw[blue] (0.6,0.6) circle (2pt);
\filldraw[red] (0.6,-0.6) circle (2pt);
\filldraw[blue] (-0.4,-0.5) circle (2pt);
\filldraw[red] (-0.3,0.4) circle (2pt);
\filldraw[blue] (0,-0.1) circle (2pt);
\filldraw[red] (0.4,0.1) circle (2pt);
\filldraw[blue] (5.2,0.1) circle (2pt);
\filldraw[red] (5.7,-0.2) circle (2pt);
\filldraw[blue] (5.3,-0.1) circle (2pt);
\filldraw[red] (5.5,0.3) circle (2pt);
\draw[decorate,decoration={snake,amplitude=0.2cm}] (-1,0) -- (3,0);
\draw[decorate,decoration={snake,amplitude=0.6cm}] (3,0) -- (6,0);
\draw [-,color=blue] (0.6,0.6) .. controls (1.5,1) and (2.5,0.8) .. (3,0);
\draw [-,color=red] (0.6,-0.6) .. controls (1.5,-1) and (2.5,-0.8) .. (3,0);
\node at (2,1.2) (i1) {$e^+$};
\node at (2,-1.2) (i2) {$e^-$};
\node at (-2.4,0) (i3) {$\gamma$};
\node at (7.2,0) (i4) {$\gamma$};
\begin{scope}[xshift=-0.5cm,yshift=4.5cm]
\node at (-0.5,1.8) (i5) {$\boxed{\IM n(u;\omega)>0}$};
\draw [draw=blue!50,fill=blue!20,thick] (0,0) circle (0.5cm);
\draw [draw=blue!50,fill=blue!20,thick] (5.5,0) circle (1cm);
\draw [->,line width=2pt] (-1.5,0) -- (-0.5,0);
\draw [->,line width=2pt] (6.5,0) -- (7.5,0);
\filldraw[black] (2.5,0) circle (2pt);
\filldraw[red] (4.9,0.6) circle (2pt);
\filldraw[blue] (4.9,-0.6) circle (2pt);
\filldraw[red] (5.3,-0.5) circle (2pt);
\filldraw[blue] (5.4,0.4) circle (2pt);
\filldraw[red] (5.7,-0.1) circle (2pt);
\filldraw[blue] (6.2,0.1) circle (2pt);
\filldraw[red] (-0.3,0.1) circle (2pt);
\filldraw[blue] (0.2,-0.2) circle (2pt);
\filldraw[red] (-0.2,-0.1) circle (2pt);
\filldraw[blue] (0,0.3) circle (2pt);
\draw[decorate,decoration={snake,amplitude=0.6cm}] (-0.5,0) -- (2.5,0);
\draw[decorate,decoration={snake,amplitude=0.2cm}] (2.5,0) -- (6.5,0);
\draw [-,color=blue] (4.9,0.6) .. controls (4,1) and (3,0.8) .. (2.5,0);
\draw [-,color=red] (4.9,-0.6) .. controls (4,-1) and (3,-0.8) .. (2.5,0);
\node at (3.5,1.2) (i1) {$e^+$};
\node at (3.5,-1.2) (i2) {$e^-$};
\node at (-1.9,0) (i3) {$\gamma$};
\node at (7.7,0) (i4) {$\gamma$};
\end{scope}
\end{tikzpicture}  
\caption{\small Heuristic pictures which illustrate the behaviour of the renormalized photon made up of modes of the bare field (the wave) and virtual cloud of $e^+e^-$ pairs. In the top diagram there are increasing Riemann
tensor components (made precise in \eqref{zaz})
along the null coordinate $u$ of the photon's propagation leading to an increase in 
virtual $e^+e^-$ pairs and an attenuation of the photon modes (dressing). In the bottom diagram there are decreasing Riemann tensor components leading to the opposite effect and an amplification of the photon modes (undressing).}
\label{f3}
\end{center}
\end{figure}

In this paper, we investigate the realisation of unitarity in photon
propagation in curved spacetime from the point-of-view of an
initial-value problem. Light-front evolution of renormalized quantum
fields from an initial null surface is studied for a variety of field
theories and initial conditions and the physical mechanisms
responsible for the observed amplification or attenuation of the
amplitude are explored. The centrepiece is the formulation of a
generalisation of the optical theorem to curved spacetime. 

The key point is that in curved spacetime, the fundamental optical
theorem relating a decay probability or cross-section and the
imaginary part of a Green function must be viewed as a {\it global\/}
result, integrated along the whole history of the decaying
particles. The corresponding {\it local\/} identity, which is the
standard theorem in flat spacetime, loses the essential positivity
needed for the identification with a probability. This vital
difference between the local and global identities arises only in
curved spacetime because of the lack, in general, of translation
invariance along the particle trajectory.

As a result, the amplitude of a renormalized quantum field propagating
through curved spacetime may be locally amplified, although integrated
along its whole past trajectory there must be a net attenuation. This
will be interpreted as a real-time dressing and undressing of the
field by its cloud of virtual pairs. Locally the curvature can undress
the field, resulting a reduced screening and an amplification of the
renormalized amplitude.  Along its entire trajectory, however, the net
dressing must remain positive. A heuristic picture of how dressing or 
undressing occurs is illustrated in Fig.\ref{f3}. 

These ideas are made precise in section 4, where an exact mathematical
formulation of the optical theorem in curved spacetime is
presented. The general theory is then illustrated with a detailed
exploration of the initial value problem and field evolution in a
variety of special cases, involving both different quantum field
theories and different initial conditions, corresponding to a
field in both the bare and dressed state.

The paper is presented as follows. We have already shown in previous
work \cite{Hollowood:2008kq,Hollowood:2009qz} 
that for massless photon propagation through curved spacetime,
the essential features of the curvature are captured by the Penrose
plane-wave limit. We therefore restrict ourselves here to the study of
quantum field theory in plane-wave backgrounds, although we shall
consider both massless and massive theories. We therefore start in
section 2 with a brief review of the nature of the Penrose limit and
the geometry of plane waves. Section 3 sets up the initial-value
problem and describes the different, physically distinct, initial
conditions that we will consider. An essential feature is the
Schwinger-Keldysh formalism
\cite{Schwinger,Mahanthappa,Keldysh,Weinberg:2005vy}
formalism for studying real-time, non-equilibrium
phenomena in QFT and the relation of the causal field equations to the
Feynman vacuum polarization or self-energy is carefully explained.
The general solution of the field equations for plane-wave backgrounds
is then given and a re-summation technique, the dynamical
renormalization group \cite{Boyanovsky:2003ui}, is applied. 
The interpretation of attenuation
and amplification of the amplitude as dressing and undressing of the
renormalized field is introduced. 

The core of the paper is section 4, where the optical theorem in
curved spacetime is presented. The relation of the local and
integrated versions of the identity are explained in detail and the
issue of unitarity and the positivity of imaginary parts of the Green
functions and refractive index is addressed. The realisation of the
optical theorem for different classes of initial conditions is also
explained.

The remainder of the paper illustrates these ideas for several quantum
field theories -- massless and massive scalar $A\phi^2$ theory in both
$d=4$ and 6 dimensions and QED. This allows us to address some 
special issues related to renormalization and also to discuss mass 
thresholds for decaying particles.

First, we study real-time renormalization phenomena in flat spacetime,
using a Laplace transform formalism, to develop insights into the
initial-value problem and the role of initial conditions. This
technique is then applied to QFT in symmetric 
(i.e.~Cahen-Wallach \cite{Cahen})
plane wave spacetimes, which are translation invariant in the
light-cone coordinate which serves as the affine parameter along the
classical null geodesics. This reveals many interesting phenomena,
including in some cases the below-threshold decay of the field $A$
into $\phi$ pairs with a rate which is non-perturbative in the
curvature. The consistency of all these results with our improved
understanding of the optical theorem and the precise nature of the
positivity constraint implied by unitarity is checked.

Finally, we consider non-translationally-invariant backgrounds, in
particular the homogeneous plane waves which arise as the Penrose
limits of null geodesics in cosmological or black hole spacetimes 
\cite{Blau:2004yi,Hollowood:2009qz}. In particular, we consider in detail the
tidal effects on a renormalized photon field as it approaches a
singularity and show explicitly how the photon can become undressed 
along such a trajectory.

The insights and techniques developed in this paper should also have
applications in a wider field. The methods of section 6 have already
been used \cite{Boyanovsky:2003ui}
to study real-time relaxation problems in quantum field theory, which
are of relevance to inflationary and early-universe cosmology where
the non-equilibrium evolution of scalar fields is important
(see e.g.~\cite{Boyanovsky:1996ab,Boyanovsky:2004gq,
Boyanovsky:2004ph}).
Plane waves and Cahen-Wallach spaces are also important as 
backgrounds in string theory and have an important role in the
AdS/CFT correspondence (see e.g.~\cite{Blau:2001ne,Blau:2002dy,
Plefka:2003nb} and references therein). The study of QFT on plane waves has been pioneered in  \cite{Gibbons:1975jb} and some more recent work motivated by string theory appears in  \cite{Brecher:2002bw,Marolf:2002bx}. Our work will answer some of the questions regarding the stability of interacting QFT in certain plane wave backgrounds posed in these latter two papers. 

\section{Penrose Limit and QFT on Plane Wave Spacetimes}

To begin, we review some essential features of the geometry of plane
wave spacetimes and how they arise in connection with photon
propagation as Penrose limits.  For a complete discussion, see our
earlier papers, especially
refs.\cite{Hollowood:2008kq,Hollowood:2009qz}.  
Some preliminary results on the
formulation of QFT on a plane wave background are then described.

\subsection{Penrose limit and the geometry of plane wave spacetimes}

To illustrate how the Penrose limit arises and motivate our
specialisation to plane wave spacetimes, consider first an interacting
scalar field theory with an $A\phi^2$ interaction. The $A$ field
(playing the role of the photon in QED) is massless while the $\phi$
field (the electron) has mass $m$. Including the one-loop vacuum
polarization, the equation of motion for the $A$ field in the presence
of a source $J$ is
\EQ{
\square\, A(x)- \int d^4x'\,
\sqrt{g(x')}\,\Pi(x,x')A(x')=J(x)\ ,
\label{aa}
}

We will define this precisely in the next section as an initial value
problem, using the Schwinger-Keldysh formalism to identify 
$\Pi(x,x')$ as the retarded vacuum polarization and $A(x)$ as the
``{\it in-in}" VEV of the quantum field. 

We consider two limits: ~(i)~ $\w \gg  L_\RR^{-1}$ ~ and 
~(ii)~ $L_\RR \gg \l_c$, ~where $\w$ is the frequency of the wave
solution for $A$, $\l_c = \frac1m$ is the Compton wavelength of the
``electron", and  $L_\RR$ is a typical curvature scale. The first is
the geometric optics condition, which allows the description of the
propagation of the free $A$ field in terms of individual rays (``photon
trajectories") following null geodesics.  We then introduce Penrose,
or adapted, coordinates $(u,V,x^a)$, $a=1,2$, in which $u$ is the affine
parameter along the null geodesic, $V$ is the associated null
coordinate, and $x^a$ are transverse spatial coordinates.
The free wave solution is then
\EQ{
A(x)=g(x)^{-1/4}\, e^{i\omega \vartheta(x)}\ ,
\label{ab}
}
where $g=-\det\, g_{\mu\nu}$ and $k^\mu=\partial^\mu \vartheta(x)$ 
is the tangent vector to a null congruence $g(k,k)=0$. Along the 
geodesics $\vartheta(x) = V$ is constant while the affine parameter 
$u$ varies; $V$ specifies the individual geodesic in the congruence.

In our previous work \cite{Hollowood:2007kt,Hollowood:2007ku,
Hollowood:2008kq,Hollowood:2009qz},
we have shown, using both worldline and conventional QFT methods,
how the second limit allows us to simplify the evaluation of the
vacuum polarization corrections to \eqref{ab}. To leading order in 
$\RR/m^2$ (where $\RR \sim  L_\RR^{-2}$ is a typical curvature scale),
the vacuum polarization is determined by geodesic fluctuations 
around the photon trajectory; in turn, this is determined entirely
by geometric quantities, specifically the Van Vleck-Morette
determinant, which describe the geometry of geodesic deviation.
But this is precisely the property of the geometry encoded in the
Penrose limit, in which the full spacetime metric in a tubular
neighbourhood of the chosen null geodesic $\c$ (with $V=x^a=0$)
is approximated by 
\EQ{
ds^2=2du\,dV+C_{ab}(u)dx^a\,dx^b\ .
\label{ac}
}

Eq.\eqref{ac} describes a gravitational plane wave, in Rosen
coordinates. An alternative form, in terms of Brinkmann coordinates,
is
\EQ{
ds^2=2du\,dv-h_{ij}(u)z^i z^j\,du^2
+dz^i\,dz^i \ ,
\label{ad}
}
where the profile function $h_{ij}(u) = R^i{}_{uju}(u)$, the
components of the Riemann tensor which occur in the Jacobi equation
describing geodesic deviation. An elegant and comprehensive
review of the Penrose limit can be found in
ref.\cite{Blau:2004yi,Blau:2006ar}, 
which explains the scaling by which the plane wave metric \eqref{ad} 
arises as the leading term in an expansion of the full metric in
Fermi null coordinates around $\c$. The relation with vacuum
polarization and photon propagation is described in full detail in our
earlier work, especially ref.\cite{Hollowood:2009qz}, which contains a 
complete review of the geometry and notation used here.\footnote{Except that 
  here we are
  following ref.\cite{Hollowood:2010bd} and using $x^a$ and $z^i$ 
  for the transverse
  coordinates in Rosen and Brinkmann respectively, rather than $Y^a$
  and $y^i$, and have changed the sign convention for $V$, $v$.}

While the relation with geodesic deviation and therefore the
motivation for using the Penrose limit is most readily seen in terms
of Brinkmann coordinates, the field-theoretic calculations are best
expressed using Rosen coordinates. The relation is given in
terms of a zweibein $E^i{}_a(u)$ for the transverse coordinates
in \eqref{ac} and \eqref{ad}:
\EQ{
C_{ab}(u) = E^i{}_a(u) \d_{ij} E^j{}_b(u) \ ,
\label{ae}
}
where $E^i{}_a$ is obtained by solving the differential equation\footnote{Note that $i,j,\ldots$  indices are raised and lowered using $\delta_{ij}$ while $a,b,\ldots$ are raised and lowered with $C_{ab}$.}
\EQ{
\ddot E_{ia}+h_{ij}E^j{}_a=0 \ .
\label{af}
}
These equations are solved subject to the requirement that $\Omega_{ij}=\dot E_{ia} E_j{}^a$ is symmetric.
The coordinates are related by 
\SP{
z^i&=E^i{}_a(u) x^a\ ,\\
v&=V + \tfrac12\dot E_{ia}(u)E^i{}_b(u)x^ax^b \ ,
\label{ag}
}
with $u$ common to both sets. The zweibein is not
unique and this property is shared by the Rosen coordinates which
are specially adapted to describing the null congruence around $\c$.

The Van Vleck-Morette determinant is defined from the geodesic interval
\EQ{
\sigma(x,x')=\frac12\int_0^1 d\tau\, 
g_{\mu\nu}(x)\dot x^\mu\dot x^\nu\ ,
\label{ah}
}
where $x^\mu=x^\mu(\tau)$ is the geodesic joining $x=x(0)$ and
$x'=x(1)$ and is given by
\EQ{
\Delta(x,x')=-\frac1{\sqrt{g(x)g(x')}}
\det\,\frac{\partial^2\sigma(x,x')}{\partial x^\mu\partial 
x^{\prime\nu}}\ .
\label{ai}
}
For a plane wave, with $g= g(u)$, the VVM determinant is a function
only of the null coordinate $u$, that is $\Delta=\Delta(u,u')$.

It is convenient when describing the mode functions on plane waves later in \eqref{ao} to introduce 
the $2\times 2$ matrix $\psi^{ab}(u)$ as the indefinite integral
\SP{
\psi^{ab}(u)=\int^u du\,\Big[C(u)^{-1}\Big]^{ab}\ ,
\label{aj}
}
together with $\psi^{ab}(u,u') = \psi^{ab}(u) - \psi^{ab}(u')$. This is related to
the VVM determinant by 
\EQ{
\Delta(u,u')=\frac1{\sqrt{g(u)g(u')}}\cdot
\frac{(u-u')^2}{\det\,\psi(u,u')}\ .
\label{ak}
}
The geodesic interval for a plane wave spacetime then takes the form
\cite{Hollowood:2008kq}
\EQ{
\sigma(x,x')=(u-u')(V-V')+
\frac{u-u'}2\psi(u,u')^{-1}_{ab}(x-x')^a(x-x')^b\ .
\label{al}
}

An equivalent way to construct and interpret the VVM determinant is by thinking of the behaviour of a spray of geodesics which emanate from the point $(u',v=0,z^i=0)$ in Brinkmann coordinates. In the space-like directions the geodesics satisfy
\EQ{
\frac{dz^i}{du^2}+h_{ij}(u)z^j=0\ .
\label{p11}
}
If the initial conditions of the spray are chosen as $z^i(u')=0$ and $dz^i(u')/du=\delta_{ij}$ then we write the solution as $z^i(u)=A_{ij}(u,u')$. The VVM matrix is then simply
\EQ{
\Delta_{ij}(u,u')=(u-u')[A^{-1}(u,u')]_{ji}\ ,
\label{pwl}
}
whose determinant is $\Delta(u,u')$.

We will be especially interested later (see section 6) in the special class
of symmetric plane waves, or Cahen-Wallach spaces. 
Here, the profile function $h_{ij}(u)$ in the Brinkmann metric
\eqref{ad} is $u$-independent. It is convenient to choose 
transverse coordinates such that $h_{ij}=\sigma^2_i\delta_{ij}$
is diagonal. Then, in Rosen coordinates
\EQ{
C_{ab}(u)=\cos^2(\sigma_au+c_a)\delta_{ab}\ .
\label{ala}
}
Notice that $\sigma_i$ can either be real or purely imaginary and $c_a$ are constants.\footnote{If one
wants to impose the null energy condition then at least one of the
$\sigma_i$ is real, say $\sigma_1$, and futhermore
$\sigma_1\geq|\sigma_2|$. However, since we are taking the background
metric to be fixed and non-dynamical there is not really a physical 
reason to impose this
condition. Later we shall look at an example with both $\sigma_i$
imaginary since it leads to a particularly simple analytic
structure. In fact the case with $\sigma_1=\sigma_2$ and both
  imaginary arises as the Penrose limit of the product geometry of
  three-dimensional de Sitter space with the real line, or
  circle. This is relevant to a version of the AdS/CFT
  correspondence \cite{Hutasoit:2009xy}.}
The VVM determinant for a symmetric plane wave is
\EQ{
\Delta(u,u')= \prod_{i=1}^2 
\frac{\sigma_i(u-u')}{\sin(\sigma_i(u-u'))}\ .
\label{alb}
}

Symmetric plane waves are one class of homogeneous plane waves.
Another is the class of singular homogeneous plane waves, where the
profile function is $h_{ij} =
\frac{1}{4u^2}\bigl(1-\alpha_i^2\bigr)\delta_{ij}$. 
Like the symmetric plane waves, which have the translation symmetry
$u\rightarrow u + c$, the singular homogeneous plane waves also have
an enhanced symmetry, since the metric is invariant under the scaling
$u\rightarrow  \lambda u$, $v\rightarrow \lambda^{-1} v$.
As a result, the VVM determinant $\Delta(u,u')$ is a function of the
single variable $u'/u$. Explicitly,
\EQ{
\Delta(u,u')=\prod_{i=1}^2\frac{\alpha_i  (u-u')(uu')^{\tfrac{\alpha_i-1}2}}
{u^{\alpha_i}-u^{\prime\alpha_i}}\ .
\label{alc}
}
This class of plane waves arises as the Penrose limit of certain
spacetimes with singularities, notably black holes and some
Robertson-Walker spacetimes, and will be considered in section 7.

\subsection{QFT on a plane wave spacetime}

To study QFT on a background spacetime, we would normally construct an
appropriate basis of mode functions describing propagation from an
initial Cauchy surface. Plane waves, however, do not admit Cauchy
surfaces. Nevertheless, as described in ref.\cite{Gibbons:1975jb}, we can
still set up an initial value problem in a light-front formalism,
choosing the null coordinate $u$ to play the role of the time
coordinate in the conventional case. 
While the surfaces $u=\text{constant}$ are not genuine Cauchy
surfaces the initial value problem is still well defined when suitable
boundary conditions in the transverse directions are specified. 

A special feature of plane waves spacetimes is that the solutions of the 
massive Klein-Gordon equation 
\EQ{
\big(\square-m^2\big)\phi(x)=0\ ,
\label{am}
}
are WKB exact \cite{Gibbons:1975jb}.
We can write a basis for the solutions in Rosen coordinates as 
\EQ{
\phi(x)=e^{-\tfrac{im^2u}{2\omega}}\Phi_p(x)\ ,
\label{an}
}
in terms of the modes
\EQ{
\Phi_p(x)=g(u)^{-1/4}\exp\left[i\omega V+ip_ax^a
-\frac i{2\omega}\psi^{ab}(u)p_ap_b\right]\ ,
\label{ao}
}
where $\psi^{ab}(u)$ is defined in \eqref{aj}.  The symmetries of the metric allow
the definition of a 3-momentum $p=(\omega,p_a)$, where $\omega$ is a
null component and the $p_a$ are spacelike.\footnote{Notice 
that in general these modes have singularities at points where the
metric is degenerate. For example, for a Cahen-Wallach space these occur when the cosine functions in \eqref{ala} vanish.
However, these are only coordinate singularities.}

The modes are classified as having positive or negative frequency with 
respect to evolution in the coordinate $u$ according to whether 
$\omega>0$ or $\omega<0$, respectively. On the null surfaces of 
constant $u$, there is a Klein-Gordon inner-product
\EQ{
\big(\Phi,\Phi')=-i\int
dV\,d^2x\,\sqrt{g(u)}\,\, \Phi^{\prime *}
\overleftrightarrow{\partial_V} \Phi \ ,
\label{ap}
}
with respect to which the modes satisfy the orthonormality property
\EQ{
\big(\Phi_{p},\Phi_{p'}\big)
=2\omega (2\pi)^3 \delta(\omega-\omega')\delta^{(2)}(p_a-p'_a) \ .
\label{aq}
}

At this point it is important to recognise, as shown by
Gibbons \cite{Gibbons:1975jb}, that for plane waves there is no
particle creation via the usual curved spacetime Bogoliubov 
transformation. The reason is that since $\partial_V$ is a
Killing vector, a positive/negative frequency mode $e^{\pm i\omega V}$ will
remain a positive/negative frequency mode. As a consequence the vacuum
state at some inital time remains empty of particles at later times.
This plays an important role in the next section in simplifying the 
application of the Schwinger-Keldysh analysis. It also means that 
the type of particle creation arising through a non-trivial Bogoliubov
relation between {\it in} and {\it out} vacua cannot be the explanation
for the increase in amplitude of the photon field observed in some
backgrounds.

The Green functions of \eqref{am} play an important role in the
analysis that follows. As usual, we define the Wightman functions
\EQ{
G_+(x,x') = \langle0|\phi(x) \phi(x')|0\rangle\ , ~~~~~~~~~
G_-(x,x') = \langle0|\phi(x')\phi(x)|0\rangle\ ,
\label{ar}
}
with mode expansions:
\EQ{
G_{\pm} = \pm {1\over (2\pi)^3} \int{d\w\over2\w} \theta(\pm\omega)
\int d^2p ~\Phi_p(x) \Phi_p(x')^\dagger e^{-\tfrac{im^2(u-u')}{2\omega}}\ .
\label{as}
}
The ``Feynman" Green function, adapting the definition to our choice of
null coordinates, is then
\EQ{
i G_\text{F}(x,x') = \langle0|T_u \,\phi(x)\phi(x')|0\rangle =
\theta(u-u') G_+(x,x') + \theta(u'-u) G_-(x,x') \ ,
\label{at}
}
where $T_u$ denotes $u$-ordering. A similar definition holds for the
anti $u$-ordered Dyson function:
\EQ{
i G_\text{D}(x,x') = \langle0|\bar T_u\, \phi(x)\phi(x')|0\rangle =
\theta(u'-u) G_+(x,x') = \theta(u-u') G_-(x,x') \ .
\label{at2}
}
Retarded and advanced Green functions, with support in the forward and
backward lightcones respectively, are also given in terms of the
Wightman functions:
\EQ{G_{\rm ret}(x,x') = - \theta(u-u') G(x,x') \ ,~~~~~~~~~~
G_{\rm adv}(x,x') = \theta(u'-u) G(x,x') \ ,
\label{au}
}
where the Pauli-Jordan, or Schwinger, function is given by the
commutator,
\EQ{
iG(x,x') = \langle0| [\phi(x), \phi(x')]|0\rangle =
G_+(x,x') - G_-(x,x')\ .
\label{av}
}

The Green functions can be written in a ``proper-time" representation 
using the expression \eqref{as} in terms of the mode functions. 
Performing the Gaussian integral over the transverse momentum $p^a$
using \eqref{aj} and \eqref{al}, and using the substitution
$T = {u-u'\over2\w}$ in the integration over $\w$, we find
\EQ{
G_+(x,x') = {i^{(1-{d\over2})}\over(4\pi)^{d\over2}} 
\sqrt{\det\D(x,x')} \int_0^\infty {dT\over T^{d\over2}}
e^{\tfrac{i\sigma(x,x')}{2T}} e^{-i m^2 T} \ ,
\label{aw}
}
for $u-u' > 0$. Note that, for future use, we have quoted the result
for $d-2$ transverse dimensions. Similar results for $G_+(x,x')$
for $u-u'<0$, and for $G_-(x,x')$, follow from the identities
$G_-(x,x') = G_+(x',x) = G_+(x,x')^*$. 

The proper-time representations for all the other Green functions
follow straightforwardly. For example,
\EQ{
iG_\text{F}(x,x') = {i^{(1-{d\over2})}\over(4\pi)^{d\over2}} 
\sqrt{\det\D(x,x')} \int_0^\infty {dT\over T^{d\over2}}
e^{\tfrac{i\sigma(x,x')}{2T}} e^{-i m^2 T}\ ,
\label{ax}
}
for all $u-u'$.

\section{Light-Front Evolution of Renormalized Quantum Fields}

In this section, we set up and study the initial-value problem
describing the evolution of a renormalized quantum field in a
plane-wave background spacetime, incorporating the effect of vacuum
polarization.

\subsection{The initial value problem}

The techniques for studying real-time non-equilibrium phenomena in
quantum field theory were established originally by Schwinger 
and Keldysh \cite{Schwinger,Mahanthappa,Keldysh,Weinberg:2005vy}.  
The essential idea is to develop an
equation of motion, obtained from a suitably defined 1PI
effective action, which describes the time evolution of the ``{\it
  in-in}" vacuum expectation value ${}_J\langle0_{in}|A|0_{in}\rangle{}_J$ of
the quantum field in the presence of a source, which can be used to
engineer appropriate initial conditions.  In our case, we will
consider evolution in the light-cone variable $u$ from an initial null 
surface, but the essential formalism is unchanged.

This should be contrasted with the more familiar 1PI effective action
derived using the conventional Feynman rules, for which the
corresponding equation of motion describes the ``{\it in-out}" VEV 
${}_J\langle0_{out}|A|0_{in}\rangle{}_J$.  The distinction is clearly
important when the vacua do not coincide, {\it i.e.}~$|0_{out}\rangle 
\neq |0_{in}\rangle$, and there is a non-trivial Bogoliubov
transformation between the states and associated particle creation.
As already emphasised \cite{Gibbons:1975jb}, this is not the case for
plane-wave backgrounds, so we expect our analysis to be independent of
which formalism we choose. We now verify this explicitly, justifying 
{\it a posteriori\/} the treatment of the refractive index problem in
curved spacetime in our earlier work.

The essential features of the Schwinger-Keldysh, or ``in-in", or
``closed-time-path", formalism are elegantly summarised in
refs.\cite{Jordan:1986ug,Calzetta:1986ey,Calzetta:1989vs} 
and we refer to these papers for further
details. For simplicity, we describe the formalism first for the QFT
of a single scalar field $\phi$. The key idea is to let the {\it in}
vacuum evolve independently under two different sources $J^{\pm}(x)$,
comparing in the far future, in the {\it out} region.  This defines a
generating functional
\EQ{
Z[J^+,J^-] &= \exp iW[J^+,J^-] \\ &= {}_{J^-} \langle
0_{in}|0_{in}\rangle_{J^+}
~\equiv~ \sum_{\psi}  {}_{J^-} \langle0_{in}|\psi_{out}\rangle 
\langle \psi_{out}|0_{in}\rangle_{J^+} \ .
\label{ba}
}
The ``{\it in-in}" expectation value of the quantum field $\phi(x)$ 
in the presence of a physical source $J(x)$ is then given by
the derivative with respect to either source, evaluated with the
sources set equal to $J$,
\EQ{
{}_J\langle0_{in}|\phi|0_{in}\rangle{}_J = \pm {\d W[J^+, J^-] \over
\d J^{\pm}}\bigg|_{J^+ = J^- = J} \ .
\label{bb}
}

The generating functional has a path integral representation in terms
of two field variables $\phi^{\pm}(x)$ as follows:
\EQ{
Z[J^+,J^-] = \int{\cal D}\phi^+ \int{\cal D}\phi^-  
\exp i \Bigl( S[\phi^+] - S[\phi^-] + J^+ \phi^+  - J^- \phi^- \Bigr)
\ ,
\label{bc}
}
where it is understood that the integrals are over field
configurations which coincide on the {\it out\/} hypersurface corresponding
to $|\psi_{out}\rangle$.  Evaluating in the free theory gives
the expression
\cite{Jordan:1986ug,Calzetta:1986ey,Calzetta:1989vs}
\EQ{
Z[J^+,J^-] &= \exp - \frac i2 \int d^4x\, \sqrt{g} \int d^4 x'\, \sqrt{g'}
\Bigl(J^+(x) G_\text{F}(x,x') J^+(x')\\ &\qquad  +  iJ^+(x) G_-(x,x')
J^-(x') + 
i J^-(x) G_+(x,x') J^+(x')  \\&\qquad\qquad+  J^-(x) G_\text{D}(x,x') J^-(x') \Bigr) \ ,
\label{bd}
}
so different derivatives of $Z$ with respect to the two sources yield
the full set $G_\text{F}$, $G_\text{D}$, $G_+$ and $G_-$ of Green functions.

Expressions \eqref{bc} and \eqref{bd} can then be used in perturbation
theory as the basis of an extended diagrammatic expansion, with
propagators $G_{++} = -G_\text{F}$,  $G_{+-} = i G_-$,  $G_{-+} = i G_+$,
$G_{--} = -G_\text{D}$ linking vertices which are ``all +" or ``all $-$", with
opposite signs of the coupling.  In turn, this allows the construction
of a 1PI effective action $\Gamma[\phi^+,\phi^-]$ by a Legendre transform 
in the usual way, written in terms of fields defined by
$\phi^{\pm} = \pm {\d W[J^+, J^-] \over \d J^{\pm}}$. 

For our problem, we initially consider the equation of motion for a
massless scalar field $A$  interacting via an $eA\phi^2$ interaction
with a scalar field $\phi$ of mass $m$ with Green functions described in
section 2.  It is straightforward to see 
that the required equation for the {\it in-in} expectation value 
$A(x) = {}_J\langle0_{in}|A(x)|0_{in}\rangle{}_J$ is
\EQ{
{\d\Gamma\over\d A^+(x)}\biggl|_{A^+ = A^- = A}  = -J
\label{be}
}
(or similarly with $A^+$ and $A^-$ interchanged), where 
$\Gamma[A^+, A^-]$ is the one-loop effective action with quadratic
part
\EQ{
\Gamma[A^+,A^-] ~&=~ -{1\over2} \int d^4 x \sqrt{g} 
\Bigl( A^+(x) \square A^+(x)  - A^-(x)  \square A^-(x) \Bigr)  \\
&~~~~-{ie^2\over4} \int d^4x \sqrt{g} \int d^4 x' \sqrt{g'} \Bigl(
A^+(x) G_\text{F}(x,x')^2 A^+(x') \\ &\qquad+ A^+(x) G_-(x,x')^2
A^-(x') + 
A^-(x) G_+(x,x')^2 A^+(x')\\ & \qquad\qquad~~~~~
+  A^- G_\text{D}(x,x')^2 A^-(x') \Bigr) \ .
\label{bf}
}
This gives an equation of the form \eqref{aa}, with the vacuum
polarization
\EQ{
\Pi_\text{SK}(x,x') = {i e^2\over2} \Bigl( G_\text{F}(x,x')^2 + G_-(x,x')^2\Bigr) \ .
\label{bg}
}

From the definitions \eqref{ar}, \eqref{at} above, we now see that 
\EQ{
\Pi_\text{SK}(x,x') = - {i e^2\over2} \theta(u-u') \Bigl(
G_+(x,x')^2 - G_-(x,x')^2 \Bigr) \ ,
\label{bh}
}
with support only for $u' < u$.  Moreover, given that the commutator
$[\phi(x),\phi(x')]$ is a $c$-number, it follows readily that 
\EQ{
\Pi_\text{SK}(x,x') = -{ie^2\over4} \theta(u-u') 
\langle 0_{in}|[\phi(x)^2,\phi(x')^2]|0_{in}\rangle \ .
\label{bi}
}
Since the commutator vanishes for spacelike separated points, it
follows that $\Pi_\text{SK}(x,x')$ vanishes for $x$ outside the 
forward light-cone of $x'$. This confirms that \eqref{aa},
with the vacuum polarization taken as $\Pi_\text{SK}(x,x')$, is 
a causal, albeit non-local, equation of motion for the 
{\it in-in\/} VEV $A(x)$ of the field.
In contrast, the usual {\it in-out\/} formalism with the Feynman
vacuum polarization
\EQ{
\Pi_\text{F}(x,x') = { i e^2\over2} G_\text{F}(x,x')^2 = -{i e^2\over2} 
\Bigl(\theta(u-u') G_+(x,x')^2 + \theta(u'-u) G_-(x,x')^2 \Bigr) \ ,
\label{bj}
}
is not manifestly causal.

Now, for propagation in a plane-wave background, the only 
non-trivial behaviour is in the $u$ direction and we can expand the 
solutions in terms of the basis \eqref{ao} of on-shell modes. 
Accordingly, for a massless field $A(x)$, we take
\EQ{
A(x)=\AMP(u)\Phi_p(x)
\label{bk}
}
and choose $\omega>0$, so that it is a positive frequency
solution.  So for the simplest case where the transverse momentum 
$p_a=0$, we just have (see \eqref{bb})
\EQ{
A(x)=\AMP(u)\Phi_{(\omega,0)}(x)=\AMP(u)g(u)^{-1/4}e^{i\omega V}\ .
\label{bl}
}
Notice that the amplitude factor $\AMP(u)$ here is 
allowed to be complex and the physical solution for $A(x)$
is the real part of \eqref{bl}. 

A key result now is that if $A(x)$ has postive frequency,
it follows from the form of \eqref{as} that
\EQ{
\int dV'\,G_-(x,x')^2 A(x')=0\ ,
\label{bm}
}
for $\omega>0$. This is a manifestation of $\omega$-conservation
at the vertex, in itself a consequence of the $\partial_V$ isometry 
of the plane-wave background.
Hence, for solutions of the kind \eqref{bk} we can replace $\Pi_\text{SK}$ 
in \eqref{aa} by the Feynman vacuum polarization $\Pi_\text{F}$ if we wish,
the difference vanishing by virtue of \eqref{bm}.  This is the
technical mechanism by which the isometry of the plane-wave
background, which ensures the absence of particle creation and the
identity of the {\it in} and {\it out} vacua, guarantees we recover the
same final results for ``photon" propagation and the refractive index 
whether we use the conventional Feynman or Schwinger-Keldysh
formalisms.

If we now exploit the symmetries of the plane-wave background 
to define the ``partial Fourier transform" of the vacuum polarization
with respect to the modes \eqref{ao}, we can reduce the equation of
motion to an effective one-dimensional problem.
Defining
\SP{
&(2\pi)^3 \d(\w-\w') \d^{(2)}(p_a - p_a^{\prime})~
\tilde\Pi_\text{SK}(u,u';\w,p_a) \\
&~~~~~~~~~~~~ =~~\int d^3x\sqrt{g}\,\int d^3x'\sqrt{g'}~
\Phi_p^*(x) \Pi_\text{SK}(x,x') \Phi_{p'}(x') \ .
\label{bn}
}
we can rewrite \eqref{aa} in the form
\EQ{
(-2i\omega\partial_u+M^2)\AMP(u)
+\int_{u_0}^udu'\,\tilde\Pi(u,u';\omega,p)\AMP(u') =0 \ .
\label{bo}
}
where for generality we have now introduced a mass $M$ for the $A$ field.
For later use, we have assumed here that the coupling constant 
is ``turned on" at some specified value $u_0$.  This will clarify 
the later discussion of initial value conditions and the optical
theorem; we can of course set $u_0 \rightarrow -\infty$ at any point.
The upper limit follows from the constraint $u-u' > 0$ from \eqref{bh}.

We can further simplify the expression for 
${\tilde\Pi}(u,u';\omega,p)$
by exploiting translation invariance in the {\it transverse\/} space
to write it in terms of an integral over the 
relative coordinates, $\hat V=V-V'$ and $\hat x^a=x^a-x^{\prime a}$. 
This gives simply:
\EQ{
\tilde\Pi_\text{SK}(u,u';\omega,p)=
\sqrt{g(u)g(u')}
\int d^3\hat x\,\Phi_{(\omega,p)}(x)^\dagger  \Pi_\text{SK}(x,x')
\Phi_{(\omega,p)}(x')\ .
\label{bp}
}

\vskip0.2cm
We now come to the choice of initial conditions. The simplest choice,
which we call ``Type I", is to assume the coupling is ``switched on" at
some initial value surface $u=0$ (so that $u_0=0$) and specify the
value of the field
$\AMP(0)$ on that surface. In this case, the equation of
motion describing the evolution of the field $\AMP(u)$ is simply
\EQ{
-2i\omega\dot\AMP(u)+M^2\AMP(u)
+\int_{0}^u du'\,\tilde\Pi_\text{SK}(u,u';\omega,p)\AMP(u')=0\ .
\label{bq}
}
Prior to switching on the coupling, the field simply evolves according
to the tree-level equation as $\AMP(u) = \AMP(0) 
\exp{-\tfrac{iM^2 u}{2\omega}}$.

With this choice, the field on the initial value surface $u=0$ is a 
bare field, and as we follow the evolution for $u>0$ it will become
dressed with a vacuum polarization cloud of virtual $\phi$ pairs 
according to \eqref{bq}. For $A\phi^2$ theory in $d=4$, where the only
UV divergence is absorbed by a mass renormalization, we can follow
this real-time field renormalization explicitly. We expect an initial
fall in $\AMP(u)$ as the bare field becomes screened by the dressing.

As with transient phenomena in general in quantum field theory, this
instantaneous ``switching-on" of the coupling may be considered rather
unphysical.  However, we can simulate these initial conditions while
allowing the coupling to be non-vanishing for all $u$ 
($u_0 \rightarrow -\infty$) by introducing an appropriate 
source $J(u)$ on the right-hand side of \eqref{aa}
chosen to cancel the contribution of the integral for
$-\infty < u' < 0$, recovering \eqref{bq}.  
However, this requires $J(u)$ to be non-vanishing also for $u>0$
and it is arguable whether the introduction 
of such a fine-tuned source for all $u$ is really any more 
physical than the original model of switching on the coupling. 
As we will discover, studying the initial transient dressing even in
flat space provides an important insight into the long-time
behaviour of the field as it evolves in time-dependent curved
spacetimes.

In the second part of this paper, we study examples of field theories
with different short-distance behaviour, specifically $A\phi^2$ in
$d=6$, which requires a UV divergent field renormalization but which is
asymptotically free, and QED, which is IR free and where 
perturbation theory in the short-time, UV regime is afflicted 
by the Landau pole. In these cases, we find a variety of problems with
the Type I initial conditions. Since our primary aim is to study the
universal long-time behaviour of the field, which we would expect to
be independent of the preparation of the initial state, we also
consider alternative initial conditions.

In ``Type II" initial conditions, which have previously been used to study
relaxation problems in non-equilibrium QFT
 (see, e.g.~ref.\cite{Boyanovsky:2003ui}),
we use the source to
hold the field $\AMP(u)$ fixed from $u_0 \rightarrow -\infty$ to
the initial surface $u=0$. In this case, the equation of motion is:
\EQ{
-2i\omega\dot\AMP(u)+M^2\AMP(u)
&+\int_{0}^u du'\,\tilde\Pi_\text{SK}(u,u';\omega,p)\AMP(u') \\
&+\int_{-\infty}^0 du'\,\tilde\Pi_\text{SK}(u,u';\omega,p)\AMP(0) 
=0\\
\label{bqq}
}
In this scenario, the field at the initial value surface $u=0$
is already renormalized and partially dressed, and as $u>0$ we watch it
relax from this state. This choice of initial conditions circumvents
the problems with short-distance physics for theories such as QED,
where the Type I conditions are not controllable in perturbation theory.

A further variant, ``Type III", is to constrain the field to evolve for $u < 0$
with a phase $\AMP(0)\exp{-\tfrac{iM^2 u}{2\omega}}$, where $M^2$ is the
renormalized mass (to be distinguished from the bare mass in the
second term in \eqref{bqq}).  The new equation of motion is the obvious
generalisation of \eqref{bqq}. In this case, the field at $u=0$ is
already renormalized and fully dressed and, as we confirm in section
6, no further evolution occurs in flat spacetime.  However, in curved
spacetime, these initial conditions are particularly appropriate and
allow us to study the effects of curvature on a fully-dressed quantum
field, even in cases requiring a divergent UV field renormalization.

\subsection{Vacuum polarization}

We now sketch the evaluation of the vacuum polarization in a 
plane-wave background in the form we need here.  Further details, 
including the equivalent calculation for QED, can be found in 
refs.\cite{Hollowood:2008kq, Hollowood:2009qz}.  We consider
initially the Feynman form,
\EQ{
\Pi_\text{F}(x,x') = -{ie^2\over2} G_\text{F}(x,x')^2 \ .
\label{br}
}

First, consider briefly the equivalent calculation in flat spacetime.
Here, $\sigma(x,x')=\tfrac12(x-x')^2$ and
$\Delta(x,x')=1$, and because of translational invariance 
the self-energy only depends on the relative position. 
We can therefore Fourier transform with respect to the
four relative coordinates $\hat x^\mu=x^\mu-x^{\prime\mu}$.
Writing the two proper-time variables as $T_1$ and $T_2$, we then
change variables from $(T_1,T_2)$ to $(T,\xi)$, where 
$T=T_1+T_2$ and $\xi=T_1/T$, 
\EQ{
\int_0^\infty \frac{dT_1\,dT_2}{(T_1T_2)^{2}}=
\int_0^\infty \frac{dT}{T^{3}}\int_0^1\frac{d\xi}{[\xi(1-\xi)]^{2}}
}
and find:
\EQ{
\tilde \Pi(p^2)=\int d^4\hat x\,e^{-i p\cdot\hat x}\Pi(\hat x) =
\frac{1}{2}\frac{e^2}{(4\pi)^2}\int_0^1d\xi\,\int_0^\infty \frac{dT}{T}
\, e^{-i(m^2+p^2\xi(1-\xi))T}\ .
\label{bs}
}
This displays a UV logarithmic divergence as the proper time 
$T\to0$.  This is a conventional UV divergence that is
cancelled by a mass renormalization for the $A$ field. 
We can evaluate the integral by Wick rotating the proper-time
$T\to-iT$ and by introducing an explicit cut-off 
$\Lambda^{-2}$ on the $T$ integral, where $\Lambda$ is a 
momentum scale:
\EQ{
\tilde\Pi(p^2)&=\frac{1}{2}\frac{e^2}{(4\pi)^2}\int_0^1
d\xi\,\int_{\Lambda^{-2}}^\infty \frac{dT}{T}
\, e^{-(m^2+p^2\xi(1-\xi))T}\\ &=
\frac{1}{2}\frac{e^2}{(4\pi)^2}\int_0^1d\xi\,\log\Big(
\frac{m^2+p^2\xi(1-\xi)}{\Lambda^2e^{-\gamma_E}}\Big)\ .
\label{bt}
}
This is the standard result for the one-loop
contribution in momentum space.

Now consider the calculation of the vacuum polarization in curved
spacetime from \eqref{bp}, initially using the Feynman form
$\Pi_\text{F}(x,x')$.  The Green functions are given in \eqref{ax}. 
The integrals over $\hat x^a$ are Gaussian, while the
integral over $\hat V$ generates a delta function:
\SP{
&\int d\hat V\,\exp\Big[i\big(\tfrac{u-u'}{2T\xi(1-\xi)}-\omega\big)\hat V\Big]
=4\pi T\xi(1-\xi)\delta\big(u-u'-2\omega\xi(1-\xi)T\big)\ ,\\
&\int d^2\hat x\,\exp\Big[\tfrac i{4T\xi(1-\xi)}(u-u')\psi(u,u')^{-1}_{ab}\hat
  x^a\hat x^b\Big] \exp\bigl[-i p_a \hat x^a\bigr] \\
&~~~~~~~~~~~~~~~~~~~=4i\pi T\xi(1-\xi)  {\sqrt{\det\psi(u,u')}\over (u-u')} 
~\exp\Bigl[-i{T\xi(1-\xi) \over (u-u')}p^a\psi_{ab}p^b\Bigr] \ . 
\label{bu}
}
Since $u-u'=2\omega\xi(1-\xi)T$, the conditions $\omega>0$ and 
$T\geq0$ mean that $u\geq u'$. This is the calculational mechanism 
already encountered in \eqref{bm} which ensures that, when acting on 
a positive frequency solution, the Feynman self-energy is only
non-vanishing when $u>u'$ and therefore gives the same result
as the retarded, Schwinger-Keldysh, self-energy. 

Imposing the delta function constraint, cancelling the 
$p^a$-dependent terms in \eqref{bu} against the identical
terms in the mode functions, and collecting terms, we find:
\EQ{
\tilde\Pi_\text{F}(u,u';\omega,p)
=-\frac{1}{2}\frac{e^2}{(4\pi)^2}\theta(u-u')\frac{\sqrt{\Delta(u,u')}}{u-u'} 
\int_0^1d\xi\,e^{-\tfrac{im^2(u-u')}{2\omega\xi(1-\xi)}}\ .
\label{bv}
}
The usual UV divergence now appears as the singularity in the 
limit $u\to u'$ and can be renormalised in the standard way.
Crucially, there are further ``geometric''
singularities when $u$ and $u'$ are conjugate points on the 
null geodesic describing the classical ``photon" trajectory.
At these points, the VVM determinant diverges. As explored
in detail in refs.\cite{Hollowood:2008kq}, it is these singularities
which give rise to the novel analytic structure of the refractive
index in curved spacetime which modifies the conventional
form of the Kramers-Kronig dispersion relation while maintaining
consistency with causality.  The correct prescription for dealing with
these singularities follows from the Feynman prescription in real
space; namely $u-u'\to u-u'-i0^+$. 

Alternatively, we can derive the Schwinger-Keldysh vacuum polarization
function $\tilde\Pi_\text{SK}(u,u';\omega,p)$ directly from \eqref{bp}
with $\Pi_\text{SK}(x,x') \rightarrow - i e^2 \theta(u-u') G_+(x,x')^2$,
immediately exploiting \eqref{bm}.  Writing the Green functions
directly in terms of the modes according to \eqref{as}, we have:
\EQ{
&\tilde\Pi_\text{SK}(u,u';\omega,p)
=~-{ie^2\over2} \theta(u-u')
\sqrt{g(u)g(u')}\int d^3\hat x\,\int\frac{d^3
p_1}{(2\pi)^3 2\omega_1}\,\int\frac{
d^3p_2}{(2\pi)^3 2\omega_2} \\
&~~\times  e^{-im^2 (u-u') \bigl({1\over 2\omega_1}~
+ {1\over2\omega_2}\bigr)} 
\Phi_{(p)}(x)^\dagger
\Phi_{p_1}(x)\Phi_{p_1}(x')^\dagger 
\Phi_{p_2}(x)\Phi_{p_2}(x')^\dagger \Phi_{(p)}(x')\ ,
\label{bw}
}
where $d^3p_i=d\omega_i\,dp_i^1\,dp_i^2$ and the integrals over $\omega_i$ are
restricted to $\omega_i\geq0$. The integrals over the relative
transverse positions $\hat x$ impose momentum conservation
\EQ{
p^a=p_1^a+p_2^a\ ,~~~~\omega=\omega_1+\omega_2\ .
\label{bx}
}
It is convenient to solve the second constraint by taking
$\omega_1=\omega\xi$ and $\omega_2=(1-\xi)\omega$ with $0\leq\xi\leq1$.
This leaves a Gaussian integral over $p_1^a$:
\EQ{
\tilde\Pi_\text{SK}(u,u';\omega,p)~&=~ -{i e^2\over2} \theta(u-u')
\big[g(u)g(u')\big]^{-1/4}
\int_0^1 d\xi\, {1\over 4\omega \xi(1-\xi)} 
e^{-{im^2 (u-u')\over 2\omega \xi(1-\xi)}} \\
&~~~~~~~~\times \int {d^2 p_1\over (2\pi)^3}
\exp\Bigl[ - {i\over 2\omega \xi(1-\xi)} (p_1 - \xi p)^a
\psi_{ab}(u,u') (p_1 - \xi p)^b \Bigr] \\
&{}\\
&=~ - \frac{1}{2}{e^2\over (4\pi)^2} 
\theta(u-u') {\sqrt{\Delta(u,u')}\over u-u'} 
\int_0^1 d\xi\, e^{-{im^2 (u-u')\over 2\omega \xi(1-\xi)}} \ ,
\label{by}
}
where we have again used \eqref{ak} to write the result 
in terms of the VVM determinant.   Naturally, this reproduces
precisely the expression \eqref{bv}.

\subsection{Field evolution and the dynamical renormalization group} 

The solution to the initial value problem, for Type I initial conditions,
is given by solving the equation of motion \eqref{bq} for the 
$u$-dependent amplitude factor $\AMP(u)$ defined in \eqref{bk}.  
We write the solution as a perturbative expansion in $e^2$ 
as:\footnote{For Type II or III initial conditions, with initial value
  surface $u=0$, the solution for $u>0$ is written as 
$$
\AMP(u) =  \AMP(0) e^{-\tfrac{iM^2u}{2\omega}}
\bigl(1 + i {\cal Q}^{(1)}(u) - i{\cal Q}^{(1)}(0) + \cdots\bigr)\ .
$$
  Note that at this order, the equations of motion for Type III and
  Type I initial conditions are identical for $u>0$. }
\EQ{
\AMP(u)=\AMP(u_0)e^{-\tfrac{iM^2(u-u_0)}{2\omega}}\big(1+i
{\cal Q}^{(1)}(u)+\cdots\big)\ .
\label{bba}
}
Implementing the constraint that $\Pi_\text{SK}(x,x')$ has support only for
$u>u'$, we find:
\EQ{
{\cal Q}^{(1)}(u)
= -\frac{1}{2\omega}\int_{u_0}^u du''\,\int_{u_0}^{u''}
du'\,\tilde\Pi_\text{SK}(u'',u';\omega,p)e^{\tfrac{iM^2(u''-u')}{2\omega}}
\ ,
\label{bbb}
}
so from \eqref{by},
\EQ{
{\cal Q}^{(1)}(u)={e^2\over(4\pi)^2}{1\over4\omega} \int_{u_0}^u
du''\,\int_{u_0}^{u''}du'\,\frac{\sqrt{\Delta(u'',u')}}{u''-u'}
\int_0^1 d\xi\,e^{i\tfrac{u''-u'}{2\omega}\big(M^2 -\tfrac{m^2}{\xi(1-\xi)}\big)}\ .
\label{bbc}
}
$\RE{\cal Q}^{(1)}(u)$ is of course UV divergent and, for the $d=4$
$A\phi^2$ theory, this divergence is absorbed as usual in a mass 
renormalization of $M^2$, as shown explicitly in section 5.

The imaginary part, $\IM{\cal Q}^{(1)}(u)$, controls the
amplification or attenuation of the amplitude and plays a vital role in
our analysis from now on. Using the symmetry property 
$\Delta(x,x') = \Delta(x',x)$, we can show that this is given by an
expression almost identical to \eqref{bbc}, but with the integration
range extended:
\EQ{
2 i\,\IM{\cal Q}^{(1)}(u)={e^2\over(4\pi)^2}{1\over4\omega} \int_{u_0}^u
du''\,\int_{u_0}^{u}du'\,\frac{\sqrt{\Delta(u'',u')}}{u''-u'}
\int_0^1 d\xi\,e^{i\tfrac{u''-u'}{2\omega}\big(M^2 -\tfrac{m^2}{\xi(1-\xi)}\big)}\ .
\label{bbd}
}
This expression, with the integration limits shown, is key to 
clarifying issues related to positivity and the optical theorem in the
next section.
Note also that $\IM{\cal Q}^{(1)}(u)$ is free of UV
divergences since the $u'$ integral avoids the singularity at 
$u'=u''$ by virtue of the Feynman prescription $u''-u'\to
u''-u'-i0^+$.  

\vskip0.2cm
Now, for backgrounds which are translation invariant in $u$
(the symmetric, or Cahen-Wallach, plane waves), the VVM
determinant will be a function only of the separation of
the two points, i.e.~$\Delta = \Delta(x-x')$. It follows that
for large $u$, ${\cal Q}^{(1)}(u) \sim u$. When $u$ is large,
therefore, the ${\cal O}(e^2)$ correction will itself 
be large and perturbation theory should break down. 
In fact, such large {\it secular\/} terms are a common occurrence
in perturbation theory and in this translation-invariant case,
the problem is overcome by a resummation which exponentiates
the dependence on ${\cal Q}^{(1)}(u)$ in \eqref{bba}. 

More generally, this kind of resummation is performed by
what is known as the 
{\it dynamical renormalization group\/} (see 
\cite{Boyanovsky:2003ui} and references therein). This is a way
of absorbing the secular terms into a renormalization of the amplitude
$\AMP(u)$ at an arbitrary time $u^*$. The quickest way to arrive
at the re-summed formula is to write in perturbation theory
\EQ{
\AMP(u)=\AMP(u^*)e^{-\frac{iM^2(u-u^*)}{2\omega}}
\big(1+i[{\cal Q}^{(1)}(u)-
{\cal Q}^{(1)}(u^*)]+\cdots\big)\ .
\label{bbe}
}
Since the right-hand side cannot depend on the arbitrary scale $u^*$ we
have, to ${\cal O}(e^2)$,
\EQ{
0=\frac{d}{du^*}\AMP(u^*) + i {M^2\over2\omega}\AMP(u^*)
-i\frac{d{\cal Q}^{(1)}(u^*)}{du^*}\AMP(u^*)\ .
\label{bbf}
}
Solving this, using the initial condition ${\cal Q}^{(1)}(u_0)=0$, 
and finally setting $u^* \rightarrow u$ gives
\EQ{
\AMP(u)=\AMP(u_0)e^{-\tfrac{iM^2(u-u_0)}{2\omega}}
e^{i{\cal Q}^{(1)}(u)}\ .
\label{bbg}
}
The effect of the resummation is to exponentiate the first order
perturbative correction. 
The importance of the dynamical renormalization group is that we can
apply it even in the non-translationally invariant cases. 

Notice now that a positive imaginary part of ${\cal Q}^{(1)}(u)$ signals an
exponential decay of the amplitude, which is the standard expectation.
For translation-invariant backgrounds, where $\IM{\cal Q}^{(1)}(u)$
is proportonal to $u$ in the long-time limit, this allows us to 
identify a constant {\it decay rate\/}.  However, this interpretation does not
extend to a general non-symmetric plane-wave background
where the usual concept of a decay rate is not well-defined.

\subsection{Refractive index and analyticity}

At this point, we make contact with our previous work
\cite{Hollowood:2008kq,Hollowood:2009qz}
on photon propagation and the refractive index for curved spacetime.

In the massless ($M=0$) case, after applying the dyanmical
renormalization group resummation, we have found the solution
for the field $A(x)$ in the form
\EQ{
A(x) = \AMP(u_0) \Phi_p(x) e^{i{\cal Q}^{(1)}(u)} \ ,
\label{bbh}
}
where 
\EQ{
{d\over du}{\cal Q}^{(1)}(u) = -{1\over2\omega}
\int_{u_0}^u du' \,\tilde\Pi_\text{SK}(u,u':\omega,p) \ .
\label{bbi}
}
Note again that whereas previously we used the Feynman, {\it in-out\/},
formalism, all our previous results for the refractive index are
identical with those found using the Schwinger-Keldysh, {\it in-in\/},
formalism.
Comparing with refs.\cite{Hollowood:2008kq,Hollowood:2009qz}, 
we identify the 
refractive index associated with the solution \eqref{bbh}
as
\EQ{
n(u;\omega) = 1 + {1\over\omega}{d\over du}{\cal Q}^{(1)}(u) 
\label{bbj}
}
and so
\EQ{
n(u;\omega) =1 - {1\over2\omega^2}
\int_{u_0}^u du'\, \tilde\Pi_\text{SK}(u,u';\omega,p) \ .
\label{bbk}
}

To complete the identification of notation, substitute the explicit
expression \eqref{bbc} for ${\cal Q}^{(1)}(u)$ and introduce the
variable $t = u-u'$. We then have
\EQ{
n(u;\omega) = 1 + {e^2\over(4\pi)^2} {1\over4\omega^2}
\int_0^1 d\xi\, 
\int_0^{u-u_0} {dt\over t}\, e^{-\frac{i m^2 t}{2\omega\xi(1-\xi)}}\,
\sqrt{\Delta(u, u-t)} \ .
\label{bbl}
}
The upper limit of the $t$ integral goes to $\infty$, as in
our previous discussions of the refractive index, when we
take the switch-on time $u_0 \rightarrow -\infty$.
Of course, this expression still requires renormalization.
This will be considered in this formalism in section 7, after 
the detailed analysis of translation invariant spacetimes in sections
5 and 6 using Laplace transform methods. 
In ref.\cite{Hollowood:2008kq}, we also introduced a variable 
$z = {m^2\over 2\omega\xi(1-\xi)}$ and wrote \eqref{bbl}
in the form:
\EQ{
n(u;\omega) = 1 + {e^2\over(4\pi)^2} {1\over4\omega^2}
\int_0^1 d\xi\,\, {\cal F}(u;z) \ ,
\label{bbm}
}
with 
\EQ{
{\cal F}(u;z) = \int_0^\infty {dt\over t}\, e^{-izt}~
\sqrt{\Delta(u,u-t)}  \ .
\label{bbn}
}
Apart from the obvious differences because here we are considering a
pure scalar theory rather than QED, these expressions match the forms
presented in \cite{Hollowood:2008kq,Hollowood:2009qz}, 
in particular section 6 of \cite{Hollowood:2008kq}.

From this point, ref.\cite{Hollowood:2008kq} discusses in 
detail how the singularities in the $t$-plane, which arise from the VVM
determinant when $u$ and $u-t$ are conjugate points along the null
geodesic describing the classical photon trajectory, integrate up to
give a novel analytic structure in the complex $\omega$-plane
for the refractive index.  In particular, the presence of new cuts 
related to the background geometry violate hermitian analyticity,
which is a standard property of S-matrix theory. The resulting
modification of standard theorems of flat-space quantum field theory,
especially the Kramers-Kronig dispersion relation, allow the
apparent paradoxes associated with low-frequency super-luminal 
phase velocities -- that is, $\RE n(u;0) < 1$ --
to be resolved while causality is maintained. In what follows, our
emphasis is different as we try to reconcile the amplification of the
field implied by a negative imaginary part of the refractive index
with unitarity.

\subsection{Dressing and undressing a quantum field}

One of the most remarkable features to emerge from our analysis of the
refractive index for QED was the discovery of cases where 
$\IM n(u;\omega)$ is negative, corresponding to an amplification of
the field as it propagates through the background curved spacetime.
To begin to develop some intuition into this phenomenon, we consider
first the weak curvature limit.  
Here, we can expand the VVM determinant to linear order in the curvature
\EQ{
\Delta(u,u')=1+\frac16R_{uu}(u)(u-u')^2-\frac1{12}
\dot R_{uu}(u)(u-u')^3+\cdots
\label{bbo}
}
where the Ricci tensor $R_{uu}=R^i{}_{uiu}$. 
Substituting this expansion into the
integral \eqref{bbi}, and after UV regularization, 
we determine the correction to the refractive index:
\EQ{
n(u;\w)=1-\frac{e^2}{(4\pi)^2}\frac1{360m^4}R_{uu}(u)-
\frac{e^2}{(4\pi)^2}\frac{i \omega}{840m^6}\dot R_{uu}(u)+\cdots\ .
\label{bbp}
}
The analogues of this expansion for scalar and spinor QED appear in 
appendix \ref{ap2} and \eqref{zaz} for the latter.
In this approximation, the ratio of the amplitude for two points 
$u_1, u_2$ on the null trajectory $\gamma$, in the long-time limit, is
\EQ{
\AMP(u_1)=\AMP(u_2)\exp\left[\frac{e^2}{(4\pi)^2}\frac{\omega^2}{840m^6}
\big(R_{uu}(u_1)-R_{uu}(u_2)\big)\right]\ .
\label{bbq}
}
If the Ricci curvature component $R_{uu}(u)$ decreases along $\gamma$,
then from the Jacobi equation (see \cite{Hollowood:2009qz})
we see that the rate of
acceleration of the nearby geodesics is positive, i.e.~the
tidal forces are increasing in the direction of stretching the
virtual cloud. In this case,  the amplitude decreases because more
virtual $\phi$ pairs are being produced: $A$ is becoming more dressed.
On the contrary, if $R_{uu}(u)$
increases along $\gamma$, i.e.~the tidal forces are increasing in
the direction of squeezing, then the amplitude increases because 
virtual pairs are recombining: 
$A$ is being ``undressed" and is returning to its bare state
\cite{Hollowood:2010bd}.\footnote{A related physical picture for
  particle creation in a background gravitational field, based on
  quantum mechanical tunnelling, has recently been presented in
  ref.\cite{Mironov:2011hp}. This shares our basic intuition of
  relating these curved-spacetime phenomena to geodesic deviation 
  through the Jacobi equation, though here we are considering one-loop 
  quantum field theoretic effects due to vacuum polarization.}
These effects are illustrated heuristically in Fig.~\ref{f3} with the 
role of the photon being played by $A$ and the $e^+e^-$ pairs by $\phi$ pairs.

We now come to the second, non-perturbative, mechanism which can
produce an imaginary part for the refractive index.
If $R_{uu}(u)$ is constant, in which case the metric \eqref{ad}
describes a symmetric plane wave, or 
Cahen-Wallach space \cite{Cahen}, then to linear order in the
curvature the amplitude is constant. In fact, $\IM n(\w)=0$ 
to all orders in the curvature expansion. However, in
the case where at least one of the eigenvalues of the 
constant $h_{ij}$ is negative, there is a positive contribution to 
$\IM n(\w)$, so a decaying amplitude, which is
non-perturbative in the curvature. The simplest case to consider is
$h_{ij}=-\sigma^2\delta_{ij}$, which is discussed in detail in section
6.  In this case, the VVM determinant is
\EQ{
\Delta(u,u')=\left[\frac{\sigma(u-u')}{\sinh\sigma(u-u')}\right]^2\ .
\label{bbr}
} 
we then have
\EQ{
n(\w)=1+\frac{e^2}{(4\pi)^2} {1\over 4\omega^2}
\int_{0}^{\infty}\frac{dt}{\sinh(\sigma t)}
\int_0^1d\xi\,e^{-\frac{im^2t}
{2\omega\xi(1-\xi)}}\ .
\label{bbs}
}
The imaginary part is then given by \eqref{bbs} by extending the contour to run
from $-\infty$ to $+\infty$ just under the real axis. 
$\IM n(\w)$ can then be computed by deforming the 
integration contour so that it picks up the residues of all the poles along the negative imaginary axis
at $t=\frac{n\pi}\sigma$, $n=1,2,\ldots$. This gives
\EQ{
\IM n(\w)=\frac{e^2}{(4\pi)^2}\frac{\pi}{\omega^2}
\int_0^1d\xi\,\Bigl[
1+e^{\frac{\pi m^2}{2\omega\sigma\xi(1-\xi)}}\Bigr]^{-1}\ ,
\label{bbt}
}
which is  non-perturbative in the curvature. In this case the
result is independent of $u$ and so it implies a constant rate of
production of $\phi$ pairs.  This process would be kinematically
disallowed in flat space, but there is no such threshold constraint in
curved spacetime. This kind of below-threshold decay  is a well-known property of curved space arising from the lack of time-translation invariance. In particular the situation in de Sitter space has been known about for a long time \cite{Nacht,Myhrvold:1983hx,Bros:2008sq,Bros:2009bz}.

This result essentially confirms the proposal of ref.\cite{Marolf:2002bx} that interacting quantum fields in a plane wave background with negative eigenvalues of $h_{ij}$ will decay. We have shown that the process is perfectly consistent with unitarity and if one starts from an initial value surface then only a finite number of $\phi^2$ pairs will be produced per unit volume. In the next section we turn to the issue of fate of the optical theorem in curved space.

\section{Optical Theorem in Curved Spacetime}

The observation that the imaginary part of the refractive index
can be negative, at least for trajectories and spacetimes where
the metric in the Penrose plane-wave limit is $u$-dependent, 
brings us to the critical issue -- how can we reconcile the increase in
the amplitude associated with $\IM n(u;\omega) < 0$ with unitarity and
the optical theorem?

The optical theorem is familiar from the discussion of scattering
amplitude in flat space: in the context of the present model, 
it states that the imaginary part of the one-loop vacuum polarisation
diagram yields the rate for the tree-level process $A\to\phi\phi$ 
and is therefore non-vanishing only above threshold. 
In curved spacetime we need a reformulation of the 
theorem that avoids the use of asymptotic paticle states. 
The initial-value problem provides a suitable formalism. 
For clarity, we consider Type I initial conditions where the
interaction is switched on at $u=u_0$.  We then calculate
the total probability for the tree-level process $A\to\phi\phi$ 
to occur between times $u_0$ and $u$ and compare with the
solution of the initial value problem for $A(x)$.
This gives a generalised version of the optical
theorem that holds in curved spacetime and directly relates the
dissipation of the $A$ field to the creation of $\phi$ pairs.

The resolution of the apparent paradox that $\IM n(u;\omega)$
can be negative lies in the fact that we have only shown that the
amplitude can increase over a local region. In order to verify  
unitarity, and relate $\IM \tilde\Pi$ to the (positive) probability 
of production of $\phi$ pairs, we have to integrate over the whole 
region from the ``switch-on" surface $u_0$ to $u$ (see eq.\eqref{cb}
below).
It turns out that while the amplitude $\AMP(u)$ can increase locally, 
compared with its value at the initial value surface it must always
decrease. The initial transient dressing of the $A$ field plays an 
important r\^ole since it is only a field that is already dressed that 
can increase its amplitude by ``undressing"; a bare field, however, 
cannot be undressed any further.

To derive the optical theorem, note first that
\EQ{
\int_{u_0}^u du^{\prime\prime}\IM n(u^{\prime\prime};\w) =
{1\over\w} \IM {\cal Q}^{(1)}(u) =
-{1\over2\w^2} 
\int_{u_0}^u du^{\prime\prime} \int_{u_0}^{u^{\prime\prime}} du'~
\IM \tilde\Pi_\text{SK}(u^{\prime\prime},u';\w,p) \ .
\label{ca}
}
where $\IM {\cal Q}^{(1)}(u)$ is given in eq.\eqref{bbd}.
We also leave $M=0$, so it is appropriate to 
use the refractive index nomenclature as in section 3.4. Otherwise, 
$\tilde\Pi_\text{SK}(u,u';\omega,p)$ in \eqref{cf},\eqref{ch} is accompanied
by a simple $M$-dependent phase, as in \eqref{bbg}.

Now consider the the transition probability for the tree-level
process $A\to\phi\phi$:
\SP{
P_{A\to\phi\phi}(u)=
\frac{e^2}{\cal N}\int\frac{d^3p_1}{(2\pi)^3 2\omega_1}\frac{d^3p_2}
{(2\pi)^3 2\omega_2}\left|\int_{u_0}^u
  d^4x\,\sqrt{g}\,A^{(0)}(x)^\dagger\phi_{p_1}(x)\phi_{p_2}(x)\right|^2\ .
\label{cb}
}
Here, $\phi_p(x)$ are the massive on-shell modes defined in \eqref{an}
and the final-state phase space integrals are over the
three-dimensional $p = (\w,p_a)$. 
The normalization factor ${\cal N} = (A^{(0)},A^{(0)}) = 
2\w \d^{(3)}(0)$ cancels against an overall integration on $(V,x^a)$. 
All the frequencies, $\omega$, $\omega_1$ and $\omega_2$ in
\eqref{cb} are positive since they refer to physical particles.

The evaluation of \eqref{cb} follows closely the calculation of 
$\tilde\Pi_\text{SK}(u,u';\omega,p)$ directly from the mode functions
in eqs.\eqref{bw} --\eqref{by}.
For convenience, recall here the definition of $\Pi_\text{SK}(x,x')$:
\EQ{
\Pi_\text{SK}(x,x') = - {i e^2\over2} \theta(u-u') \Bigl(
G_+(x,x')^2 - G_-(x,x')^2 \Bigr) \ ,
\label{cc}
}
the expression for the Wightman functions in terms of the modes
$\phi_p(x)$:
\EQ{
G_{\pm}(x,x') = \pm \int\frac{d^3p}{(2\pi)^3 2\omega}~
\theta(\pm\omega)\,\phi_p(x) \phi_p(x')^\dagger \ .
\label{cd}
}
and the observation \eqref{bm} that 
$\int dV' G_-(x,x')^2 A^{(0)}(x')= 0$ for positive frequency $\omega$.
It then follows that
\EQ{
P_{A\to\phi\phi}(u) =
{e^2\over{\cal N}}\,\int_{u_0}^u du^{\prime\prime}\sqrt{g^{\prime\prime}}
\int_{u_0}^u du^{\prime}\sqrt{g^{\prime}}\,
\int d^3 x^{\prime\prime} \int d^3 x^{\prime}\,
\Phi_{p}(x^{\prime\prime})^\dagger G_+(x^{\prime\prime},x^\prime)^2
\Phi_{p}(x)  \ .
\label{ce}
}
Separating out an overall volume factor using
translation invariance in the transverse space, as in \eqref{bp}, 
and using the properties $G_+(x,x')^* = G_-(x,x')$
and $G_+(x',x) = G_-(x,x')$,
we find:
\EQ{
P_{A\to\phi\phi}(u) =  -{2\over\w} \int_{u_0}^u du^{\prime\prime}\, 
\int_{u_0}^{u^{\prime\prime}} du' ~ 
\IM \tilde \Pi_\text{SK}(u^{\prime\prime},u';\w,p) \ ,
\label{cf}
}
where note again that we could equally have written $\tilde\Pi_\text{F}$ above,
and so
\EQ{
P_{A\to\phi\phi}(u) = 4\w \int_{u_0}^u du^{\prime\prime}
\IM n(u^{\prime\prime};\w) \ .
\label{cg}
}
This is the statement of the optical theorem in curved spacetime.
Unitarity is respected and the integral over the whole trajectory
of $\IM n(u;\w)$ is positive. This shows how unitarity prevents 
a local amplification of the amplitude from becoming 
unbounded.\footnote{
To emphasise the importance of integrating over the entire region
from the ``switch-on' surface $u_0$ in these expressions, consider the
change $\Delta P(u_2|u_1)$ in the total probability of pair production
between two null surfaces $u_1$ and $u_2$. From \eqref{cf}
this is:
$$
\Delta P(u_2|u_1) = -{2\over\omega}\int_{u_1}^{u_2} du^{\prime\prime}\,
\int_{u_0}^{u^{\prime\prime}}du^\prime\, 
\IM \tilde\Pi_\text{SK}(u^{\prime\prime},u^{\prime};\omega,p) \ .
$$
Contrast this with the integral $I(u_2|u_1)$ defined as 
$\Delta P(u_2|u_1)$ but with the lower limit of the $u^\prime$
integral set to $u_1$.  This is a positive quantity, since by the
above construction it can be written as a $|\cdots|^2$ as in 
$\eqref{cb}$, but it does {\it not\/} represent the probability
of pair production between $u_1$ and $u_2$. In fact, while
$P(u_1)$, $P(u_2)$ and $I(u_2|u_1)$ are all positive quantities, 
because of the extra region of integration from the switch-on 
surface $u_0$ to $u_1$, $\Delta P(u_2|u_1)$ itself need not be
positive.}

We can also write a local version, defining $\Gamma(u) = \partial_u
P_{A\to\phi\phi}(u)$ as the ``instantaneous rate" of pair production.
Then,
\EQ{
\Gamma(u) = 4\w \IM n(u;\w) = - {2\over\w} \int_{u_0}^u du'~
\IM \tilde\Pi_\text{SK}(u,u';\w,p) \ .
\label{ch}
}
Unlike the integrated form, there is no positivity constraint on \eqref{ch}
in general. This explains why there is no conflict with unitarity in the 
examples where we have found $\IM n(u;\w)<0$. However, in cases where 
we have translation invariance along the photon trajectory, as in 
flat spacetime, $n(\w)$ becomes $u$-independent in the large $u$
limit beyond the transient region 
(equivalently $u_0\rightarrow -\infty$; see \eqref{bbl}) 
and $P_{A\to\phi\phi}(u)$ is then
proportional to $u$, so unitarity implies $\IM n(\w)>0$ and
$\Gamma>0$ can be properly interpreted as the rate of 
$A\rightarrow\phi\phi$.
 
\vskip0.2cm
To summarise, we have shown how the effect of gravitational tidal
forces on vacuum polarization can alter the dressing of a photon 
as it propagates through space. In particular, we have seen how 
the imaginary part $\IM n(u;\w)$ of the position-dependent refractive 
index can be negative as well as positive, corresponding to 
``undressing' of the photon rather than the conventional dressing.
Two mechanisms were identified. The first, of order 
$\w\partial_u\RR/m^4$, admits an intuitive interpretation in
terms of curvature variations along the $A$ field (``photon") trajectory
altering the balance of the bare field with its virtual $e^+e^-$
cloud, with increasing stretching (squeezing) giving rise to more
dressing (undressing).  The second, which is non-perturbative in
$\w^2\RR/m^4$, can occur even when the curvature is constant and
is related to the existence of conjugate points on the photon's
null geodesic.

Nevertheless, unitarity is still respected and the optical theorem 
still holds in curved spacetime. In its integrated form \eqref{cf},
\eqref{cg}, it relates the total probability for $e^+e^-$ pair
production to the integral of $\IM n(u;\w)$ along the photon
trajectory, which is manifestly positive.
Except in the special case of translation invariance
along the null geodesic, the corresponding local form \eqref{ch}
has no positivity constraint and the usual interpretation of $\IM n(u;\w)$
as the rate of pair production can break down. However, it does
describe the variation of the amplitude and its interpretation in
terms of dressing and undressing of the photon field. In a sense,
photon propagation in curved spacetime resembles the initial
transient phase in flat spacetime, with the characteristic features
of an oscillating amplitude and below-threshold decay.

\section{Real-time Field Renormalization}

In the second part of this paper, we illustrate these general
principles with a range of examples featuring different quantum field
theories, initial conditions and plane-wave backgrounds.

Most of the discussion will focus on spacetimes with $u$-translation
invariance.  In this case, a particularly powerful method of analysing
the equation of motion for $\AMP(u)$ is provided by the Laplace
transform, which automatically re-sums chains of one-loop vacuum
polarization diagrams and provides a convenient formalism in which to
exploit their analytic structure.

After introducing the Laplace transform method, the remainder of this
section specialises to flat spacetime. As well as providing a
relatively simple example to develop our techniques, it turns out that
by studying real-time field renormalization in the transient region in
flat spacetime, we develop a lot of insight into the evolution of a
renormalized quantum field in a curved spacetime background.

\subsection{Laplace Transform and $\boldsymbol u$-translation invariance}

To introduce the Laplace transform method \cite{Boyanovsky:2003ui}, 
consider first the
$A\phi^2$ theory in $d=4$.  To simplify notation for these
$u$-translation invariant cases, we define the kernel function in the
equation of motion as $\Sigma(u,u') = \tilde\Pi(u,u';\omega,p)$
and note that $\Sigma(u,u') = \Sigma(t)$, where as in section 3.4
we let $t= u-u'$.

The equation of motion, with Type I boundary conditions and 
$u_0 = 0$, is then (see \eqref{bq}):
\EQ{
-2i\omega \dot\AMP(u) + M^2 \AMP(u) + \int_0^u dt\,\Sigma(t)
\AMP(u-t) = 0\ .
\label{da}
}
Since the final term is a convolution, this can be solved by 
introducing the Laplace transform,
\EQ{
\tilde\AMP(s)=\int_0^\infty du\,e^{-su}\AMP(u)\ ,~~~~
\tilde\Sigma(s)=\int_0^\infty dt\,e^{-st}\Sigma(t)\ .
\label{db}
}
The transform of \eqref{da} is simply
\EQ{
-2i\omega\big(s\tilde\AMP(s)-\AMP(0)\big)+M^2\tilde\AMP(s) 
+\tilde\Sigma(s)\tilde\AMP(s)=0
\label{dc}
}
and so
\EQ{
\tilde\AMP(s) =  {2i\omega \AMP(0) \over 2i\omega s - M^2 
-\tilde\Sigma(s)} \ .
\label{dd}
}
The inverse transform is then implemented by the usual Bromwich integral
\EQ{
\AMP(u)=\int_{c-i\infty}^{c+i\infty}\frac{ds}{2\pi i}\,
\frac{2i\omega\AMP(0)e^{su}}{2i\omega s-M^2-\tilde\Sigma(s)}\ ,
\label{de}
}
where the contour lies to the right of any singularities in the
complex $s$ plane.

Now consider the renormalization of the theory.
The analytic structure for the inverse transform depends on properties
of the vacuum polarization kernel $\tilde\Sigma(s)$ and is
model-dependent. Suppose first (this will be the below threshold case
in flat space) that there is an isolated pole
at $s=s_0$ given from \eqref{dd} as the solution of 
\EQ{
2i\omega s_0-M_B^2-\tilde\Sigma_B(s_0)=0\ ,
\label{df}
}
where we have explicitly indicated the bare quantities in
\eqref{de} with a subscript.  We renormalize 
by identifying the mass counterterm, 
$M_B^2 =  M^2+\delta M^2$, in such a way that $M$ is the physical mass
defined as the position of the single particle pole, i.e.
\EQ{
2i\omega s_0=M^2\ .
\label{dg}
}
It follows from \eqref{df} that the counterterm is
\EQ{
\delta M^2=-\tilde\Sigma_B\Big(\frac{M^2}{2i\omega}\Big)\ ,
\label{dh}
}
which is real if we are below threshold.
The expression $\AMP(u)$ in \eqref{de} then takes the same form
with the mass interpreted as the renormalized mass $M^2$ and
the kernel replaced by the subtracted form
\EQ{
\tilde\Sigma(s)=\tilde\Sigma_B(s)-\tilde\Sigma_B\Big(\frac{M^2}{2i\omega}\Big)
\ .
\label{di}
}
Field renormalization will be considered in section 5.3.

\subsection{$\boldsymbol{A\phi^2}$ in $\boldsymbol{d=4}$:~transient evolution in flat spacetime}

In a $u$-translation invariant spacetime, the Fourier transform 
of the kernel $\Sigma(t)$ is simply the usual vacuum polarization with
momentum $p = (\omega, -is -\frac12C_{ab}p^a p^b, p^a)$,
which implies $p^2 = -2i\omega s$.
So in flat spacetime, we can immediately read off the expression for
the bare $\tilde\Sigma(s)$ from \eqref{bt}:
\EQ{
\tilde\Sigma_B(s)=\Pi_B(-2i\omega s)=
\frac{1}{2}\frac{e^2}{(4\pi)^2}\int_0^1d\xi\,\log\Big(
\frac{m^2-2i\omega\xi(1-\xi)s}{\Lambda^2e^{-\gamma_E}}\Big)\ .
\label{dj}
}
Implementing the mass renormalization subtraction \eqref{di}, we
therefore have
\EQ{
\tilde\Sigma(s)=
\frac{1}{2}\frac{e^2}{(4\pi)^2}\int_0^1d\xi\,\log\left(\frac{
m^2-2i\omega s\xi(1-\xi)}{m^2-M^2\xi(1-\xi)}\right)
\ .
\label{dk}
}

\begin{figure}[t]
\begin{minipage}[t]{.47\textwidth}
\begin{center} 
\begin{tikzpicture}[scale=1]
\draw (-2.7,1.8) rectangle (-0.7,2.7);
\node at (-2,2.5) (j1) {below};
\node at (-1.7,2.1) (j1) {threshold};
\draw [->] (-2.5,0) -- (2.5,0);
\draw [->] (0,-2.5) -- (0,2.5);
\filldraw[black] (0,-0.5) circle (2pt);
\filldraw[black] (0,-1) circle (2pt);
\draw[decorate,
decoration={snake},color=blue] (0,-2.5) -- (0,-1);
\node at (2.5,2.5) (i1) {$\boxed{s}$};
\node at (0.5,-0.5) (i2) {$\frac{M^2}{2i\omega}$};
\node at (-0.6,-1) (i3) {$\frac{2m^2}{i\omega}$};
\end{tikzpicture}
     \end{center}
  \end{minipage}
  \hfill
  \begin{minipage}[t]{.47\textwidth}
    \begin{center}
\begin{tikzpicture}[scale=1]
\draw (-2.7,1.8) rectangle (-0.7,2.7);
\node at (-2,2.5) (j1) {above};
\node at (-1.7,2.1) (j1) {threshold};\draw [->] (-2.5,0) -- (2.5,0);
\draw [->] (0,-2.5) -- (0,2.5);
\filldraw[black] (-0.3,-1.8) circle (2pt);
\filldraw[black] (0,-1) circle (2pt);
\draw[decorate,
decoration={snake},color=blue] (-0.8,-2.5) -- (-0.8,-1.8) -- (0,-1);
\node at (2.5,2.5) (i1) {$\boxed{s}$};
\node at (1.7,-1.8) (i2) {$\frac{M^2}{2i\omega}-\frac\pi{2\omega}\rho(M)$};
\node at (0.5,-1) (i3) {$\frac{2m^2}{i\omega}$};
\draw [->] (i2) -- (-0.1,-1.8);
\end{tikzpicture} 
    \end{center}
  \end{minipage}
  \caption{\small The analytic structure of the Laplace transform
  in the below thresold (left) and above threshold (right)
  situations. The diagrams show the particle pole at
  $s=\frac{M^2}{2i\omega}$ and the 2-particle threshold branch point
  and associated cut at $s=\frac{2m^2}{2i\omega}$. In the case above
  threshold we have deformed the cut to expose the simple pole on the
 un-physical sheet.}
\label{f2}
\end{figure}
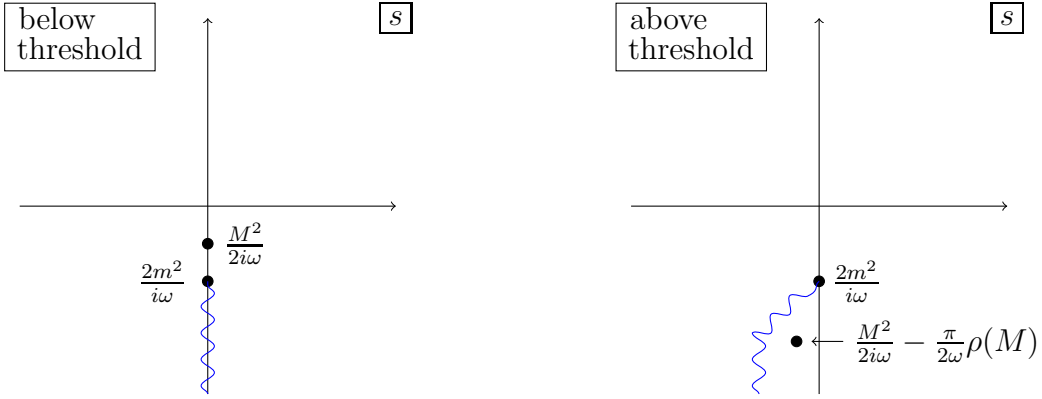
The analytic structure of
the Laplace transform is illustrated in the left part of Fig.~\ref{f2}. $\tilde\Sigma(s)$ itself has a 
branch cut at $s=-\tfrac{2im^2}\omega$, equivalent to
$p^2=-4m^2$, which is the threshold
for the production of pairs of $\phi$ particles. It is convenient
in the following discussion using the spectral density 
to introduce the variable $\nu^2 = -p^2 = 2i\omega s$.
Notice that with this definition, the $\nu$ plane is a
double cover of the $s$ plane. 
Consequently, whereas in the $s$-plane there is only a
single 2-particle cut beginning at the branch-point 
$s=-\tfrac{2im^2}{\omega}$, in $\nu$ plane there are two
threshold branch points at $\nu=\pm2m$ and we take the cuts to lie
along $-\infty\leq\nu\leq-2m$ and $2m\leq\nu\leq\infty$.

The spectral density, which plays a key role in our analysis,
is then given as usual by the imaginary part of the vacuum
polarization kernel,
\EQ{
\rho(\nu)= - \frac1\pi \IM\tilde\Sigma\Bigl(
\frac{\nu^2}{2i\omega}\Bigr)\ ,
\label{dkk}
}
defined implicitly with $\nu\rightarrow \nu +i0^+$, 
i.e.~as the limit from just above the cut.
For $A\phi^2$ theory in $d=4$, it is related to $\tilde\Sigma$
by the once-subtracted dispersion relation,
\EQ{
\tilde\Sigma\Bigl({\nu^2\over2i\omega}\Bigr) =
\tilde\Sigma(0) - \nu \int_{-\infty}^\infty\,
{d\nu'\over\nu'}\, {\rho(\nu')\over\nu' - \nu -i0^+}
\label{dkkk}
}

From the one-loop expression \eqref{dk}, we find the imaginary part:
\EQ{
\IM\tilde\Sigma\Bigl({\nu^2\over2i\omega}\Bigr) =
-\frac{1}{2}{e^2\over(4\pi)^2}\theta(\nu-2m) \pi \int_{\xi_-}^{\xi+} d\xi\ ,
\label{dl}
}
where $\xi_{\pm}$ are the roots of 
$\xi^2 - \xi + {m^2\over\nu^2} = 0$.
The spectral function for $A\phi^2$ theory in $d=4$ is therefore:
\EQ{
\rho(\nu)=\frac{1}{2}\frac{e^2}{(4\pi)^2}\theta(\nu-2m)
\sqrt{1-\frac{4m^2}{\nu^2}}\ ,
\label{dm}
}
valid for $\nu>0$. The $\nu$ dependence of this expression
explains the need for the subtraction in \eqref{dkkk} to ensure 
convergence of the integral for large $\nu$.
For negative $\nu$, we have $\rho(-\nu)=-\rho(\nu)$.
Notice that the imaginary part of $\tilde\Sigma(s)$ and the spectral 
density are UV finite quantities. 

\vskip0.2cm
Below threshold
$M<2m$, the contribution from the pole and cut are
disentangled and the result is simply a sum of the two contributions,
\EQ{
\AMP(u)=\AMP_\text{pole}(u)+\AMP_\text{cut}(u)\ .
\label{dn}
}
The single particle pole contributes
\EQ{
\AMP_\text{pole}(u)=
\frac{\AMP(0) e^{s_0 u} }{1-\tfrac1{2i\omega}\tilde\Sigma'(s_0)}\,=
\frac{\AMP(0) e^{-\frac{iM^2u}{2\omega}} }
{1+\frac{e^2}{(4\pi)^2}
\frac{1}{2M^2}f(\tfrac{4m^2}{M^2})}\ ,
\label{do}
}
where, using \eqref{dk} and evaluating the $\xi$ integral, we have
\EQ{
f(z)=\frac{z}{\sqrt{z-1}} \arctan{\frac{1}{\sqrt{z-1}}}
-1\ ,
\label{dp}
}
with $z = {4m^2\over M^2}$.
So the contribution from the pole is simply the classical solution
with a quantum corrected amplitude.
The contribution from the cut is:
\EQ{
\AMP_\text{cut}(u) = - 2\AMP(0)
\int_{2m}^{\infty} d\nu\,
\frac{\nu\rho(\nu) e^{-\frac{i\nu^2u}{2\omega} } }
{\big[\nu^2-M^2-\RE\tilde\Sigma\bigl({\nu^2\over2i\omega }\bigr)
\big]^2
+\pi^2\rho(\nu)^2}\ ,
\label{dq}
}

We can immediately check consistency by evaluating the 
sum rule\footnote{This follows from
  the fact that at $u=0$ we can pull the contribution from the pole
  and the cut off to a large circle at infinity. In this limit,
  $2i\omega s-M^2-\tilde\Sigma(s)\to 2i\omega s$ and the so the
  contribution is a simple pole at infinity with residue $\AMP(0)$.}
\EQ{
\Big[1+
\frac{e^2}{(4\pi)^2}\frac{1}{2M^2}f\Big(\frac{4m^2}{M^2}\Big)\Big]^{-1}  - 2
\int_{2m}^{\infty} d\nu\,
\frac{\nu\rho(\nu)}
{\big[\nu^2-M^2-
\RE\tilde\Sigma\bigl({\nu^2\over2i\omega s}\bigr)
\big]^2
+\pi^2\rho(\nu)^2}=1\ .
\label{dr}
}
It is also important to notice that the solution that
we have found using the Laplace transform remains perturbative 
throughout the evolution. In other words, in the inverse transform 
\eqref{de} we can consistently expand in powers of of the coupling:
\EQ{
\AMP(u)=\AMP(0)\int_{c-i\infty}^{c+i\infty}\frac{ds}{2\pi i}\,e^{su}
\left[\frac1{s-\tfrac{M^2}{2i\omega}}+
\frac1{\big(s-\tfrac{M^2}{2i\omega}\big)^2}\frac{\tilde\Sigma(s)}{2i\omega}
+\cdots\right]\ .
\label{ds}
}
The pole at $s=\tfrac{M^2}{2i\omega}$ yields the terms
\EQ{
\AMP(0)e^{-\tfrac{iM^2u}{2\omega}}\Bigl[1+
\frac1{2i\omega}   
\tilde\Sigma'\Big(\frac{M^2}{2i\omega}\Big)
+\cdots\Bigr]\ ,
\label{dt}
}
which are precisely the terms of order ${\cal O}(e^2)$ in the expansion of
\eqref{do}.  At this order, the cut contribution is
\EQ{
\AMP_\text{cut}(u)=\AMP(0) {e^2\over(4\pi)^2}
\int_{2m}^{\infty} d\nu\,{\nu\over(\nu^2 - M^2)^2}
\sqrt{1 - {4m^2\over\nu^2}} \, e^{-{i\nu^2 u\over2\omega}} \ .
\label{du}
}
Evaluating this integral and comparing with $f(\tfrac{4m^2}{M^2})$,
we readily verify the sum rule \eqref{dr} explicitly at 
${\cal O}(e^2)$.
 
Since the spectral density is positive, for positive $\nu$, the
contribution to $\AMP(u)$ from the cut is positive. Although it
cannot be evaluated analytically, we can calculate its large $u$
behaviour, which is dominated by the form of the integrand near
the branch point $\nu = m$. This gives the long-time 
behaviour,
\EQ{
\AMP_\text{cut}(u) \thicksim
\Bigl({M^2 u\over 2\omega}\Bigr)^{-3/2}\, e^{- i{2m^2 u \over\omega} } \ ,
\label{dv}
}
up to $u$-independent prefactors.\footnote{In the large $u$ limit, 
we use the fact that 
 $\int_{2m}^\infty
  d\nu\,\nu e^{-\tfrac{i\nu^2u}{2\omega}}
\sqrt{\nu-2m}$ is given approximately by
  $\tfrac{\sqrt\pi}2u^{-3/2}\exp\big(-\tfrac{2im^2u}{\omega}-
\tfrac{3i\pi}2\big)$.}

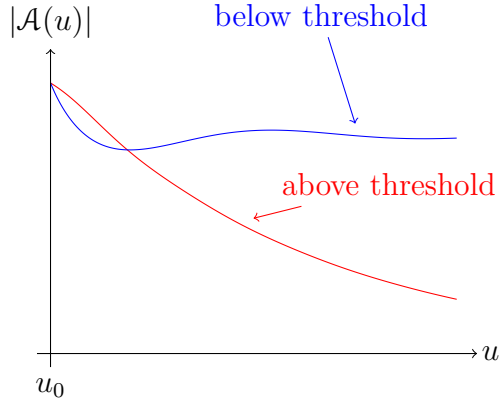
\begin{figure}[t]
    \begin{center}
\begin{tikzpicture}[scale=0.9]
\begin{scope}[xscale=1,yscale=4]
\draw[color=red] plot[smooth] coordinates {(0., 1.)  (0.06, 0.990379)  (0.12, 0.980109)  (0.18, 
  0.969116)  (0.24, 0.957418)  (0.3, 0.945069)  (0.36, 
  0.932142)  (0.42, 0.918725)  (0.48, 0.904912)  (0.54, 
  0.890802)  (0.6, 0.876492)  (0.66, 0.862077)  (0.72, 
  0.847645)  (0.78, 0.833277)  (0.84, 0.819044)  (0.9, 
  0.805005)  (0.96, 0.791209)  (1.02, 0.777695)  (1.08, 
  0.764488)  (1.14, 0.751605)  (1.2, 0.739052)  (1.26, 
  0.726828)  (1.32, 0.714924)  (1.38, 0.703325)  (1.44, 
  0.692013)  (1.5, 0.680964)  (1.56, 0.670156)  (1.62, 
  0.659565)  (1.68, 0.649167)  (1.74, 0.63894)  (1.8, 
  0.628866)  (1.86, 0.618929)  (1.92, 0.609111)  (1.98, 
  0.599415)  (2.04, 0.589822)  (2.1, 0.580334)  (2.16, 
  0.570951)  (2.22, 0.561677)  (2.28, 0.552514)  (2.34, 
  0.543469)  (2.4, 0.534547)  (2.46, 0.52576)  (2.52, 
  0.517109)  (2.58, 0.508602)  (2.64, 0.500245)  (2.7, 
  0.492039)  (2.76, 0.483988)  (2.82, 0.476092)  (2.88, 
  0.468347)  (2.94, 0.460754)  (3., 0.453307)  (3.06, 
  0.446001)  (3.12, 0.438831)  (3.18, 0.43179)  (3.24, 
  0.424871)  (3.3, 0.418066)  (3.36, 0.411373)  (3.42, 
  0.40478)  (3.48, 0.398287)  (3.54, 0.391886)  (3.6, 
  0.385576)  (3.66, 0.379355)  (3.72, 0.37322)  (3.78, 
  0.367171)  (3.84, 0.361207)  (3.9, 0.355329)  (3.96, 
  0.349538)  (4.02, 0.343839)  (4.08, 0.338225)  (4.14, 
  0.332705)  (4.2, 0.327275)  (4.26, 0.32194)  (4.32, 
  0.316697)  (4.38, 0.311546)  (4.44, 0.306488)  (4.5, 
  0.30152)  (4.56, 0.296641)  (4.62, 0.291848)  (4.68, 
  0.28714)  (4.74, 0.282513)  (4.8, 0.277965)  (4.86, 
  0.273492)  (4.92, 0.269091)  (4.98, 0.264761)  (5.04, 
  0.260496)  (5.1, 0.2563)  (5.16, 0.252164)  (5.22, 0.248091)  (5.28,
   0.244078)  (5.34, 0.240125)  (5.4, 0.236229)  (5.46, 
  0.232394)  (5.52, 0.228618)  (5.58, 0.224899)  (5.64, 
  0.22124)  (5.7, 0.217639)  (5.76, 0.214098)  (5.82, 
  0.210616)  (5.88, 0.207192)  (5.94, 0.203828)  (6., 0.200521)};
  \begin{scope}[yshift=-3cm]
\begin{scope}[yscale=4]
\draw[color=blue] plot[smooth] coordinates {(0., 1.)  (0.06, 0.991312)  (0.12, 0.983742)  (0.18, 
  0.977072)  (0.24, 0.971179)  (0.3, 0.965973)  (0.36, 
  0.961383)  (0.42, 0.95735)  (0.48, 0.953822)  (0.54, 
  0.950755)  (0.6, 0.948109)  (0.66, 0.945847)  (0.72, 
  0.943938)  (0.78, 0.94235)  (0.84, 0.941057)  (0.9, 
  0.940033)  (0.96, 0.939254)  (1.02, 0.938697)  (1.08, 
  0.938341)  (1.14, 0.938168)  (1.2, 0.938157)  (1.26, 
  0.938293)  (1.32, 0.938557)  (1.38, 0.938936)  (1.44, 
  0.939414)  (1.5, 0.939977)  (1.56, 0.940614)  (1.62, 
  0.94131)  (1.68, 0.942055)  (1.74, 0.942839)  (1.8, 
  0.943651)  (1.86, 0.944482)  (1.92, 0.945323)  (1.98, 
  0.946167)  (2.04, 0.947006)  (2.1, 0.947833)  (2.16, 
  0.948643)  (2.22, 0.94943)  (2.28, 0.950188)  (2.34, 
  0.950913)  (2.4, 0.951603)  (2.46, 0.952252)  (2.52, 
  0.952859)  (2.58, 0.953421)  (2.64, 0.953937)  (2.7, 
  0.954404)  (2.76, 0.954822)  (2.82, 0.955191)  (2.88, 
  0.95551)  (2.94, 0.95578)  (3., 0.956)  (3.06, 0.956172)  (3.12, 
  0.956298)  (3.18, 0.956377)  (3.24, 0.956413)  (3.3, 
  0.956406)  (3.36, 0.95636)  (3.42, 0.956275)  (3.48, 
  0.956155)  (3.54, 0.956003)  (3.6, 0.955819)  (3.66, 
  0.955609)  (3.72, 0.955373)  (3.78, 0.955116)  (3.84, 
  0.954839)  (3.9, 0.954546)  (3.96, 0.954239)  (4.02, 
  0.953921)  (4.08, 0.953595)  (4.14, 0.953263)  (4.2, 
  0.952929)  (4.26, 0.952594)  (4.32, 0.952263)  (4.38, 
  0.951932)  (4.44, 0.951606)  (4.5, 0.951295)  (4.56, 
  0.95099)  (4.62, 0.950699)  (4.68, 0.950421)  (4.74, 
  0.950158)  (4.8, 0.94991)  (4.86, 0.94968)  (4.92, 0.949469)  (4.98,
   0.949277)  (5.04, 0.949105)  (5.1, 0.948953)  (5.16, 
  0.948822)  (5.22, 0.948713)  (5.28, 0.948624)  (5.34, 
  0.948556)  (5.4, 0.948509)  (5.46, 0.948483)  (5.52, 
  0.948476)  (5.58, 0.948489)  (5.64, 0.948521)  (5.7, 
  0.948571)  (5.76, 0.948639)  (5.82, 0.948721)  (5.88, 
  0.94882)  (5.94, 0.948932)  (6., 0.949058)};
  \end{scope}
\end{scope}
\end{scope}
\draw[->] (-0.2,0) -- (6.3,0);
\draw[->] (0,-0.2) --(0,4.5);
\node at (0,4.9) (i1) {$|\AMP(u)|$};
\node at (6.5,0) (i2) {$u$};
\node at (0,-0.5) (i3) {$u_0$};
\node[color=red] at (5,2.5) (i4) {above threshold};
\draw[->,color=red] (i4) -- (3,2);
\node[color=blue] at (4,5) (i5) {below threshold};
\draw[->,color=blue] (i5) -- (4.5,3.4);
\end{tikzpicture}
\end{center}
\caption{\small The amplitude of the field as a function of $u$
  for the below and above threshold cases for a
  representative choice of parameters. The case above threshold
  illustrates exponential decay while the case below threshold
  oscillates transiently before going asymptotically
  to a constant which gives the finite wave-function
  renormalization .}\label{f1}
\end{figure}

The combined result for $\AMP(u)$ is plotted numerically in 
Fig.~\ref{f1}.
For large $u$, the cut contribution to the amplitude itself
(factoring out the overall $e^{-i{M^2u\over2\omega}}$ phase), 
oscillates with a scale $u_O = {2\omega\over4m^2 - M^2}$ 
while decaying as a power law on the 
scale $u_\text{D} = {2\omega\over M^2}$,
and ultimately $\AMP(u)$ converges to the constant 
$\AMP_\text{pole}(u)$. Note that the scale of the oscillations
increases as $M^2$ approaches (from below) the threshold
$4m^2$. We recognise the ratio 
\EQ{
Z=\left|\frac{\AMP(\infty)}{\AMP(0)}\right|
 =\left[
1-\frac{\partial\Pi(p)}{\partial p^2}\Big|_{p^2=-M^2}\right]^{-1}
=\frac{1}{1+
\frac{e^2}{(4\pi)^2}
\frac{1}{2M^2}f(\tfrac{4m^2}{M^2})}
\ ,
\label{dw}
}
as the wave-function renormalization factor evaluated in the
equilibrium theory. In particular, due to the positivity of the
spectral density the contribution from the cut
is positive and therefore the sum rule \eqref{dr} implies $Z<1$.
So intuitively what is happening is that 
we are seeing the dressing of
the field $A$ in real time by the creation of $\phi$ pairs even
though we are below the 
threshold for decay. The fact that such a decay cannot happen
energetically is by-passed here because we are looking over a finite
region of time and consequently the energy has an associated
uncertainty which allows the below-threshold process to occur.
It is important that this effect does not involve a constant rate but is
just a transient effect. 

\vskip0.2cm
We now discuss what happens above threshold, when $M>2m$. 
In this case the position of the single particle pole \eqref{df}
$s_0$ moves off the imaginary axis. To ${\cal O}(e^2)$
\EQ{
s_0=\frac{M^2}{2i\omega}+\frac1{2\omega}\IM
\tilde\Sigma\Big(\frac{M^2}{2i\omega}\Big)
+\cdots\ .
\label{dx}
}
Notice that the mass counterterm is now specified more precisely as 
\EQ{
\delta M^2=-\RE
\tilde\Sigma\Big(\frac{M^2}{2i\omega}\Big)\ ,
\label{dz}
}

Since the 2-particle cut arises form a square-root branch point, the
function $\tilde\Sigma(s)$ is defined on a 2-sheeted cover of the
$s$-plane with the Bromwich contour on what we call the upper sheet.
A closer analysis reveals that with the 2-particle cut lying along
the negative imaginary axis in the $s$-plane, the single particle pole
lies on the lower sheet. Hence, the Bromwich integral only receives a
contribution from the cut. However, in order to determine the large
$u$ behaviour we can deform the cut to the left of the pole as
illustrated in the right-hand side of Fig.~\ref{f2} and then
the pole {\it does\/} contribute 
\EQ{
\AMP_\text{pole}(u)=\AMP(0)e^{\tfrac{M^2u}{2i\omega}-\tfrac\pi{2\omega}
\rho(M)u} 
\left[1 + \frac1{2i\omega}\tilde\Sigma'\Big(\frac{M^2}{2i\omega}\Big) 
+ \ldots\right]\ ,
\label{dzz}
}
to order ${\cal O}(e^2)$. So above threshold, for large $u$
the $A$ decays into $\phi$ pairs with a characteristic life-time
\EQ{
\Gamma=\frac{\pi}{2\omega}\rho(M)=\frac{e^2}{(4\pi)^2}
\frac{\pi}{4\omega}
\sqrt{1-\frac{4m^2}{M^2}}\ .
\label{dzzz}
}
The behaviour of the field is illustrated in Fig.~\ref{f1}.
Again note that as $M^2$ approaches threshold (from above),
the lifetime becomes large.

\subsection{$\boldsymbol{A\phi^2}$ in $\boldsymbol{d=6}$:~ field renormalization and 
initial conditions}

We now consider the initial value problem for a theory, $A\phi^2$ in
$d=6$,\footnote{The extra dimension for $d=6$ are taken to be a trivial extension of
the transverse space, with the 6-dimensional plane wave metric being simply
$ds^2 = 2 du dV + C_{ab}(u) dx^a dx^b$
with $a,b = 1, \ldots, 4$.}
which requires a UV divergent field (wave-function)
renormalization. The theory is asymptotically free and short distance
physics remains perturbative. Nevertheless, we would expect difficulties
with applying the Type I initial conditions due to the UV divergence
when the interaction is switched on instantaneously at the initial
value surface $u = u_0 = 0$. Indeed, this is what happens, and we
illustrate these difficulties before moving on to a solution of the
initial value problem for Type II and Type III initial conditions, where 
$u_0 \rightarrow -\infty$.

Incorporating a field renormalization to ${\cal O}(e^2)$ in the equation
of motion \eqref{bq}, we have
\EQ{
-2i\omega Z\dot\AMP(u)+M_B^2 Z\AMP(u)
+\int_{0}^u du'\,\tilde\Pi_{B}(u,u';\omega,p)\AMP(u')=0\ .
\label{ddaa}
}
The solution for $\tilde\AMP(s)$ is then
\EQ{
\tilde\AMP(s) =  {2i\omega Z \AMP(0) \over (2i\omega s - M_B^2 )Z
-\tilde\Sigma_B(s)} \ .
\label{ddbb}
}

The Laplace transform kernel $\tilde\Sigma_B(s)$ is again obtained from
the flat space $\Pi_B(p^2)$ by the substitution $p^2 \rightarrow
-2i\omega s$. In dimensional regularization, this gives
\EQ{
\tilde\Sigma_B(s) = - {e^2\over(4\pi)^3}
\int_0^1 d\xi\, \bigl[m^2 - 2i\omega s\xi(1-\xi)\bigr]\,
\biggl({1\over d-6} + {1\over2}\log\biggl[{m^2 - 2i\omega s
\xi(1-\xi)\over 4\pi \mu^2 e^{-\gamma}}\biggr]\,\biggr) \ ,
\label{dda}
}
for the bare kernel.
Physical mass renormalization is as before,
\EQ{
\tilde\Sigma_r(s)=\tilde\Sigma_B(s)-
\tilde\Sigma_B\Big(\frac{M^2}{2i\omega}\Big)
\ ,
\label{ddb}
}
where $M$ is the renormalized mass,
and the field renormalization corresponds to the
further subtraction 
\EQ{
\tilde\Sigma(s) = \tilde\Sigma_r(s) - (Z-1)\bigl(
2i\omega s - M^2\bigr) \ ,
\label{ddc}
}
where in $\overline{\rm MS}$,
$Z = 1 + {1\over6} {e^2\over(4\pi)^3}\bigl({1\over d-6} 
+ \log 4\pi -\gamma\bigr)$.
Note that this introduces a scheme dependence into 
$\tilde\Sigma(s)$, though of course physical results must be
independent of this choice.
We therefore find the following expression for $\tilde\AMP(s)$:
\EQ{
\tilde\AMP(s) =  {2i\omega Z \AMP(0) \over 2i\omega s - M^2 
-\tilde\Sigma(s)} \ ,
\label{ddcc}
}
where the renormalized kernel in $\overline{\rm MS}$ is:
\EQ{
\tilde\Sigma(s) = -{1\over2} {e^2\over(4\pi)^3} &\int_0^1 d\xi\,
\Bigl( \bigl[m^2 - 2i\omega s \xi(1-\xi)\bigr] 
\log\bigl[m^2 - 2i\omega s\xi(1-\xi)\bigr]\mu^{-2} \\
&~~~-
 \bigl[m^2 - M^2 \xi(1-\xi)\bigr] 
\log\bigl[m^2 - M^2\xi(1-\xi)\bigr]\mu^{-2} \Bigr) \ .
\label{ddd}
}

The physical mass renormalization condition we are using ensures
that the renormalized kernel satisfies the condition
$\tilde\Sigma(\tfrac{M^2}{2i\omega}) = 0$. In the following section 
on curved spacetime, we will find it convenient to use a RG scheme 
where the freedom to add a further finite counterterm in $Z$ is used
to impose the additional condition 
$\tilde\Sigma'(\tfrac{M^2}{2i\omega}) = 0$.
This is achieved by the definition\footnote{We assume here that 
the subtractions are real, which will be the case below threshold
where $M^2<4m^2$. In general, the subtractions are specified
as the real parts of $\tilde\Sigma_B(\tfrac{M^2}{2i\omega})$
and $\tilde\Sigma_B^{\prime}(\tfrac{M^2}{2i\omega})$.}
\EQ{
\tilde\Sigma(s) = \tilde\Sigma_B(s) -  
\tilde\Sigma_B\Big(\frac{M^2}{2i\omega}\Big)  -
\Big(s - \frac{M^2}{2i\omega}\Big)
\tilde\Sigma_B^{\prime}\Big(\frac{M^2}{2i\omega}\Big) 
\ .
\label{dddd}
}

The spectral function, which is independent of renormalization issues, 
is identified as before, and a short calculation
gives:
\EQ{
\rho(\nu) ~&=~- {1\over\pi}
\IM\tilde\Sigma\Big(\frac{\nu^2}{2i\omega}\Big)
\\
&=~ {1\over12}{e^2\over(4\pi)^3} \theta(\nu - 2m)
\nu^2 \Bigl(1 - {4m^2\over\nu^2}\Bigr)^{3/2} \ .
\label{dde}
}
This time, because of the extra $\nu^2$ factor in \eqref{dde}
compared to \eqref{dm}, the relation to the kernel is via a
twice-subtracted dispersion relation,
\EQ{
\tilde\Sigma\Big(\frac{\nu^2}{2i\omega}\Big)=
\tilde\Sigma(0) + \nu^2 \tilde\Sigma'(0)
- \nu^3 \int_{-\infty}^\infty\, {d\nu'\over\nu^{\prime 3}}
{\rho(\nu)\over \nu' - \nu - i0^+} \ .
\label{ddf}
}
This may again be checked explicitly by performing the integrals over
$\nu$ and $\xi$ in \eqref{ddf} and \eqref{ddd}.\footnote{
The integral over the spectral function on the rhs of \eqref{ddf}
is explicitly 
$$
{1\over6}{e^2\over(4\pi)^3}\Bigl[
(z-1)^{3/2}\arctan {1\over\sqrt{z-1}} + {4\over3} - z \Bigr]\ ,
$$ 
for $z>1$,
where here $z = {4m^2\over\nu^2}$.}

The first indication of problems with Type I initial conditions is the
remaining presence of the $Z$ factor in the numerator of \eqref{ddcc}.
Essentially, this is indicating that in a theory requiring field
renormalization, and in the absence of a physical short-distance
cut-off, the initial value at $u_0 = 0$ must be divergent if
the long-time evolution is to remain finite. 
The analytic structure is the same as for the $A\phi^2$ theory in $d=4$, and the inverse Laplace transform again separates, below
threshold, into distinct $\AMP_\text{pole}(u)$ and
$\AMP_\text{cut}(u)$ contributions. For Type I initial conditions,
the analysis of the pole contribution is very similar to the 4-dim
theory. The new problem arises with the cut contribution where,
substituting the spectral function, we find to ${\cal O}(e^2)$:
\EQ{
\AMP_\text{cut}(u) = {1\over6} \AMP(0){e^2\over(4\pi)^3}
\int_{2m}^\infty d\nu\,
{\nu^3\over(\nu^2 - M^2)^2} \Bigl(1 - {4m^2\over\nu^2}
\Bigr)^{3/2}\, e^{-i\nu^2 u\over2\omega} \ .
\label{ddg}
}
Contrast this with the corresponding expression for $A\phi^2$ in $d=4$, eq.\eqref{du}. Here, the $\nu$ integral is log divergent for
large $\nu^2$, due to the extra $\nu^2$ power dependence in
$\rho(\nu)$. In turn, this can be traced back to the behaviour of the
kernel for large $s$, where from \eqref{ddd} we see that
$\tilde\Sigma(s) \sim s\log s$ rather than 
$\tilde\Sigma(s) \sim \log s$ in the $d=4$ theory.

To evade these problems, which are intimately related to the divergent
short distance behaviour, we choose instead to analyse
the theory using Type II or Type III initial conditions. This shifts the divergent
physics off to the switch-on surface at $u_0\rightarrow -\infty$, so
that we start the initial value problem from $u=0$ with a renormalized
field, already at least partially dressed.

The Laplace transform of the equation of motion \eqref{bqq}
for Type II initial conditions is:
\EQ{
-2i\omega Z\bigl(s\tilde\AMP(s) - \AMP(0)\bigr) + 
M_B^2 Z \tilde\AMP(s) + \tilde\Sigma_B(s)\tilde\AMP(s) - 
\AMP(0){1\over s} \bigl(\tilde\Sigma_B(s) - \tilde\Sigma_B(0)\bigr)
=0 \ ,
\label{ddh}
}
Solving this, and writing in terms of renormalized quantities,
we have
\EQ{
\tilde\AMP(s) ={ \AMP(0)\over s}\biggl[
1 + {\tilde\Sigma(0) + M^2 \over
2i\omega s - M^2  - \tilde\Sigma(s)}\biggr] \ .
\label{ddi}
}
It is straightforward to check directly in this expression how the
subtractions \eqref{ddb}, \eqref{ddc} implement the usual mass and 
field renormalizations. In particular, note how all factors of 
$Z$ are absorbed by the subtraction \eqref{ddc}.  

Once again, this has a pole at $s = s_0$  (see \eqref{df}) and a cut
with branch point at $2i\omega s = 4m^2$ from \eqref{ddd}.
Note that, for $M^2 \neq 0$, there is no pole at $s=0$ despite the 
$1/s$ factor in \eqref{ddi}.  The inverse Laplace transform is
evaluated as usual, and we find
\EQ{
\AMP(u) = \AMP_\text{pole}(u) + \AMP_\text{cut}(u) \ ,
\label{ddj}
}
where now
\EQ{
\AMP_\text{pole}(u) = \AMP(0) e^{-{iM^2 u\over2\omega}}\,
\frac{1 + {\tilde\Sigma(0)\over M^2}}{1 - {1\over2i\omega}
\tilde\Sigma'(\tfrac{M^2}{2i\omega})} \ ,
\label{ddk}
}
and
\EQ{
\AMP_\text{cut}(u) = 2 \AMP(0)\,
\Bigl(1 + {\tilde\Sigma(0)\over M^2}\Bigr) \,M^2\,
\int_{2m}^\infty {d\nu\over\nu}\, 
{\rho(\nu) e^{-i{\nu^2 u\over2\omega}} \over
\big[\nu^2-M^2-\RE\tilde\Sigma(\tfrac{M^2}{2i\omega}) \big]^2
+\pi^2\rho(\nu)^2} \ ,
\label{ddl}
}
Note the differences from the Type I case -- the extra 
$\Bigl(1 + {\tilde\Sigma(0)\over M^2}\Bigr)$ factor, which affects the
asymptotic ratio $\AMP(u)/\AMP(0)$, and the extra $M^2/\nu^2$
factor in the cut contribution, which plays the role of a subtraction
in a normal dispersion relation in providing the necessary convergence
factor for the integral over $\nu$.

It is instructive to evaluate these expressions at 
${\cal O}(e^2)$.
For the pole contribution, we have:
\EQ{
\AMP_\text{pole} (u) &= \AMP(0) e^{-{iM^2 u \over2\omega}}\, 
\left[1 + {\tilde\Sigma(0)\over M^2} +
{1\over2i\omega}
\tilde\Sigma'\Big(\frac{M^2}{2i\omega}\Bigr)\right]
\\ &=\AMP(0) e^{-{iM^2 u \over2\omega}}\,
\left[1 + {1\over12} {e^2\over(4\pi)^3} 
h\Big(\frac{4m^2}{M^2}\Big)\right]\ ,
\label{ddm}
}
where
\EQ{
h(z) = 3z \sqrt{z-1} \arctan {1\over\sqrt{z-1}} 
- 3z + 1 \ .
\label{ddn}
}
Since $h(z) <0$ for all $z>1$ 
(i.e.~below threshold, $M^2 < 4m^2$),
the effect of the pole contribution is to give a $u$-independent
reduction in the amplitude.  Note that the renormalization scheme
ambiguity cancels in the combination 
$\tilde\Sigma(0) + \tfrac{M^2}{2i\omega} 
\tilde\Sigma'(\tfrac{M^2}{2i\omega})$, as is easily checked from 
\eqref{ddc}; in particular, this ensures
the disappearance of any $\mu$-dependence in $h(z)$.

At ${\cal O}(e^2)$, substituting the explicit form \eqref{dde}
of the spectral function, we find the cut contribution is 
\EQ{
\AMP_\text{cut}(u)= \AMP(0) {1\over6} {e^2\over(4\pi)^3} M^2 
\int_{2m}^\infty d\nu\, {\nu\over(\nu^2 - M^2)^2}
\Bigl(1 - {4m^2\over\nu^2}\Bigr)^{3/2}\, 
e^{-{i\nu^2 u\over2\omega}} \ .
\label{ddo}
}
For $u=0$, the integral over $\nu$
can be performed analytically and, together with \eqref{ddk}, confirms
the sum rule $\AMP_\text{pole}(0) + \AMP_\text{cut}(0) = \AMP(0)$
required for consistency.

The physical picture emerging from \eqref{ddm} and \eqref{ddo} closely
resembles that already described for the $A\phi^2$ theory in $d=4$.
Evaluating the cut contribution for large $u$, we find
\EQ{
\AMP_\text{cut}(u) \thicksim \Big({M^2 u\over 2\omega}\Big)^{-5/2}\,
e^{-\frac{i2m^2 u}{\omega}} \ ,
\label{ddp}
}
similar to \eqref{dv} but with a faster power law decay. 
The overall behaviour is similar to Fig.~\ref{f1} in the below
threshold case. Once again, factoring out the overall phase
$e^{-{iM^2\over2\omega}}$, the amplitude oscillates on a scale 
$u_O = {2\omega\over 4m^2 - M^2}$, while the power law decay is on a
scale $u_\text{D} = {2\omega\over M^2}$. Although the field at the initial
value surface is renormalized, rather than bare as with Type I initial 
conditions, once the source holding the amplitude fixed at 
$\AMP(0)$ for all $u<0$ is removed, the field relaxes in real time,
tending asymptotically to the constant value given by 
$\AMP_\text{pole}(u)$. This further, finite, dressing of the field, as
well as the transient behaviour for small $u$ is of course consistent
with the general unitarity constraints encoded in the optical theorem.

\vskip0.2cm
It is also instructive to analyse this theory with the Type III
initial conditions introduced in section 3.1. The Laplace transform of
the equation of motion in this case is (compare \eqref{ddh}):
\EQ{
-2i\omega Z\bigl(s\tilde\AMP(s) - \AMP(0)\bigr) &+ 
M_B^2 Z \tilde\AMP(s) + \tilde\Sigma_B(s)\tilde\AMP(s)  \\&-
{\AMP(0)\over s - {M^2\over 2i\omega}} \Big(\tilde\Sigma_B(s) - 
\tilde\Sigma_B\Big(\frac{M^2}{2i\omega}\Big)\Big)
=0 \ ,
\label{ddq}
}
where the inclusion of the phase with the renormalized mass 
makes a crucial modification to the final term.
The solution is 
\EQ{
\tilde\AMP(s) = {\AMP(0)\over s - {M^2\over2i\omega}}\biggl[
1 + {\tilde\Sigma(\tfrac{M^2}{2i\omega})  \over
2i\omega s - M^2  - \tilde\Sigma(s)}\biggr] \ .
\label{ddr}
}
Again, we see how all the renormalization counterterms are absorbed
into the subtractions in the renormalized kernel.

However, for flat spacetime -- though {\it not} in curved spacetime as we
shall shortly see -- there is a further simplification. Here, it follows
immediately from the mass and field renormalization conditions 
\eqref{ddb} and \eqref{ddc} that
$\tilde\Sigma(\tfrac{M^2}{2i\omega}) = 0$. So then, we simply have
\EQ{
\tilde\AMP(s) = {\AMP(0)\over s - {M^2\over2i\omega}}
\label{dds}
}
and only the simple pole remains. The inverse Laplace transform gives
\EQ{
\AMP(u) = \AMP(0) e^{-\frac{iM^2 u}{2\omega}} \ .
\label{ddt}
}
The physical explanation is simple. Since for these Type III conditions,
the field at the initial value surface has been prepared in a fully
dressed, renormalized state, it simply continues with that evolution 
for $u> 0$.  This is, however, special to flat spacetime. As we see in
the next section, the subsequent evolution in curved spacetime can be
highly non-trivial.

\subsection{Quantum electrodynamics}

Our final example is quantum electrodynamics which, 
as well as requiring a UV divergent
field renormalization, is not asymptotically free.  As we shall see, 
this introduces further problems with Type I initial conditions.
We quote results for QED with both scalar and spinor ``electrons', with 
non-zero mass $m$.

The source-free equation of motion for QED in curved spacetime
takes the form
\EQ{
\frac1{\sqrt{g}}\partial_\nu\big(\sqrt{g}g^{\mu\lambda}g^{\nu\sigma}
F_{\lambda\sigma}\big)+\int d^4x'\,\sqrt{g(x')}\,
\Pi^{\mu\nu}(x,x')A_\nu(x')=0\ .
\label{ddda}
}
In a plane-wave background, the solutions of the classical Maxwell 
equations are
\EQ{
\Phi_{p,\mu}^{(i)}(x)=\delta_{\mu a}E^i{}_{a}(u)\Phi_p(x)\ .
\label{dddb}
}
Here, $E_{ia}(u)$ is the zweibein for the transverse metric
$C_{ab}(u)$ defined in \eqref{ae}. The index, $i=1,2$ (which are
associated to the transverse Brinkmann coordinates) labels the two
physical polarization states. 
To solve the initial-value problem, we then make a similar ansatz 
to the scalar $A\phi^2$ theory (compare \eqref{bk}):
\EQ{
A_\mu^{(i)}(x)= \AMP_{ij}(u) \Phi_{p,\mu}^{(j)}\ .
\label{dddc}
}

Specialising to flat spacetime for the remainder of this
section, where the polarization dependence is trivial
and $\AMP_{ij}(u) = \AMP(u)\delta_{ij}$,
the full equation of motion reduces to the following
equation for a single complex amplitude  $\AMP(u)$:
\EQ{
-2i\omega\dot\AMP(u)+\int_{u_0}^u du'\,
\Sigma(u,u')\AMP(u')= 0\ ,
\label{dddd2}
}
subject to the various initial conditions considered above. 
The Laplace transform of the kernel
$\tilde\Sigma(s)$ is determined from the usual momentum-space
vacuum polarization tensor $\Pi(p^2)$ as in \eqref{dj} with the 
substitution $p^2 \rightarrow -2i\omega s$:
\EQ{
\tilde\Sigma(s)=\Pi(-2i\omega s) \ ,
\label{ddde}
}
where for scalar QED, in dimensional regularisation,
\EQ{
\Pi_B(p^2)= -p^2\left[
\frac{\alpha}{6\pi} {1\over (d-4)} + 
\frac{\alpha}{4\pi}\int_0^1d\xi\,(1-2\xi)^2\log\Big(\frac{m^2+\xi(1-\xi)p^2}
{4\pi\mu^2e^{-\gamma_E}}\Big)\right]\ .
\label{dddf}
}
Note again that we are free to use the usual Feynman vacuum
polarization here since it gives the same result in \eqref{dddd2}
as the Schwinger-Keldysh form.

Gauge invariance ensures there is no mass renormalization 
and the field renormalization in  $\overline{\rm MS}$ is implemented by
the subtraction
\EQ{
\tilde\Sigma(s) ~&=~ \tilde\Sigma_B(s) - (Z-1)2i\omega s \\
&=~ 2i\omega s~ {\alpha\over4\pi}  \int_0^1 d\xi \,(1-2\xi)^2
\log\Big(\frac{m^2 -2i\omega s\xi(1-\xi)}
{\mu^2}\Big)\ .
\label{dddg}
}
Note that $\tilde\Sigma(0)=0$.
The spectral function is readily evaluated as before and 
we find\footnote{
The equivalent results for spinor QED are
$$
\tilde\Sigma(s) = ~ 2i\omega s~ {2\alpha\over\pi}  
\int_0^1 d\xi \, \xi (1-\xi)
\log\Big(\frac{m^2 -2i\omega s\xi(1-\xi)}
{\mu^2}\Big)
$$
and
$$
\rho(\nu)=
\frac{\alpha}{3\pi} \theta(\nu-2m) \nu^2
\Big(1 +\frac{2m^2}{\nu^2}\Big)
\Big(1-\frac{4m^2}{\nu^2}\Big)^{1/2} \ .
$$
}
\EQ{
\rho(\nu)=-\frac1\pi\IM\tilde\Sigma\Big(\frac{\nu^2}{2i\omega}\Big)=
\frac{\alpha}{12\pi} \theta(\nu-2m)\nu^2
\Big(1-\frac{4m^2}{\nu^2}\Big)^{3/2}\ .
\label{dddh}
}

With Type I initial conditions, the solution to the initial value
problem (compare \eqref{ddcc}) is given by the inverse Laplace
transform 
\EQ{
\tilde\AMP(s)= { 2i\omega Z \AMP(0)\over
2i\omega s-\tilde\Sigma(s)} \ .
\label{dddi}
}
Once again, the analytic structure of the solution consists of a
single particle pole, this time at $s=0$, and the 2-particle threshold 
cut starting at $2i\omega s=-4m^2$.
In addition, though, there is the infamous Landau pole
$s=s_L$ determined by the solution of the equation,
\EQ{
1-\frac{\alpha}{4\pi}
\int_0^1d\xi\,(1-2\xi)^2
\log\big(\frac{m^2 - 2i\omega s_L\xi(1-\xi)}{\mu^2}\big)
=0\ .
\label{dddj}
}
Notice that this pole is at a non-perturbatively large Euclidean value
of the momentum $p^2\sim e^{{12\pi}/\alpha}$. 

The contribution from the simple pole at $s=0$ is just a constant
\EQ{
\AMP_\text{pole}(u)=\frac{Z\AMP(0)}{1 - {1\over 2i\omega}
\tilde\Sigma'(0)}\ .
\label{dddk}
}
Again, the explicit presence of $Z$ the fundamental 
problem of Type I conditions in a theory requiring a UV divergent
field renormalization, although once again we see that to 
${\cal O}(e^2)$, the right-hand side of \eqref{dddk} is scheme independent.  

The cut contribution, at ${\cal O}(\alpha)$ is 
\EQ{
\AMP_\text{cut}(u)= 2 \AMP(0)  \int_{2m}^\infty {d\nu\over\nu^3}\,
\rho(\nu) e^{-\tfrac{i\nu^2 u}{2\omega}} \ .
\label{dddl}
}
With the spectral function \eqref{dddh}, $\rho(\nu) \sim \nu^2$ for
large $\nu$ and the cut contribution is logarithmically divergent.
The power counting responsible for this is of course linked
to the presence of UV divergences.

Finally, the contribution from the Landau pole is the rapidly
oscillating function
\EQ{
\AMP_\text{LP}(u) = \frac{e^{s_L u}}\alpha \ ,
\label{dddm}
} 
since $s_L$ is purely imaginary. 
This contribution is non-perturbatively large and contaminates the 
solution for $\AMP(u)$ at large $u$. This reflects the fact that not only
does QED require a UV divergent field renormalization but it is not
asymptotically free, so the short-distance physics is
non-perturbative.

These problems are specific to Type I initial conditions and arise
because the interaction is being turned on at the initial value
surface, $u_0 = 0$.\footnote{
Another way to see how problems arise, not specific to flat spacetime,
is to recall from \eqref{bbi} that 
$$
{\cal Q}^{(1)}(u)=-\frac1{2\omega}\int_{u_0}^u du''\,\int_{u_0}^{u''}
du'\,\Sigma(u'',u')\ .
$$
In QED, the imaginary part involves the integral
$$
\int_{u_0}^u du''\,\int_{u_0}^u du'\,\frac1{(u''-u'-i0^+)^2}\ .
$$
which is divergent. 
In contrast, for the scalar $A\phi^2$ theory in $d=4$, 
the same integral has one less
power in the denominator and is convergent.}
Instead, we can analyse the theory with initial conditions with
$u_0 \rightarrow -\infty$. Since gauge invariance ensures the
photon in QED is massless, in this theory there is no distinction 
between Type II and Type III initial conditions.
From \eqref{ddi}, we immediately have the solution
\EQ{
\tilde\AMP(s)=\frac{\AMP(0)}{s}\left(1+\frac{\tilde\Sigma(0)}{2i\omega
  s-\tilde\Sigma(s)}\right)\ .
\label{dddn}
}
However, in flat spacetime (though {\it not\/} in curved spacetime)
$\tilde\Sigma(0)=0$ and so the solution of the
initial-value problem is trivial: $\tilde\AMP(s)$ only has the simple
pole at $s=0$ and 
$\AMP(u)=\AMP(0)$ for all $u$. Even the Landau pole is absent.
The photon field is set up at the
initial value surface $u=0$ already in a renormalized, fully-dressed
state.  No further dressing can take place in flat spacetime, so the
evolution for $u>0$ is trivial.  However, this initial condition will
be particularly appropriate in the following section when we analyse
QFTs in curved spacetime, since it allows us to distinguish clearly
the effects of curvature on dressing from the short-distance transient 
phenomena present even in flat spacetime.

\section{Field Propagation in Symmetric Plane Waves}

We now return to curved spacetime and in this section consider field
evolution in a special class of plane-wave spacetimes where the metric
is $u$-translation invariant.  These are the symmetric plane waves,
or Cahen-Wallach spaces.

The $u$-translation symmetry means that the intial value problem can
be analysed using all the formalism of the Laplace transform method
discussed in the previous section.  We also note the relation with the
general formalism for the refractive index and optical theorem
described in sections 3 and 4 and our previous work 
\cite{Hollowood:2008kq,Hollowood:2009qz}.

\subsection{The inital value problem and renormalization in curved
  spacetime}

Since we are interested here in the effects of curvature on the
evolution of the field and not on transient phenomena, we consider
initial conditions with $u_0 \rightarrow -\infty$, in particular 
Type III. 

The Laplace transform analysis of the equation of motion goes through
exactly as before and so, for massive scalar $A\phi^2$ theory, we find
\EQ{
\tilde\AMP(s) = {\AMP(0) \over s - \frac{M^2}{2i\omega}}
\biggl[ 1 + {\tilde\Sigma(\tfrac{M^2}{2i\omega}) \over
2i\omega s - M^2 - \tilde\Sigma(s)} \biggr]
\label{ea}
}
written entirely in terms of the renormalized quantities.
The subtractions which implement mass and field renormalization are
the same as in the flat spacetime theory since the UV divergences are
curvature independent. However, this has important implications for
the final results for $\AMP(u)$.  Implementing the physical mass
renormalization and the scheme choice \eqref{dddd} for the field
renormalization (adapted for curved spacetime)
the required subtraction in the kernel is
\EQ{
\tilde\Sigma(s) = \tilde\Sigma_B(s) - 
\RE \tilde\Sigma_B^{\text{flat}}\Big(\frac{M^2}{2i\omega}\Big)  -
\Big(s - \frac{M^2}{2i\omega}\Big) 
\RE \tilde\Sigma_B^{\text{flat}}{}'\Big(\frac{M^2}{2i\omega}\Big)\ .
\label{eb}
}
While this ensures the renormalization conditions
$\tilde\Sigma^{\text{flat}}(\tfrac{M^2}{2i\omega}) = 0$ and
$\tilde\Sigma^{\text{flat}}{}'(\tfrac{M^2}{2i\omega})= 0$
in flat spacetime, crucially this is no longer true in curved spacetime
and as a result the expression \eqref{ea} for $\tilde \AMP(s)$ for
Type III initial conditions becomes non-trivial.

The simple pole in the inverse Laplace transform
\EQ{
\AMP(u)=\AMP(0) \int_{c-i\infty}^{c+i\infty}\frac{ds}{2\pi i}\,
e^{su}\,
{1\over s - \frac{M^2}{2i\omega}} 
\biggl[1 +\frac{\tilde\Sigma(\tfrac{M^2}{2i\omega})}
{2i\omega s-M^2-\tilde\Sigma(s)}\biggr] \,
\label{ec}
}
is shifted to $s_0$, where now
\EQ{
2i \omega s_0 = M^2 + \tilde\Sigma(\tfrac{M^2}{2i\omega}) \ .
\label{ed}
}
Evaluating the pole contribution, we therefore find
\EQ{
\AMP_{\text{pole}}(u) = \AMP(0) e^{-{iM^2 u\over 2\omega}}
e^{-\tfrac{i}{2\omega}\tilde\Sigma\big(\tfrac{M^2}{2i\omega}\big) u}
\Bigl[1 - \frac{1}{2i\omega} \tilde\Sigma'\Big(\frac{M^2}{2i\omega}\Big)
+ \ldots~\Bigr] \ ,
\label{ef}
}
to ${\cal O}(e^2)$ in the pre-factor.  
As before, the cut contribution goes to zero for large $u$,
so \eqref{ef} gives the asymptotic solution for
the amplitude $\AMP(u)$.  
This should be compared with \eqref{dzz} for $\AMP(u)$ in 
flat spacetime above threshold.
Clearly, \eqref{ef} opens up the possibility of particle decay
in curved spacetime even below threshold.  Also note
the similarity with \eqref{bbg}, which shows that the Laplace
transform method has automatically performed the DRG 
resummation of secular terms into the exponent.

In what follows, we use the leading order approximation
\EQ{
\AMP(u) = \AMP(0) e^{-{iM^2 u\over 2\omega}}
e^{-\tfrac{i}{2\omega}\tilde\Sigma\big(\tfrac{M^2}{2i\omega}\big) u} \ ,
\label{eg}
}
to study the evolution of the field amplitude in a variety of
symmetric plane-wave spacetimes.  The identifications with 
the general formalism in sections 3.3 and 3.4 are evident.

The kernel $\tilde\Sigma(s)$ for $A\phi^2$ in $d=4$
is immediately read off from 
\eqref{bv}.  The bare kernel is
\EQ{
\tilde\Sigma_B(s) = - \frac{1}{2}{e^2\over(4\pi)^2} \int_0^1 d\xi\,
\int_0^\infty {dt\over t} \sqrt{\Delta(t)} 
e^{-\frac{i m^2 t}{2\omega \xi(1-\xi)}} 
e^{-st} \ ,
\label{eh}
}
where we simply write $\Delta(t)$ for the VVM determinant
$\Delta(u,u-t)$, which is $u$-independent.  The integral over $t$ is
defined by the Feynman prescription $t \rightarrow t-i0^+$, which
ensures that the contour may normally be evaluated by rotating 
into the Euclidean region, $t \rightarrow -it + 0^+$.
The integral is divergent at $t=0$, corresponding to the usual
UV divergence. Although this can be removed by a mass 
renormalization alone, we choose to use the scheme 
\eqref{eb} and introduce a (finite) field renormalization as well.
The renormalized kernel is then
\EQ{
\tilde\Sigma(s) = -\frac{1}{2} {e^2\over(4\pi)^2} \int_0^1 d\xi
&~\left[~\int_0^\infty {dt\over t} 
e^{\frac{i}{2\omega}\big(M^2 - \tfrac{m^2}{\xi(1-\xi)}\big)t}
\sqrt{\Delta(t)} e^{- \bigl(s - \tfrac{M^2}{2i\omega}\bigr)t}\right.\\
&\left.- \RE\int_0^\infty {dt\over t} 
e^{\frac{i}{2\omega}\big(M^2 - \tfrac{m^2}{\xi(1-\xi)}\big)t}
\Bigl(1 - \bigl(s - \tfrac{M^2}{2i\omega}\bigr)t\Bigr)\right] \ .
\label{ei}
}

This expression clearly simplifies at $s=\tfrac{M^2}{2i\omega}$, 
so the crucial exponent in \eqref{eg} is given by
\EQ{
\RE\tilde\Sigma(\tfrac{M^2}{2i\omega}) = 
- \frac{1}{2}{e^2\over(4\pi)^2} \RE\int_0^1 d\xi\,
\int_0^\infty {dt\over t} 
e^{\frac{i}{2\omega}\big(M^2 - \tfrac{m^2}{\xi(1-\xi)}\big)t}
\Bigl[\sqrt{\Delta(t)} -1 \Bigr] 
\label{ej}
}
and
\EQ{
\IM \tilde\Sigma(\tfrac{M^2}{2i\omega}) = 
{i\over4} {e^2\over(4\pi)^2} \int_0^1 d\xi\,
\int_{-\infty}^\infty {dt\over t} 
e^{\frac{i}{2\omega}\big(M^2 - \tfrac{m^2}{\xi(1-\xi)}\big)t}
~\sqrt{\Delta(t)}  \ .
\label{ejj}
}
Note that there is no subtraction in the imaginary part.  To derive
\eqref{ejj}, we need the property $\Delta^*(-t) = \Delta(t)$ of the 
VVM determinant.  The location of the integration contour is
important.  According to the Feynman prescription, it lies just {\it below\/} 
the real $t$-axis, in particular evading the pole at $t=0$.

The real part of the kernel $\tilde\Sigma(\tfrac{M^2}{2i\omega})$
modifies the phase of $\AMP(u)$, while the imaginary part determines
the amplitude. Note that both are explicitly $\omega$-dependent in
curved spacetime, in contrast to the flat space case.
In our previous papers, we have studied this frequency-dependence
extensively for QED with massless photons, where it determines 
the refractive index as described in section 3. All this analysis, 
including the all-important analyticity properties, goes through in 
the same way here for the massive $A\phi^2$ theory with only
straightforward modifications. In particular, section 7 of
ref.\cite{Hollowood:2008kq} gives explicit calculations for QED
in symmetric plane waves which complement the discussion that follows
and can usefully be read in parallel.

Here, we wish to concentrate on the imaginary part of
$\tilde\Sigma(\tfrac{M^2}{2i\omega})$, which determines the evolution
of the amplitude $|\AMP(u)|$ itself, and in particular to investigate
the effect of curvature on the decay thresholds.  For these
$u$-translation invariant spacetimes, the amplitude $|\AMP(u)|$
can only decrease for large enough $u$ (beyond the transient region), 
in line with the general constraints of the optical theorem.
This is consistent with positivity of the spectral density
$\rho(M;\omega) = -\tfrac{1}{\pi} \IM \tilde\Sigma(\tfrac{M^2}{2i\omega})$.
We have,
\EQ{
|\AMP(u)| = |\AMP(0)| e^{-\tfrac{\pi}{2\omega} \rho(M) u} \ ,
\label{ek}
}
with the decreasing amplitude corresponding to real
$A\rightarrow\phi\phi$ decay with the well-defined rate 
$\tfrac{\pi}{2\omega} \rho(M;\omega)$. Moreover, the discussion in section 3
and 4 shows that this decay rate can only be non-perturbative in the
curvature. However, we will find that for certain classes of
plane-wave background, this decay can take place below the usual
flat-space threshold.

\subsection{Particle decay in symmetric plane-wave spacetimes}

We now study the rate of $A\rightarrow\phi\phi$ 
decay in symmetric plane-wave backgrounds
according to the formula \eqref{ek} with the spectral function 
$\rho(M;\omega)$ given by
\EQ{
\rho(\nu;\omega) = - {i\over4\pi} {e^2\over(4\pi)^2} \int_0^1 d\xi\,
\int_{-\infty}^{\infty} {dt\over t}\, e^{-i\hat zt} \sqrt{\Delta(t)} \ .
\label{el}
}
Here, we have introduced the notation 
$\hat z = \tfrac{1}{2\omega} \bigl(\tfrac{m^2}{\xi(1-\xi)} - \nu^2\bigr)$,
generalising that in section 3.4.

As a preliminary check, we recover the result already obtained in the
flat space limit, where the VVM determinant $\Delta(t) = 1$.
For $\hat z >0$, the contour can be closed in the lower-half complex
$t$-plane. Since there are no singularities there, the integral
vanishes and we simply find $\rho(M) = 0$, corresponding to the below
threshold case where there is no decay. For $\hat z<0$, the contour is
closed in the upper-half plane and picks up the pole at $t=0$, giving
\EQ{
\rho(M) =  \frac{1}{2}{e^2\over(4\pi)^2} \int_{\xi_-}^{\xi_+} d\xi\ ,
\label{em}
}
where $\xi_{\pm}$ are the upper and lower solutions of the quadratic
$\xi(1-\xi) - m^2/M^2 = 0$, and we recover \eqref{dzzz} in the
above threshold case.

\vskip0.2cm
We now consider the three classes of symmetric plane wave in turn:
\vskip0.2cm

\noindent(i)~{\it Conformally flat symmetric plane wave}:

The VVM determinant for a conformally flat symmetric plane wave,
for which $\sigma_1 = \sigma_2 = \sigma$ (with $\sigma$ real),
is
\EQ{
\Delta(u,u')=\left[\frac{\sigma(u-u')}{\sin\sigma(u-u')}\right]^2\ ,
\label{en}
}
where $u-u' = t$.  Inserting into \eqref{eh} and rotating the contour
$t\rightarrow -it$, we have
\EQ{
\tilde\Sigma(s)=
-\frac{\sigma}{2}\frac{e^2}{(4\pi)^2}\int_0^1d\xi\,\int_{0}^\infty dt\,
\frac{e^{-{\hat z} t}}{\sinh(\sigma t)}\ ,
\label{eo}
}
using $\nu^2 = 2i\omega s$ in the definition of $\hat z$.

Using the renormalization prescription \eqref{eb}, we can evaluate the
renormalized kernel exactly and find\footnote{In the notation of
  section 3 and ref.\cite{Hollowood:2008kq}, \eqref{ep} involves
$$
{\cal F}(\hat z) = \psi\big(\tfrac{1}{2} +\tfrac{\hat
  z}{2\sigma}\big) - \RE \log \tfrac{\hat z}{2\sigma} \ ,
$$
where $\hat z = \tfrac{1}{2\omega}\big(\tfrac{m^2}{\xi(1-\xi)} - M^2\big)$.
This should be compared directly with eq.(7.6) of
ref.\cite{Hollowood:2008kq}. This reference also illustrates
the $\omega$-dependence of the QED analogues of \eqref{ep}
for conformally flat, Ricci flat and general plane-wave spacetimes.}
\EQ{
\tilde\Sigma(s)
= \frac{1}{2}\frac{e^2}{(4\pi)^2}\int_0^1d\xi\,
\biggl[
&\psi\Big(\frac12-\frac{is}{2\sigma}
+\frac{m^2}{4\sigma\omega\xi(1-\xi)}\Big)  \\
&-\RE\log\Big(\frac{m^2}{4\sigma\omega\xi(1-\xi)}
-\frac{M^2}{4\sigma\omega}\Big)
+\RE{2i\omega s - M^2\over M^2 - \tfrac{m^2}{\xi(1-\xi)}}~
\biggr]  \ .
\label{ep}
}
The final term is just the optional finite field renormalization
factor in \eqref{eb}.  We can now see explicitly  
how the analytic structure of the kernel is changed in curved spacetime.
Since the di-gamma function $\psi(x)$ has simple poles at
$x=0,-1,-2,\ldots$ with residue -1, it follows that $\tilde\Sigma(s)$ 
now has an infinite series of branch points at
\EQ{
s=-\frac{2im^2}{\omega}- i(2p-1)\sigma\ ,~~~~~p=1,2,\ldots\ .
\label{eq}
}
Consequently, the 2-particle branch cut of the flat space case has become an
infinite sequence of branch cuts with branch points down the negative
imaginary axis. 

It is useful to understand how this arises by considering 
a less direct method of evaluation.
If we expand the denominator in \eqref{eo} in powers of 
$e^{-\sigma t}$ and then perform the $t$ integral, using a 
small $t$ cut-off $\delta$, we find
\EQ{
\tilde\Sigma_B(s)=
-\frac{1}{2}\frac{e^2}{(4\pi)^2}\int_0^1d\xi ~
\sum_{p=1}^\infty
\frac{2\sigma e^{-({\hat z}+(2p-1)\sigma)\delta}}
{{\hat z}+(2p-1)\sigma} 
\label{er}
}
and \eqref{ep} can be recovered from the summation after
renormalization.  In this method, we see clearly the origin of 
the poles, which become the branch points \eqref{eq} 
after the $\xi$ integration.

The analytic structure of $\tilde\Sigma(\tfrac{M^2}{2i\omega})$ therefore 
comprises a sequence of branch cuts with branch points at 
\EQ{
M^2 = 4m^2 \Bigl(1 + (2p-1)\frac{\omega\sigma}{2m^2}\Bigr) 
~~~~~~~~~~~~p=1,2,\ldots \ .
\label{es}
}
These new curvature-dependent branch points therefore depend
on the parameter $\omega\sqrt{\RR}/m^2$ which, as we have frequently
encountered, is a characteristic scale for non-perturbative phenomena
for fields propagating in curved spacetime.
As we take the flat space limit, the branch points all
converge on the single threshold branch point at $M^2 = 4m^2$.

The spectral density is found from the imaginary part of 
$\tilde\Sigma(s)$ by integrating around the simple poles of the 
digamma function. This gives
\EQ{
\rho(\nu;\omega)=\frac{e^2}{(4\pi)^2}\sum_{n=0}^{n_0}\Big|
\frac{d\xi}{dn}\Big|\ ,
\label{et}
}
where $\xi(n)$ is the root of the equation
\EQ{
n=-\frac12+\frac{\nu^2}{4\omega\sigma}-\frac{m^2}{4\omega\sigma\xi(1-\xi)}\ ,
\label{eu}
}
with $0\leq\xi(n)\leq\tfrac12$, and $n_0$ is the smallest integer
\EQ{
n_0\in{\mathbb Z}\ ,\qquad n_0\geq
-\frac12+\frac{\nu^2-4m^2}{4\omega\sigma}\ .
\label{ev}
}
One can check that in the flat space limit $\sigma\to0$, the sum over
$n$ becomes a continuum and recovers the result \eqref{dm}. 

The long-distance behaviour of the initial value problem then follows
from \eqref{ek}. The solution for $|\AMP(u)|$
dissipates with a decay rate $\Gamma=\tfrac\pi\omega\rho(M)$ 
which is non-vanishing above a threshold
\EQ{
M^2>4m^2+2\omega\sigma\ .
\label{ew}
}
There is then a series of further curvature-dependent thresholds at
\EQ{
M^2>4m^2+2\omega\sigma(2n+1)\ ,\qquad n\in{\mathbb Z}>0\ ,
\label{ex}
}
at which the decay rate $\Gamma$ jumps discontinuously.
Note that the threshold is raised relative to the flat space value
and that in this case there is no below-threshold decay.
We can therefore conclude that in the conformally-flat plane
wave backgrounds, the curvature is suppressing the 
$A\rightarrow \phi\phi$ decays.

\noindent(ii)~{\it Ricci flat symmetric plane wave}:

If we take $\sigma_1=\sigma$ and $\sigma_2=i\sigma$, the
plane-wave is Ricci flat and the VVM determinant is
\EQ{
\Delta(u,u')=\frac{(\sigma (u-u'))^2}{\sin\sigma( u-u')\sinh\sigma (u-u')}\ .
\label{ey}
}
In this case the integrand in \eqref{ef} has branch points on the 
imaginary $t$-axis as well as the real axis.

In this case, the integral does not have a simple explicit solution as in 
\eqref{ep}. However, we can readily find the analytic structure 
of $\tilde\Sigma(s)$ using the method above by expanding the integrand
in a double sum coming from the $\sin\sigma t$ and $\sinh\sigma t$ 
and then perform the $t$ integral: 
\EQ{
\tilde\Sigma_B(s)=
-\frac{1}{2}\frac{e^2}{(4\pi)^2}\int_0^1d\xi ~
\sum_{p,q=1}^\infty
c_{pq}\frac{\sigma
  e^{-({\hat z}+(2p-2iq-1)\sigma)\hat\delta}}
{{\hat z}+(2p-2iq-1)\sigma}\ ,
\label{ez}
}
for some coefficients $c_{pq}$. It follows that 
in this case, $\tilde\Sigma(s)$ has branch points at
\EQ{
s=-\frac{im^2}{2\omega}-i(2p-2iq-1)\sigma\ ,~~~~~~~p,q=1,2,\ldots\ .
\label{eza}
}

The spectral density can be written in the form (see \eqref{ejj}
\EQ{
\rho(\nu;\omega) =
-{1\over\pi}\IM\tilde\Sigma\Big(\frac{M^2}{2i\omega}\Big) =
 {\sigma\over4} {e^2\over(4\pi)^2} \int_0^1d\xi~
\int_{-\infty}^{\infty}dt~ e^{-i{\hat z}t}~
{1\over \sqrt{\sin \sigma t \sinh \sigma t}} \ .
\label{ezb}
}
The branch points on the imaginary $t$-axis give rise to a
non-vanishing spectral density $\rho(\nu;\omega)$ even below the threshold
$\nu = 2m$. To see this, we evaluate the $t$-integral 
by deforming the contour so that it wraps around the 
negative imaginary axis. The imaginary part of 
$\tilde\Sigma(\tfrac{M^2}{2i\omega})$ in the limit 
$4m^2 - \nu^2 \gg \sigma\omega$ is dominated 
by  the contribution around the
first banch point at $t=-i\tfrac\pi{\sigma}$:
\EQ{
\rho(\nu;\omega)&\simeq
\frac{e^2}{(4\pi)^2}{\sqrt{2\sigma}\over4\pi}
\int_0^1d\xi\,\int_{\tfrac\pi{\sigma}}^\infty
dt\,\frac{e^{-(\hat z
+\tfrac{\sigma}2)t}}{\sqrt{t-\tfrac\pi{\sigma}}}\\
&=\frac{e^2}{(4\pi)^2}{1\over2\sqrt{2\pi}}
\int_0^1d\xi\,e^{-(\hat z+\frac{\sigma}2)\tfrac\pi{\sigma}}\ .
\label{ezc}
}
In the same limit the $\xi$ integral is then
dominated by the saddle-point at $\xi=\tfrac12$: 
\EQ{
\int_0^1d\xi\,e^{-\frac{\pi m^2}{2\sigma\omega\xi(1-\xi)}}
\simeq \sqrt{\frac{\sigma\omega}{8m^2}} 
e^{-\tfrac{2\pi m^2}{\sigma\omega}} .
\label{ezd}
}
Consequently, in this limit we have
\EQ{
\rho(\nu;\omega)\simeq
\frac{e^2}{(4\pi)^2}{1\over4\pi}
\sqrt{\frac{\pi\omega\sigma}{4m^2}}
\,e^{-\tfrac{\pi(4m^2-\nu^2)}{2\sigma\omega}-\tfrac{\pi}{2}}\ .
}
As a consequence, the large $u$ behaviour of
the solution $\AMP(u)$ of the initial-value problem dissipates
with a characteristic decay rate
\EQ{
\Gamma=\frac{\pi}{2\omega}\rho(M;\omega)\simeq
{1\over8}\frac{e^2}{(4\pi)^2}
\sqrt{\frac{\pi\sigma}{4m^2\omega}}
\,e^{-\tfrac{\pi(4m^2-M^2)}{2\sigma\omega}-\tfrac{\pi}{2}}
\ ,
\label{eze}
}
valid when $4m^2-M^2\gg \sigma\omega$.

So for a Ricci-flat plane-wave background, the curvature
induces below-threshold decays of $A\rightarrow\phi\phi$,
with a decay rate which is non-perturbative in the curvature. 

\noindent(iii)~{\it Null energy violating symmetric plane wave}:

In this example we take $\sigma_1=\sigma_2=i\sigma$ so that 
the VVM determinant is
\EQ{
\Delta(u,u')=\left[\frac{\sigma(u-u')}{\sinh\sigma(u-u')}\right]^2\ .
\label{ezf}
}
This background violates the null-energy condition so would not be 
considered a valid solution of Einstein's equations. However, we are
free to consider it as an example of a fixed background. It also
admits a valuable check of our use of the Penrose limit to study
field propagation in curved spacetimes. This is explained in the
appendix \ref{a1}.

In this case, the integrand has simple poles on the imaginary $t$ axis at
$t=-i\frac{n\pi}\sigma$, $\p\in{\mathbb Z}$. In the below threshold regime,
$z>0$, the contribution is from the poles at $t=-i\frac{n\pi}\sigma$,
$n\in{\mathbb Z}>0$, while in the above thresold region, $z<0$, the
contribution is from the poles at  $t=i\frac{n\pi}\sigma$,
$n\in{\mathbb Z}\geq0$. In both cases we can sum up the contribution
into a common analytic function valid for either $z<0$ or
$z>0$ yielding an expression for the spectral density which is valid
for all $\nu$:
\EQ{
\rho(\nu;\omega)=\frac{1}{2}\frac{e^2}{(4\pi)^2}\int_0^1d\xi\,\frac1{
1+e^{\tfrac\pi{2\omega\sigma}\big(\tfrac{m^2}{\xi(1-\xi)}-\nu^2\big)}}
\ .
\label{ezg}
}
It is straightforward to check that the flat-space limit is correctly
recovered since as $\sigma\to 0$ the integral only receives support
from the region $\xi_0\leq\xi\leq1-\xi_0$, where $\xi_0$ is the
smallest root of the equation $\xi(1-\xi)-m^2/\nu^2=0$. 

This expression for the spectral density $\rho(M;\omega)$ shows that the
decay $A\to\phi\phi$ can take place even below threshold, with a rate 
once again non-perturbatively small in the curvature.

\subsection{Quantum Electrodynamics}

Now consider quantum electrodynamics in a symmetric plane wave
background.  The bare kernel for scalar QED\footnote{The equivalent
expression for spinor QED is given in eq.(5.28) of 
ref.\cite{Hollowood:2009qz}.
Note that in refs.\cite{Hollowood:2008kq, Hollowood:2009qz} 
there is an overall sign
error in the refractive index for scalar QED, arising from omitting
the relative minus sign between the scalar and spinor loop in the
vacuum polarization (corrected in arXiv versions). 
In particular, as we have proved here, for the
symmetric plane wave examples, $\IM n(\omega)$ is always positive,
whether for scalar or spinor QED.}
was evaluated in ref.\cite{Hollowood:2008kq}, and it follows
directly that the renormalized, Laplace transform kernel is:
\EQ{
\tilde\Sigma_{ij}(s)=\frac{\alpha}{\pi}\omega\int_0^1 d\xi\,
\xi(1-\xi)~
\int_0^\infty 
\frac{dt}{t^2}\,ie^{-\tfrac{im^2t}{2\omega\xi(1-\xi)}}
\Big[\Delta_{ij}(t)\sqrt{\Delta(t)}e^{-st} 
- 1 - st\Bigr]  \ .
\label{ezj}
}
using the renormalization condition \eqref{eb}.
This ensures $\tilde\Sigma(0)=\tilde\Sigma'(0)=0$ in flat spacetime, 
though not in curved spacetime, where
\EQ{
\RE\tilde\Sigma_{ij}(0)=\frac{\alpha}{\pi}\omega\int_0^1 d\xi\,
\xi(1-\xi)~\int_0^\infty 
\frac{dt}{t^2}\,ie^{-\tfrac{im^2t}{2\omega\xi(1-\xi)}}
\big[\Delta_{ij}(t)\sqrt{\Delta(t)} - \delta_{ij} \bigr] \\
\label{ezkk}
}
and
\EQ{
\IM\tilde\Sigma_{ij}(0)=\frac{1}{2}\frac{\alpha}{\pi}\omega\int_0^1
d\xi\,\xi(1-\xi)~\int_{-\infty}^\infty 
\frac{dt}{t^2}\, e^{-\tfrac{im^2t}{2\omega\xi(1-\xi)}}
~\Delta_{ij}(t)\sqrt{\Delta(t)}\ ,
\label{ezk}
}

With Type III initial conditions (equivalent to Type II for massless 
photons), the field $\AMP(u)$ is given by the inverse Laplace 
transform as
\EQ{ 
\AMP(u)=\AMP(0)\int_{c-i\infty}^{c+i\infty}\frac{ds}{2\pi i}\,
\frac{e^{su}}{s}\left(1+\frac{\tilde\Sigma(0)}
{2i\omega s-\tilde\Sigma(s)}\right)\ ,
\label{ezl}
}
where we have suppressed the matrix indices. This is
regular at $s=0$, the curvature shifting the simple pole to  
\EQ{
s_0=\frac1{2i\omega}\tilde\Sigma(0)\ .
\label{ezm}
}
The contribution from the pole is therefore
\EQ{
\AMP_\text{pole}(u)=\AMP(0)e^{\tfrac{u}{2i\omega}\tilde\Sigma(0)}
\Big[1 - \frac{1}{2i\omega}\tilde\Sigma'(0)\Big] \ .
\label{ezn}
}
The contribution from the Landau pole is
now, however, highly supressed:
\EQ{
\AMP_\text{LP}(u)\thicksim e^{-\tfrac{4\pi}\alpha}e^{i|s_L|u}\ ,
\label{ezo}
}
so this choice of initial conditions gives a formulation of the
initial-value problem that circumvents the problems that arise from 
the Landau pole and leads to a consistent perturbative expansion.
Once again, for large $u$ the contribution from the cut
goes to zero, so asymptotically \eqref{ezn} gives the full
result for $\AMP(u)$.

Recalling the series of formulae in section 3.4, we see by comparison 
with the exponent in\eqref{ezn} that the refractive index is given to
leading order by 
\EQ{
n_{ij}(\omega) = \delta_{ij} - \frac{1}{2\omega^2} \tilde\Sigma_{ij}(0) \ .
\label{ezp}
}
That is,
\EQ{
n_{ij}(\omega) = \delta_{ij} - \frac{\alpha}{\pi} \frac{1}{2\omega} \int_0^1 d\xi~
\xi(1-\xi)~ {\cal F}_{ij}(z) \ ,
\label{ezq}
}
where
\EQ{
{\cal F}_{ij}(z) = \int_0^\infty {dt\over t^2}~i e^{-izt} 
\big[\Delta_{ij}(t) \sqrt{\Delta(t)} -\delta_{ij}\big] \ .
\label{ezr}
}
This reproduces the result found originally in
ref.\cite{Hollowood:2008kq}.
In that paper, the analytic properties of $n(\omega)$ are explored 
and an extensive discussion is given 
(see especially section 7 of ref.\cite{Hollowood:2008kq})
of the frequency dependence 
of the refractive index in symmetric plane wave backgrounds,
showing how conventional dispersion relations are violated in curved
spacetime while causality is maintained.  Just as for the scalar 
$A\phi^2$ theory described in the last section, we find 
examples of backgrounds, notably the Ricci flat plane waves, where 
$\IM n(\omega) = \frac{\pi}{2\omega^2}\rho(0)$
is non-vanishing, showing that the curvature induces the
(necessarily below threshold) decay of the photon into 
electron-positron pairs.

\section{Homogeneous Plane Waves and Singularities}

Finally, we consider the initial value problem in a class of
backgrounds without $u$-translation symmetry. In this case,
$\IM n(u;\omega)$ can be negative and it is interesting to see
explicitly how this is reconciled with the optical theorem.
In particular, we shall consider singular homogeneous plane waves,
which we have already studied in detail in
ref.\cite{Hollowood:2009qz}. These arise as Penrose limits 
in the near singularity region of black holes and in cosmological
FRW spacetimes with an initial singularity.

The profile function for a singular homogeneous plane wave
(see section 2) is $h_{ij}(u) = \frac{1-\alpha_i^2}{4}
\frac{1}{u^2}\delta_{ij}$
for some constants $\alpha_i$.  The $\alpha_i$ characterise the
nature of the singularity and display a remarkable universality
\cite{Blau:2004yi,Hollowood:2009qz} due to their relation with the
Szekeres-Iyers classification of power-law singularities 
\cite{Szekeres:1993qe,Celerier:2002yi}.
As an example, the near-singularity Penrose limit 
of a null geodesic with non-vanishing angular momentum in 
the Schwarzschild metric has this form with 
$\alpha_1=\tfrac15$ and $\alpha_2=\tfrac75$. 
This implies $h_{ij} = \frac{6}{25} \frac{1}{u^2}~ 
{\rm diag}(1,-1)$.
Note that, like the original spacetime, the Penrose limit is
Ricci flat. In  general the parameters $\alpha_i$ are real
and $\geq0$, or imaginary. 

The VVM matrix can easily be extracted from the equation for the geodesics in \eqref{p11}:
\EQ{
\frac{dz^i}{du^2}+\frac{1-\alpha_i^2}{4u^2}z^i=0\ .
}
Since the equation is homogeneous, the solution for the geodesic spray $z^i(u)=A_{ij}(u,u')$, defined above \eqref{pwl},  is
\EQ{
A_{ij}(u,u')=\alpha_i^{-1}(uu')^{(1-\alpha_i)/2}(u^{\alpha_i}-u^{\prime \alpha_i})\delta_{ij}\ ,
}
which gives the VVM matrix as
\EQ{
\Delta_{ij}(u,u')=\frac{\alpha_i(u-u')(uu')^{\tfrac{\alpha_i-1}2}}
{u^{\alpha_i}-u^{\prime\alpha_i}}\delta_{ij}\ .
\label{fa}
}
As noted in section 2, the singular homogeneous plane waves have an
enhanced scaling symmetry, as a result of which the VVM matrix
is a function only of the ratio $r = u'/u$.  In particular, the VVM
determinant is simply
\EQ{
\Delta(u,u') = \frac{\alpha_1 \alpha_2(1 - r)^2 r^{-p} }{
\bigl(1-r^{\alpha_1}\bigr) \bigl(1 - r^{\alpha_2}\bigr) } \ ,
\label{fb}
}
where $p = 1 - \frac{(\alpha_1 +\alpha_2)}{2}$. Note $p < 1$ for 
$\alpha_i > 0$.  The leading behaviour for large and small $r$ 
follows immediately:
\EQ{
\Delta(r) \sim \sqrt{\alpha_1 \alpha_2}~ r^p   
~~~~~(r\rightarrow\infty) \ , ~~~~~~~~~~
\Delta(r) \sim \sqrt{\alpha_1 \alpha_2} ~r^{-p}  
~~~~~(r\rightarrow 0) \ ,
\label{fbb}
}
while $\Delta(r) = 1$ in the limit $r\rightarrow 1$.

We shall focus here on the imaginary part of the refractive
index.  Recall from sections 3.3, 3.4 that this is related to the
field amplitude through the formula 
$\AMP(u) = \AMP(u_0) \bigl(1 + i {\cal Q}^{(1)}(u) \bigr)$
(before applying the DRG), where 
\EQ{
\IM  {\cal Q}^{(1)}(u)  = \omega \int_{u_0}^u du^{\prime\prime} 
\IM n(u^{\prime\prime};\omega) \ .
\label{fc}
}
We consider both scalar $A\phi^2$ theory and QED which, as we
see, exhibit distinct and interesting physical effects.

\noindent{\it $A\phi^2$ theory:}

For the purely scalar $A\phi^2$ theory in $d=4$,
\EQ{
\IM n(u;\omega) = {e^2\over (4\pi)^2} {1\over 4\omega^2}
\int_0^1 d\xi\, \IM{\cal F}(u;z) \ ,
\label{fd}
}
where
\EQ{
\IM {\cal F}(u;z) = \IM \int_0^{u-u_0} {dt\over t} e^{-i z t} 
\sqrt{\Delta(u,u-t)} \ .
\label{fe}
}
As usual, $z = \tfrac{1}{2\omega}
\bigl(\tfrac{m^2}{\xi(1-\xi)} - M^2\bigr)$ and we
restrict to the below-threshold case where $z$ is real
and positive.  
Note that here we have kept $u_0$ explicit, as in \eqref{bbi}.
In general, the refractive index $n(u;u_0;\omega)$ is now a function
of both $u$ and the initial value $u_0$, though to simplify notation
we suppress the $u_0$ dependence and just continue to write
$n(u;\omega)$ as before.

The analysis of cosmological FRW spacetimes requires us to
start from an initial value surface $u=u_0$ with $u_0$ finite 
and positive, rather than letting $u_0 \rightarrow -\infty$ as
we can do in the black hole case. This raises some important 
subtleties, especially with regard to renormalization, which we
consider first.

It follows from the description of renormalization in the context of
the Laplace transform method in section 5 that the vacuum polarization
$\tilde \Pi(u,u';\omega,p)$, incorporating the subtraction
corresponding to mass renormalization, is:
\EQ{
\tilde \Pi(u,u';\omega,p) = \tilde \Pi_B(u,u';\omega,p) 
+ \delta M^2 \delta(u,u') \ .
\label{fe2}
}
From the definition $\delta M^2 =
-\RE\tilde\Sigma_B^{\text{flat}}\bigl(\tfrac{M^2}{2i\omega}\bigr)$,
together with the explicit expression \eqref{eh} for
$\tilde\Sigma_B(s)$, we therefore have
\EQ{
\tilde \Pi(u,u';\omega,p) = -{1\over2}{e^2\over(4\pi)^2} 
\int_0^1 d\xi~
\biggl[{\sqrt{\Delta(u,u')}\over u-u'} e^{-iz(u-u')}~  -
\RE \int_0^\infty {dt\over t} e^{-izt} \d(u,u')~\biggr] \ .
\label{ff}
}
The refractive index itself is 
\EQ{
n(u;\omega) = 1 - {1\over2\omega^2}\int_{u_0}^u du' ~
\tilde\Pi(u,u';\omega,p) 
\label{fg}
}
and so, changing variable to $t = u-u'$ in the first term, we find
\EQ{
n(u;\omega) = 1 + {e^2\over(4\pi)^2} {1\over4\omega^2} 
\int_0^1 d\xi~
\biggl[ \int_0^{u-u_0} {dt\over t} e^{-izt} \sqrt{\Delta(u,u-t)} ~
- \RE \int_0^\infty {dt\over t} e^{-izt}~\biggr] \ .
\label{fh}
}

The UV divergence at $t=0$ cancels between the two terms in
\eqref{fh}. The subtraction, by definition, is real and so the
expression \eqref{fd} for $\IM n(u;\omega)$ is unaffected by
renormalization. The crucial point, though, is that the upper 
limits on the $t$-integrals of the bare and subtraction terms in 
\eqref{fi} are {\it not\/} the same for finite $u_0$. In the flat
spacetime limit and with $u_0\rightarrow -\infty$, we confirm as
expected that $n(u;\omega) = 1$. However, even in flat spacetime, 
this difference in limits implies that $n(u;\omega)$ has a non-trivial,
$u$-dependent behaviour when we start from an initial value surface 
at finite $u_0$. In turn, this implies a non-vanishing 
${\cal Q}^{(1)}(u)$ and 
$u$-dependent evolution of the amplitude $\AMP(u)$.

Of course, this is precisely the transient behaviour we have already
studied using the Laplace transform method in section 5. It is
interesting, however, to see this behaviour reproduced by 
the general expression \eqref{fh} in the case of flat spacetime, 
and in fact very similar results follow in the FRW cosmology 
considered below.

We can evaluate the integral explicitly to give\footnote{
Note that the integrals require care near the pole at $t=0$. The
correct prescription matches that explained following \eqref{eh},
with the $t$ contour lying below the axis and picking up a
contribution from the pole. For example,
$$
\IM \int_0^{u-u_0} {dt\over t} e^{-izt} = {\pi\over2} -
\int_0^{u-u_0} {dt\over t} \sin(zt) \ .
$$
\label{fn1}}
\EQ{
\IM n(u;\omega) ~=~  {e^2\over(4\pi)^2} {1\over4\omega^2}~\int_0^1d\xi\,
\Bigl[ {\pi\over2} - {\rm Si}(\hat u - \hat u_0) \Bigr]\ ,
\label{fi}
}
introducing the notation $\hat u = zu$. ${\rm Si}$ is the sine integral.
Similarly,
\EQ{
\IM {\cal Q}(u) & ~=~ {1\over2}{e^2\over(4\pi)^2} \int_0^1d\xi\,
{1\over \frac{m^2}{\xi(1-\xi)} - M^2} ~\\ &\times
\Bigl[(\hat u - \hat u_0) \big(
{\pi\over2} - {\rm Si}(\hat u - \hat u_0) \big) 
- \cos(\hat u - \hat u_0) + 1 ~\Bigr] \ .
\label{fj}
}

\begin{figure}[t]
  \begin{minipage}[t]{.47\textwidth}
    \begin{center}
\begin{tikzpicture}[scale=0.9]
\begin{scope}[xscale=0.8,yscale=1.6]
\draw[->] (-0.2,0) -- (7.5,0);
\draw[->] (0,-0.2) --(0,2);
\draw[color=red] plot[smooth] coordinates {(0, 1.5708) (0.07, 1.27229)  (0.14, 0.982668)  (0.21, 0.710326)  (0.28, 
  0.462749)  (0.35, 0.246113)  (0.42, 
  0.0649795)  (0.49, -0.0779023)  (0.56, -0.181689)  (0.63,  
-0.247416)  (0.7, -0.277856)  (0.77, -0.277284)  (0.84, -0.251152)   
(0.91, -0.205705)  (0.98, -0.147572)  (1.05, -0.0833441)  (1.12,  
-0.019179)  (1.19, 0.0395431)  (1.26, 0.0884962)  (1.33, 
  0.124599)  (1.4, 0.146109)  (1.47, 0.152622)  (1.54, 
  0.14498)  (1.61, 0.125094)  (1.68, 0.0957073)  (1.75, 
  0.0601148)  (1.82, 
  0.0218589)  (1.89, -0.0155703)  (1.96, -0.0490103)  (2.03,  
-0.0758582)  (2.1, -0.0942437)  (2.17, -0.103133)  (2.24, -0.102361)   
(2.31, -0.0925877)  (2.38, -0.0751989)  (2.45, -0.0521444)  (2.52,  
-0.0257449)  (2.59, 0.00152223)  (2.66, 0.0272398)  (2.73, 
  0.0492503)  (2.8, 0.0658251)  (2.87, 0.0757884)  (2.94, 
  0.0785902)  (3.01, 0.0743232)  (3.08, 0.0636849)  (3.15, 
  0.0478907)  (3.22, 0.0285471)  (3.29, 
  0.00749782)  (3.36, -0.0133446)  (3.43, -0.0321651)  (3.5,  
-0.0473981)  (3.57, -0.0578518)  (3.64, -0.0627961)  (3.71,  
-0.0620077)  (3.78, -0.0557699)  (3.85, -0.0448298)  (3.92,  
-0.0303166)  (3.99, -0.0136302)  (4.06, 0.003689)  (4.13, 
  0.0200987)  (4.2, 0.0341882)  (4.27, 0.0447964)  (4.34, 
  0.0511043)  (4.41, 0.052695)  (4.48, 0.0495772)  (4.55, 
  0.0421712)  (4.62, 0.0312594)  (4.69, 0.0179073)  (4.76, 
  0.00336223)  (4.83, -0.0110605)  (4.9, -0.0240946)  (4.97,  
-0.0346317)  (5.04, -0.0418135)  (5.11, -0.0450999)  (5.18,  
-0.0443073)  (5.25, -0.039615)  (5.32, -0.03154)  (5.39, -0.0208826)   
(5.46, -0.00864918)  (5.53, 0.00404118)  (5.6, 0.0160576)  (5.67, 
  0.0263569)  (5.74, 0.0340722)  (5.81, 0.0385856)  (5.88, 
  0.0395752)  (5.95, 0.0370372)  (6.02, 0.0312782)  (6.09, 
  0.0228811)  (6.16, 0.0126477)  (6.23, 
  0.00152337)  (6.3, -0.00948936)  (6.37, -0.0194201)  (6.44,  
-0.027415)  (6.51, -0.0328096)  (6.58, -0.0351827)  (6.65,  
-0.0343886)  (6.72, -0.0305637)  (6.79, -0.024109)  (6.86,  
-0.0156491)  (6.93, -0.0059728)  (7., 0.00403979)};
\end{scope}
\node at (0,3.7) (i1) {$\IM n(u;\omega)$};
\node at (6.3,0) (i2) {$u$};
\node at (0,-0.5) (i3) {$u_0$};
\end{tikzpicture}
    \end{center}
  \end{minipage}
  \hfill
  \begin{minipage}[t]{.47\textwidth}
    \begin{center}
\begin{tikzpicture}[scale=0.9]
\begin{scope}[xscale=0.8,yscale=2.7]
\draw[->] (-0.2,0) -- (7.5,0);
\draw[->] (0,-0.1) --(0,1.2);
\draw[color=red] plot[smooth] coordinates {(0., 1.)  (0.07, 0.82946)  (0.14, 0.694294)  (0.21, 0.592927)  (0.28,
   0.522824)  (0.35, 0.480627)  (0.42, 0.462334)  (0.49, 
  0.463499)  (0.56, 0.479464)  (0.63, 0.50558)  (0.7, 0.53743)  (0.77,
   0.571024)  (0.84, 0.602955)  (0.91, 0.630527)  (0.98, 
  0.651817)  (1.05, 0.665701)  (1.12, 0.671823)  (1.19, 
  0.670523)  (1.26, 0.662725)  (1.33, 0.6498)  (1.4, 0.633407)  (1.47,
   0.615335)  (1.54, 0.597345)  (1.61, 0.581031)  (1.68, 
  0.567704)  (1.75, 0.55831)  (1.82, 0.553382)  (1.89, 
  0.55303)  (1.96, 0.556959)  (2.03, 0.564528)  (2.1, 
  0.574825)  (2.17, 0.586765)  (2.24, 0.59919)  (2.31, 
  0.610971)  (2.38, 0.621105)  (2.45, 0.628792)  (2.52, 
  0.633486)  (2.59, 0.634936)  (2.66, 0.633183)  (2.73, 
  0.628547)  (2.8, 0.621581)  (2.87, 0.613014)  (2.94, 
  0.603679)  (3.01, 0.594436)  (3.08, 0.586097)  (3.15, 
  0.579359)  (3.22, 0.574746)  (3.29, 0.572576)  (3.36, 
  0.572938)  (3.43, 0.575697)  (3.5, 0.580514)  (3.57, 
  0.586881)  (3.64, 0.594177)  (3.71, 0.601722)  (3.78, 
  0.60884)  (3.85, 0.614918)  (3.92, 0.619456)  (3.99, 
  0.622107)  (4.06, 0.622702)  (4.13, 0.621258)  (4.2, 
  0.617971)  (4.27, 0.613193)  (4.34, 0.607393)  (4.41, 
  0.601117)  (4.48, 0.594935)  (4.55, 0.58939)  (4.62, 
  0.584954)  (4.69, 0.581986)  (4.76, 0.580705)  (4.83, 
  0.581174)  (4.9, 0.583303)  (4.97, 0.586857)  (5.04, 
  0.59148)  (5.11, 0.596736)  (5.18, 0.602141)  (5.25, 
  0.607213)  (5.32, 0.611512)  (5.39, 0.614679)  (5.46, 
  0.616461)  (5.53, 0.616736)  (5.6, 0.615518)  (5.67, 
  0.612952)  (5.74, 0.609296)  (5.81, 0.604903)  (5.88, 
  0.600177)  (5.95, 0.595546)  (6.02, 0.591417)  (6.09, 
  0.588145)  (6.16, 0.586)  (6.23, 0.585145)  (6.3, 0.58563)  (6.37, 
  0.587379)  (6.44, 0.590213)  (6.51, 0.593855)  (6.58, 
  0.597966)  (6.65, 0.602172)  (6.72, 0.606097)  (6.79, 
  0.609401)  (6.86, 0.611803)  (6.93, 0.613108)};
\end{scope}
\node at (0,3.7) (i1) {$|\AMP(u)/\AMP(u_0)|$};
\node at (6.3,0) (i2) {$u$};
\node at (0,-0.6) (i3) {$u_0$};
\node at (-0.4,2.7) (i4) {$1$};
\node at (-0.4,0) (i5) {$0$};
\end{tikzpicture}
   \end{center}
  \end{minipage}
  \caption{\small The left-hand diagram shows $\IM n(u;\omega)$
as a function of $u$ in flat spacetime. Note that 
$\IM n(u;\omega)$ is non-zero
at the initial value surface $u_0$ and can take negative values.
The right-hand diagram shows the evolution of the field amplitude
$|\AMP(u)|$ showing the characteristic transient behaviour as the bare
field becomes dressed in real time.}
\label{figFLAT}
\end{figure}
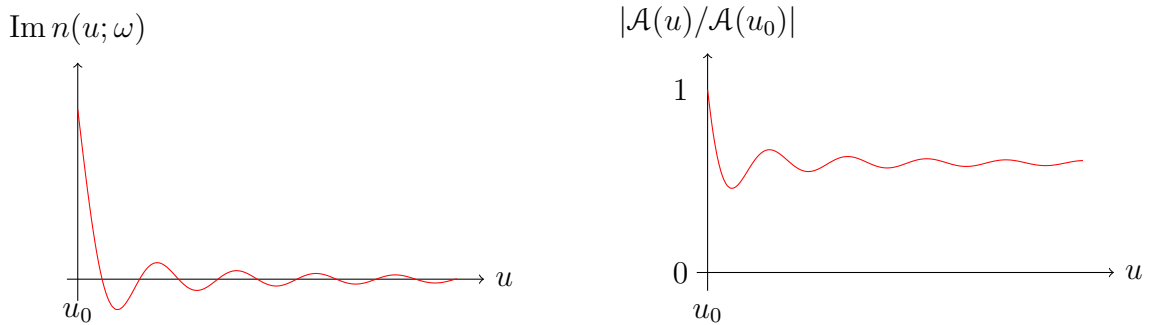
Plots of $\IM n(u;\omega)$ and the amplitude 
$|\AMP(u)| = |\AMP(u_0)| \bigl(1 - \IM {\cal Q}^{(1)}(u)\bigr)$
are shown in Fig.~\ref{figFLAT}.
These display all the features previously described in section 5 --
an oscillatory transient behaviour where $|\AMP(u)|$ falls as the bare
field becomes dressed, before settling to a fixed value in the
asymptotic large-$u$ region where $\IM n(u;\omega) \rightarrow 0$.
Note that $\IM n(u;\omega)$ is locally negative in the transient
phase, corresponding to the oscillatory behaviour of $|\AMP(u)|$,
while its integral $\IM {\cal Q}^{(1)}(u)$ remains positive, in accordance 
with the optical theorem. 

\vskip0.2cm

\noindent{\it Quantum Electrodynamics:}

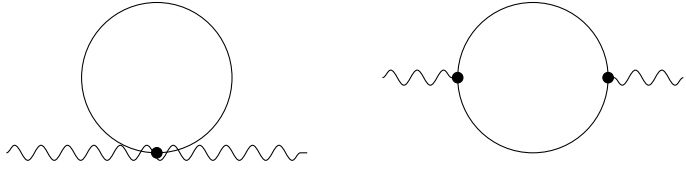
\begin{figure}[t]
\begin{center}
\begin{tikzpicture}[scale=1]
\filldraw[black] (0,0) circle (2pt);
\draw (0,1) circle (1cm);
\draw[decorate,decoration={snake,amplitude=0.1cm}] (-2,0) -- (2,0); 
\begin{scope}[xshift=5cm,yshift=1cm]
\filldraw[black] (-1,0) circle (2pt);
\filldraw[black] (1,0) circle (2pt);
\draw (0,0) circle (1cm);
\draw[decorate,decoration={snake,amplitude=0.1cm}] (-2,0) -- (-1,0); 
\draw[decorate,decoration={snake,amplitude=0.1cm}] (2,0) -- (1,0); 
\end{scope}
 \end{tikzpicture}  
\end{center}
\caption{\small The two Feynman diagrams contributing to the
  vacuum polarization in scalar QED.}
\label{figFeynman}
\end{figure}

The analysis for QED follows the same lines, but there are
important differences, both technical and in the resulting physics.
In scalar QED, there are two contributions to the vacuum polarization,
corresponding to the Feynman diagrams in Fig.~\ref{figFeynman}.

For the Schwinger-Keldysh (retarded) vacuum polarization, we have 
previously shown \cite{Hollowood:2008kq} that
\EQ{
\tilde\Pi_{ij}(u,u';\omega,p)= - {\alpha\over\pi} \omega
\int_0^1 d\xi \,\xi(1-\xi) &\biggl[~
i~ {e^{-i z (u-u')}\over (u-u')^2} \theta(u-u') ~
\Delta_{ij}(u,u') \sqrt{\Delta(u,u')} \\
&~~~~~~~~~~~~+
i \delta_{ij}\int_0^\infty {dt\over t^2} e^{-i z t} ~\delta(u,u')~\biggr]\ ,
\label{fk}
}
where the term proportional to $\delta(u,u')$ comes from the first
Feynman diagram in Fig.~\ref{figFeynman}, and as usual 
$z = \tfrac{m^2}{2\omega\xi(1-\xi)}$. The two transverse polarizations
are labelled by the Brinkmann coordinates $i,j = 1,2$ and 
$\tilde\Pi_{ij}$ can be taken as diagonal with no loss of generality.

Given the relation \eqref{bbk} of the refractive index to the 
vacuum polarization, 
\EQ{
n_{ij}(u;\omega) =\delta_{ij} - {1\over2\omega^2}
\int_{u_0}^u du'\, \tilde\Pi_{\text{SK},ij}(u,u';\omega,p) \ ,
\label{fll}
}
we therefore have
\EQ{
n_{ij}(u;\omega) = \delta_{ij} - {\alpha\over\pi} {1\over 2\omega}
\int_0^1 d\xi\, \xi(1-\xi)~ {\cal F}_{ij}(u;z) \ ,
\label{fl}
}
with
\EQ{
{\cal F}_{ij}(u;z) = \int_0^{u-u_0}{dt\over t^2}~i e^{-i z t} 
\Delta_{ij}(u, u-t) \sqrt{\Delta(u,u-t)} -\delta_{ij}
\int_0^\infty{dt\over t^2} ~i e^{-i z t} \ .
\label{fm}
}
Notice that the limits on the $t$-integrals in the two terms are
different in general. Only when we take
$u_0\rightarrow -\infty$ (as in our previous papers) do they both
become the same, as the dependence on the initial value surface 
disappears.

It is clear that the second contribution acts very much like the mass
counterterm subtraction for $A\phi^2$ theory in \eqref{fh}, but with
important differences. Although it removes the potential singularity
at $t=0$, it is not a counterterm and is {\it not\/} simply the real
part as in \eqref{fh}. It is essential in QED to maintain gauge
invariance, and it is gauge invariance which keeps the photon
massless, with no mass renormalization. Also note the $t^{-2}$
power in the integrals, which arises through power counting and is
ultimately related to the need for field (wave-function)
renormalization in QED. These differences make a subtle, but vitally
important, difference in the behaviour of the field amplitude in QED,
even in flat spacetime.

Consider first QED in flat spacetime.  Combining the two terms in 
\eqref{fm}, we have ${\cal F}_{ij}(u;z) = {\cal F}(u;z) \delta_{ij}$ where
\EQ{
{\cal F}(u;z) = \int_{u-u_0}^\infty {dt\over t^2}~i e^{-i z t} =
- {1\over (u-u_0)}~\int_1^\infty {d\hat t\over {\hat t}^2}~i 
e^{- i z (u-u_0)\hat t} \ ,
\label{fn}
}
with the change of variable $t = (u-u_0)\hat t$.
Notice the occurrence of the vital factor $1/(u-u_0)$, which  
appears because of
the different power-counting in QED compared to $A\phi^2$ in $d=4$.
For the imaginary part (note that both vacuum polarization diagrams
are contributing),
\EQ{
\IM {\cal F}(u;z) = - {1\over(u-u_0)}~\int_1^\infty 
{d\hat t\over {\hat t}^2}~\cos\bigl(z (u-u_0) \hat t\bigr) \ .
\label{fo}
}
The integral can be done analytically and we find the following
expression for the refractive index
\EQ{
\IM n(u;\omega) = {\alpha\over\pi} {m^2\over4\omega^2}\int_0^1d\xi\,
\Bigl[~{\rm Si} (\hat u - \hat u_0) - {\pi\over2} + 
{1\over  \hat u - {\hat u}_0}\cos (\hat u - \hat u_0)~\Bigr] \ ,
\label{fp}
}
where $\hat u=zu$.

\begin{figure}[t]
  \begin{minipage}[t]{.47\textwidth}
    \begin{center}
\begin{tikzpicture}[scale=0.9]
\begin{scope}[xscale=0.8,yscale=14]
\draw[->] (-0.4,0) -- (6.2,0);
\draw[->] (-0.20,-0.2) --(-0.2,0.2);
\draw[color=red] plot[smooth] coordinates {(0., 0.158721)  (0.06, -0.0363973)  (0.12, -0.135358)  (0.18,  
-0.182326)  (0.24, -0.19898)  (0.3, -0.197279)  (0.36, -0.184393)   
(0.42, -0.164896)  (0.48, -0.141837)  (0.54, -0.117313)  (0.6,  
-0.0927877)  (0.66, -0.0692897)  (0.72, -0.0475299)  (0.78,  
-0.0279819)  (0.84, -0.0109374)  (0.9, 0.00345546)  (0.96, 
  0.0151615)  (1.02, 0.0242358)  (1.08, 0.0308045)  (1.14, 
  0.035049)  (1.2, 0.0371919)  (1.26, 0.037485)  (1.32, 
  0.0361985)  (1.38, 0.0336116)  (1.44, 0.0300039)  (1.5, 
  0.0256479)  (1.56, 0.0208032)  (1.62, 0.0157104)  (1.68, 
  0.0105874)  (1.74, 0.00562585)  (1.8, 
  0.00098868)  (1.86, -0.00319111)  (1.92, -0.00681106)  (1.98,  
-0.00979896)  (2.04, -0.0121117)  (2.1, -0.0137336)  (2.16,  
-0.0146742)  (2.22, -0.0149651)  (2.28, -0.0146571)  (2.34,  
-0.0138168)  (2.4, -0.0125228)  (2.46, -0.0108623)  (2.52,  
-0.00892733)  (2.58, -0.00681152)  (2.64, -0.00460678)  (2.7,  
-0.00240051)  (2.76, -0.000273038)  (2.82, 0.00170443)  (2.88, 
  0.00347147)  (2.94, 0.00497965)  (3., 0.0061932)  (3.06, 
  0.00708933)  (3.12, 0.00765801)  (3.18, 0.00790144)  (3.24, 
  0.00783307)  (3.3, 0.00747643)  (3.36, 0.0068636)  (3.42, 
  0.00603358)  (3.48, 0.00503049)  (3.54, 0.00390173)  (3.6, 
  0.00269618)  (3.66, 0.00146236)  (3.72, 
  0.000246851)  (3.78, -0.000907227)  (3.84, -0.0019615)  (3.9,  
-0.00288338)  (3.96, -0.00364686)  (4.02, -0.00423306)  (4.08,  
-0.00463046)  (4.14, -0.0048349)  (4.2, -0.00484933)  (4.26,  
-0.00468331)  (4.32, -0.00435236)  (4.38, -0.00387705)  (4.44,  
-0.00328206)  (4.5, -0.00259508)  (4.56, -0.00184573)  (4.62,  
-0.00106434)  (4.68, -0.00028094)  (4.74, 0.000475847)  (4.8, 
  0.00117972)  (4.86, 0.00180751)  (4.92, 0.00233988)  (4.98, 
  0.00276181)  (5.04, 0.00306291)  (5.1, 0.00323764)  (5.16, 
  0.00328522)  (5.22, 0.00320948)  (5.28, 0.00301854)  (5.34, 
  0.00272429)  (5.4, 0.00234185)  (5.46, 0.00188886)  (5.52, 
  0.00138476)  (5.58, 0.000850058)  (5.64, 
  0.000305487)  (5.7, -0.000228686)  (5.76, -0.00073339)  (5.82,  
-0.00119137)  (5.88, -0.00158776)  (5.94, -0.00191046)  (6.,  
-0.00215052)};
\end{scope}
\node at (0,3.5) (i1) {$\IM n(u;\omega)$};
\node at (5.6,0) (i2) {$u$};
\node at (-0.5,-0.5) (i3) {$u_0$};
\node at (2,2.6) (i6) {divergence};
\draw[->] (i6) -- (0.1,2.2);
\end{tikzpicture}
    \end{center}
  \end{minipage}
  \hfill
  \begin{minipage}[t]{.47\textwidth}
    \begin{center}
\begin{tikzpicture}[scale=0.9]
\begin{scope}[xscale=0.8,yscale=16,yshift=-0.8cm]
\draw[color=red] plot[smooth] coordinates {(0., 1.)  (0.06, 0.837345)  (0.12, 0.780695)  (0.18, 
  0.763216)  (0.24, 0.763985)  (0.3, 0.7738)  (0.36, 0.78789)  (0.42, 
  0.803546)  (0.48, 0.81916)  (0.54, 0.833763)  (0.6, 
  0.846792)  (0.66, 0.857947)  (0.72, 0.86711)  (0.78, 
  0.874284)  (0.84, 0.879562)  (0.9, 0.883092)  (0.96, 
  0.885065)  (1.02, 0.885691)  (1.08, 0.885194)  (1.14, 
  0.883797)  (1.2, 0.881718)  (1.26, 0.879162)  (1.32, 
  0.876317)  (1.38, 0.873353)  (1.44, 0.870416)  (1.5, 
  0.867629)  (1.56, 0.86509)  (1.62, 0.862875)  (1.68, 
  0.861035)  (1.74, 0.859602)  (1.8, 0.858586)  (1.86, 
  0.85798)  (1.92, 0.857763)  (1.98, 0.857901)  (2.04, 
  0.858349)  (2.1, 0.859057)  (2.16, 0.859969)  (2.22, 
  0.861027)  (2.28, 0.862176)  (2.34, 0.863358)  (2.4, 
  0.864524)  (2.46, 0.865628)  (2.52, 0.866631)  (2.58, 
  0.867502)  (2.64, 0.868217)  (2.7, 0.868761)  (2.76, 
  0.869126)  (2.82, 0.869311)  (2.88, 0.869324)  (2.94, 
  0.869178)  (3., 0.868889)  (3.06, 0.868481)  (3.12, 
  0.867978)  (3.18, 0.867407)  (3.24, 0.866795)  (3.3, 
  0.86617)  (3.36, 0.865556)  (3.42, 0.864977)  (3.48, 
  0.864454)  (3.54, 0.864004)  (3.6, 0.86364)  (3.66, 0.86337)  (3.72,
   0.863199)  (3.78, 0.863129)  (3.84, 0.863155)  (3.9, 
  0.863272)  (3.96, 0.863468)  (4.02, 0.863733)  (4.08, 
  0.864051)  (4.14, 0.864407)  (4.2, 0.864785)  (4.26, 
  0.86517)  (4.32, 0.865545)  (4.38, 0.865898)  (4.44, 
  0.866214)  (4.5, 0.866483)  (4.56, 0.866699)  (4.62, 
  0.866854)  (4.68, 0.866945)  (4.74, 0.866973)  (4.8, 
  0.866939)  (4.86, 0.866848)  (4.92, 0.866706)  (4.98, 
  0.86652)  (5.04, 0.8663)  (5.1, 0.866057)  (5.16, 0.8658)  (5.22, 
  0.865541)  (5.28, 0.865289)  (5.34, 0.865054)  (5.4, 
  0.864844)  (5.46, 0.864667)  (5.52, 0.864528)  (5.58, 
  0.86443)  (5.64, 0.864377)  (5.7, 0.864368)  (5.76, 
  0.864401)  (5.82, 0.864474)  (5.88, 0.864582)  (5.94, 0.86472)  (6.,
   0.864881)};
\end{scope}
\draw[->] (-0.4,-1) -- (6.3,-1);
\draw[->] (-0.2,-1.2) -- (-0.2,4);
\node at (0,4.8) (i1) {$|\AMP(u)|$};
\node at (6.7,-1) (i2) {$u$};
\node at (-0.2,-1.6) (i3) {$u_0$};
\node at (2,3.6) (i6) {divergence};
\draw[->] (i6) -- (0.1,3.2);
\end{tikzpicture}
   \end{center}
  \end{minipage}
  \caption{\small This shows (left) $\IM n(u;\omega)$ for QED
in flat spacetime as a function of $u$. Note that for $u\sim
u_0$, $\IM n(u;\omega)$ is positive and divergent.  The right-hand
diagram shows the equivalent behaviour for the amplitude $|\AMP(u)|$,
showing the expected divergence on the initial value surface.}
\label{figQEDFlat}
\end{figure}
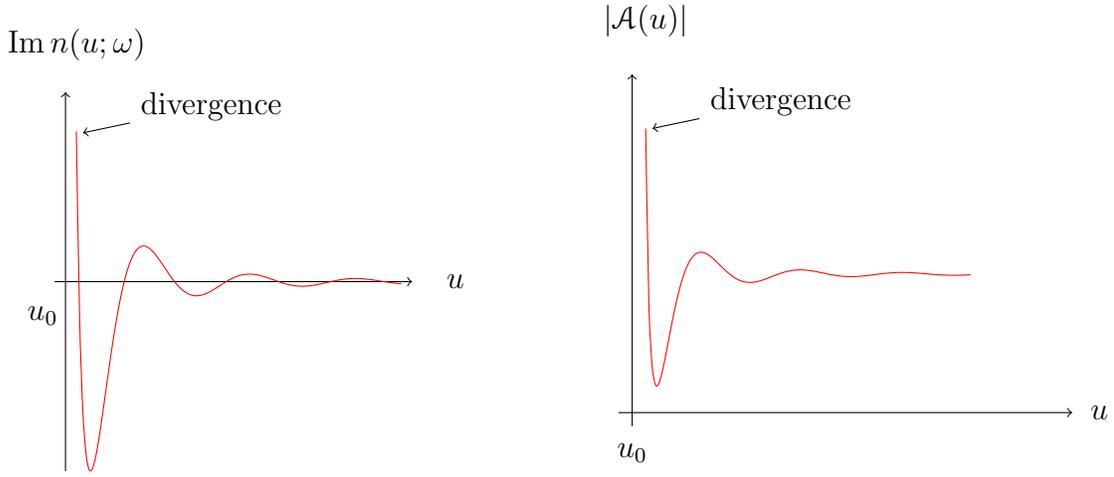

This is plotted in Fig.~\ref{figQEDFlat}. Again we see the
characteristic oscillations which occur when the initial value surface
is taken at finite $u_0$. The key point, however, is that for QED,
$\IM n(u;\omega)$ diverges as $1/(u - u_0)$ for $u$ close to the
initial value surface.  For small $(u-u_0)$,  $\IM n(u;\omega)$ is
positive, so this is consistent with the observation in section 6 of a
divergent initial dressing in a theory like QED which requires wave
function renormalization.

We can see this more explicitly by comparing the amplitude 
$\AMP(u)$ at some small value of $u$ with its value $\AMP(u_1)$ for
some reference value $u_1$ sufficiently removed from the initial value
surface. Evaluating $\IM Q(u)$ from
\EQ{
\IM Q(u_1) - \IM Q(u) = \omega \int_u^{u_1} du^{\prime\prime}~
\IM n(u^{\prime\prime};\omega) \ ,
\label{fq}
}
with $\IM n(u;\omega)$ as in \eqref{fp}, we find
\EQ{
{|\AMP(u)| \over |\AMP(u_1)|} = 
1 - \bigl(\IM Q(u) - \IM Q(u_1)\bigr) \ ,
\label{fr}
}
with
\EQ{
\IM Q(u) - \IM Q(u_1) ~&=
{\alpha\over2\pi}\int_0^1d\xi\,\xi(1-\xi) \biggl(
\Bigl[ (\hat u - {\hat u}_0) {\rm Si} (\hat u - {\hat u}_0)  
+ {\rm Ci} (\hat u - {\hat u}_0) \\ &+ \cos (\hat u - {\hat u}_0)
- {\pi\over2}(\hat u - {\hat u}_0) \Bigr]
- \Bigr[ ~{\hat u} \rightarrow  {\hat u}_1~\Bigr]\biggr).
\label{fs}
}
This is shown in Fig.~\ref{figQEDFlat}. For small $(u-u_0)$, the
dominant term arises from the ${\rm Ci}$ function, which diverges 
logarithmically, and we find 
\EQ{
{|\AMP(u)| \over |\AMP(u_1)|} = 1 - {\alpha\over12\pi}~
\log (u - {u}_0) \ .
\label{ft}
}
We can exponentiate this using the dynamical renormalization group and
conclude that for small $(u - u_0)$,
\EQ{
{|\AMP(u)| \over |\AMP(u_1)|} ~\sim~ 
(u - u_0)^{-\tfrac{\alpha}{12\pi}} \ .
\label{fu}
}
The physical picture, as inferred in section 6, is that in order to
obtain a finite evolution of the photon field in QED, we have to start 
from a divergent value on the initial value surface itself. This is
the expected behaviour for a theory requiring field (wave-function)
renormalization. Nevertheless, we can still study the evolution of the
field amplitude in QED, either by simply taking $u_0\rightarrow
-\infty$ as in our previous work, or by comparing the ratio of
$|\AMP(u)|$ to its value at a fixed reference point away from the
initial value surface where the interaction is assumed to be switched
on. Of course, this is the essential idea behind the Type II or III
initial value conditions introduced in section 6.

\subsection{Cosmological FRW spacetimes}

As a first example, consider a spatially-flat FRW spacetime.
As shown in ref.\cite{Hollowood:2009qz}, the Penrose limit 
for null geodesics with no angular momentum in the transverse
space is a conformally-flat, singular homogeneous plane wave 
with profile function\footnote{The coefficient $\gamma$ is related
to the usual $w$ parameter in the FRW equation of state 
$p = w\rho$ by $\gamma = \tfrac{2}{3(1 + w)}$, so that 
$\gamma = \frac23, \frac12$ or $\infty$ for a matter, radiation or
cosmological constant dominated universe respectively.
The FRW scale factor is $a(t) \sim t^\gamma$.  
The interesting example of the Milne universe considered
in ref.\cite{Hollowood:2009qz} with $\gamma = 1$ corresponds 
to $w = -\frac13$, at the boundary between a decelerating and 
accelerating universe. The 
$\tilde\alpha\equiv\alpha_1=\alpha_2$ (to avoid confusion with the fine structure constant) parameter in the profile function is
$\tilde\alpha = \bigl|{1-\gamma\over 1+ \gamma}\bigr|$.
The null coordinate $u$ in the Penrose limit is related to 
$t$ in the FRW metric by $u \sim t^{\gamma + 1}$.
The flat space limit is $\gamma=0$, corresponding to 
$\tilde\alpha = 1$.}
\EQ{
h_{ij} = {\gamma\over(\gamma + 1)^2} {1\over u^2} \delta_{ij} \ .
\label{ffa} 
}
Spatially flat FRW spacetimes are special in that the Penrose 
limit is a singular homogeneous plane wave for all $u$, not
simply in the near-singularity limit $u \simeq 0$.

We are considering an initial value problem where the interaction
is turned on at some finite value $u_0>0$, since clearly we have to
start away from the initial singularity.  We then study the development 
of the field amplitude along the null geodesic as the universe
evolves. 

\noindent{\it $A\phi^2$ theory:}

Specialising first to $A\phi^2$ theory with $M=0$, and making the convenient
change of variable from $t=u-u'$ to $\hat t = 1 - \tfrac{u'}{u}
= 1 -r$, we find from \eqref{fe},
\EQ{
\IM {\cal F}(u;z) = \IM \int_{0-i\epsilon}^{1-\tfrac{u_0}{u}} 
{d\hat t\over \hat t}~ e^{-i z u \hat t} F(\hat t) \ ,
\label{ffb}
}
where $z = \tfrac{m^2}{2\omega \xi(1-\xi)}$ and
\EQ{
F(\hat t) = {\tilde\alpha\hat t (1 - \hat t)^{\tfrac{\tilde\alpha-1}{2}} \over
1 - (1 - \hat t)^{\tilde\alpha}} \ .
\label{ffc}
}
The integral requires careful treatment of the pole at $\hat t = 0$
(see footnote \ref{fn1}) and follows the prescription described following
\eqref{eh}. There, in flat spacetime, the $t$-integral ran from
$-\infty$ to $\infty$ with the contour running below the 
real axis -- this ensured that below threshold ($z>0$) the contour
could be closed in the lower half-plane so that $\IM n(u;\omega)$
vanished. Here, the same $i\epsilon$ prescription in \eqref{ffb}
picks up a contribution $\tfrac{1}{2}\pi i \,{\rm Res}(\hat t = 0)$
from the pole at $t=0$, and since $F(0) = 1$, we find
\EQ{
\IM {\cal F}(u;z) = {\pi\over2} -
\int_0^{1-\tfrac{u_0}{u}} {d\hat t\over \hat t}~
\sin(z u \hat t) F(\hat t) \ .
\label{ffd}
}

\begin{figure}[t]
    \begin{center}
\begin{tikzpicture}[scale=0.9]
\begin{scope}[xscale=1,yscale=40]
\draw[->] (-0.2,0) -- (7.5,0);
\draw[->] (0,-0.06) --(0,0.06);
\draw[color=red] plot[smooth] coordinates {(0., -2.90309*10^-7)  (0.06, -0.00107675)  (0.12, -0.00448971)   
(0.18, -0.00947875)  (0.24, -0.0153174)  (0.3, -0.0214352)  (0.36,  
-0.0273686)  (0.42, -0.032735)  (0.48, -0.0372209)  (0.54,  
-0.0405786)  (0.6, -0.0426261)  (0.66, -0.0432469)  (0.72,  
-0.0423898)  (0.78, -0.0400675)  (0.84, -0.0363533)  (0.9,  
-0.0313764)  (0.96, -0.0253156)  (1.02, -0.0183913)  (1.08,  
-0.0108562)  (1.14, -0.00298519)  (1.2, 0.00493525)  (1.26, 
  0.0126183)  (1.32, 0.0197874)  (1.38, 0.0261867)  (1.44, 
  0.0315907)  (1.5, 0.0358119)  (1.56, 0.0387079)  (1.62, 
  0.0401861)  (1.68, 0.0402068)  (1.74, 0.0387844)  (1.8, 
  0.035986)  (1.86, 0.0319288)  (1.92, 0.026775)  (1.98, 
  0.0207258)  (2.04, 0.0140133)  (2.1, 
  0.00689149)  (2.16, -0.000372946)  (2.22, -0.00751096)  (2.28,  
-0.014261)  (2.34, -0.0203788)  (2.4, -0.0256459)  (2.46, -0.029878)   
(2.52, -0.0329311)  (2.58, -0.0347066)  (2.64, -0.0351547)  (2.7,  
-0.0342751)  (2.76, -0.0321175)  (2.82, -0.0287784)  (2.88,  
-0.0243973)  (2.94, -0.0191515)  (3., -0.0132482)  (3.06,  
-0.00691736)  (3.12, -0.000402104)  (3.18, 0.00604999)  (3.24, 
  0.0121965)  (3.3, 0.017809)  (3.36, 0.0226817)  (3.42, 
  0.0266391)  (3.48, 0.0295415)  (3.54, 0.0312911)  (3.6, 
  0.0318342)  (3.66, 0.0311635)  (3.72, 0.0293173)  (3.78, 
  0.0263784)  (3.84, 0.0224699)  (3.9, 0.0177502)  (3.96, 
  0.012407)  (4.02, 0.00664995)  (4.08, 
  0.00070198)  (4.14, -0.00520875)  (4.2, -0.0108577)  (4.26,  
-0.0160326)  (4.32, -0.020541)  (4.38, -0.0242179)  (4.44,  
-0.0269312)  (4.5, -0.028587)  (4.56, -0.0291322)  (4.62,  
-0.0285568)  (4.68, -0.0268935)  (4.74, -0.024216)  (4.8,  
-0.0206362)  (4.86, -0.0162992)  (4.92, -0.0113779)  (4.98,  
-0.00606589)  (5.04, -0.000569893)  (5.1, 0.00489808)  (5.16, 
  0.010129)  (5.22, 0.0149245)  (5.28, 0.0191049)  (5.34, 
  0.0225153)  (5.4, 0.0250317)  (5.46, 0.0265652)  (5.52, 
  0.0270655)  (5.58, 0.0265221)  (5.64, 0.0249646)  (5.7, 
  0.0224612)  (5.76, 0.0191158)  (5.82, 0.0150637)  (5.88, 
  0.0104662)  (5.94, 0.00550432)  (6., 0.000371573)};
\begin{scope}[yscale=2.5]
\draw[color=blue] plot[smooth] coordinates {(0., -0.0000665627)  (0.06, -0.0107896)  (0.12, -0.0168271)  (0.18,  
-0.0199519)  (0.24, -0.0211303)  (0.3, -0.0209458)  (0.36,  
-0.0197841)  (0.42, -0.0179208)  (0.48, -0.0155668)  (0.54,  
-0.0128911)  (0.6, -0.0100344)  (0.66, -0.00711622)  (0.72,  
-0.00423855)  (0.78, -0.00148824)  (0.84, 0.00106189)  (0.9, 
  0.00335246)  (0.96, 0.00533716)  (1.02, 0.00698239)  (1.08, 
  0.0082669)  (1.14, 0.00918129)  (1.2, 0.00972732)  (1.26, 
  0.00991709)  (1.32, 0.009772)  (1.38, 0.00932165)  (1.44, 
  0.00860243)  (1.5, 0.00765617)  (1.56, 0.00652866)  (1.62, 
  0.00526813)  (1.68, 0.00392374)  (1.74, 0.00254406)  (1.8, 
  0.00117583)  (1.86, -0.000137439)  (1.92, -0.00135649)  (1.98,  
-0.00244732)  (2.04, -0.00338187)  (2.1, -0.00413865)  (2.16,  
-0.00470292)  (2.22, -0.00506696)  (2.28, -0.00522986)  (2.34,  
-0.00519724)  (2.4, -0.00498077)  (2.46, -0.00459751)  (2.52,  
-0.00406911)  (2.58, -0.00342091)  (2.64, -0.00268098)  (2.7,  
-0.0018791)  (2.76, -0.00104568)  (2.82, -0.000210828)  (2.88, 
  0.000596686)  (2.94, 0.00135027)  (3., 0.00202628)  (3.06, 
  0.00260473)  (3.12, 0.00306971)  (3.18, 0.00340983)  (3.24, 
  0.00361837)  (3.3, 0.00369332)  (3.36, 0.0036373)  (3.42, 
  0.00345728)  (3.48, 0.00316415)  (3.54, 0.00277228)  (3.6, 
  0.00229886)  (3.66, 0.00176322)  (3.72, 0.00118615)  (3.78, 
  0.000589105)  (3.84, -6.52617*10^-6)  (3.9, -0.000580135)  (3.96,  
-0.00111257)  (4.02, -0.00158676)  (4.08, -0.00198815)  (4.14,  
-0.00230522)  (4.2, -0.00252971)  (4.26, -0.00265679)  (4.32,  
-0.00268519)  (4.38, -0.00261703)  (4.44, -0.00245774)  (4.5,  
-0.00221572)  (4.56, -0.00190197)  (4.62, -0.00152966)  (4.68,  
-0.00111358)  (4.74, -0.000669608)  (4.8, -0.000214145)  (4.86, 
  0.000236499)  (4.92, 0.000666637)  (4.98, 0.00106173)  (5.04, 
  0.00140888)  (5.1, 0.0016972)  (5.16, 0.00191814)  (5.22, 
  0.00206575)  (5.28, 0.00213678)  (5.34, 0.00213073)  (5.4, 
  0.00204983)  (5.46, 0.00189884)  (5.52, 0.00168486)  (5.58, 
  0.001417)  (5.64, 0.00110604)  (5.7, 0.000764001)  (5.76, 
  0.000403674)  (5.82, 
  0.0000381931)  (5.88, -0.000319444)  (5.94, -0.000656831)  (6.,  
-0.000962562)};
\end{scope}
\end{scope}
\node at (0,3.5) (i1) {$\IM n(u;\omega)$};
\node at (7.8,0) (i2) {$u$};
\node at (-0.4,-0.3) (i3) {$u_0$};
\node[color=red] at (2,2.5) (i4) {$A\phi^2$};
\draw[->,color=red] (i4) -- (1.8,1.8);
\node[color=blue] at (4,2.5) (i5) {QED};
\draw[->,color=blue] (i5) -- (4.5,0.1);
\end{tikzpicture}
\end{center}
\caption{\small The curvature contribution
to $\IM n(u;\omega)$ plotted as a function of $u-u_0$, 
for $A\phi^2$  and QED , for a fixed 
choice of $u_0$ in a FRW spacetime with $\tilde\alpha = \frac13$. 
Note that $\IM n(u;\omega)$ 
can be locally negative, corresponding to a curvature-induced
``undressing" of the quantum field.}
\label{figFRWcurved}
\end{figure}
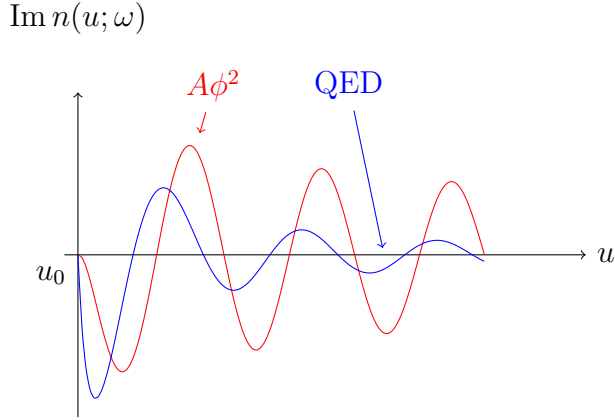

The refractive index is then
\EQ{
\IM n(u;\omega) = {e^2\over (4\pi)^2} {1\over 4\omega^2}~
\int_0^1 d\xi~\biggl[{\pi\over2} - \int_0^{1-\tfrac{u_0}{u}} 
{d\hat t\over \hat t}~\sin(z u \hat t) F(\hat t) \biggr] \ .
\label{ffe}
}
The integrals over $\hat t$ and $\xi$ can now be performed
numerically for various values of the FRW parameter $\tilde\alpha$. In fact, the function $F(\hat t)$ is very flat over most of the range
$t = 0$ to $1$.  The refractive index therefore has the same
qualitative behaviour as shown in Fig.~\ref{figFLAT}. 
In particular, $\IM n(u;\omega)$ falls from an initially positive
value and oscillates as $u$ increases away from $u_0$, and is locally
negative. In those regions, ${\cal Q}^{(1)}(u)$ decreases and in turn the 
amplitude $|\AMP(u)|$ increases. 
In Fig.~\ref{figFRWcurved}, we subtract the 
transient flat-spacetime result and plot the curvature-induced
contribution to $\IM n(u;\omega)$, viz.
\EQ{
\IM n(u;\omega)\Big|_{\text{curved}} 
= -{e^2\over (4\pi)^2} {1\over 4\omega^2}~
\int_0^1 d\xi~\int_0^{1-\tfrac{u_0}{u}} 
{d\hat t\over \hat t}~\sin(z u \hat t) \big[F(\hat t) - 1\big]\ .
\label{ffg}
}

This shows clearly how the time-varying gravitational tidal forces
induce both a local dressing and undressing of the renormalized 
quantum field. The amplitude $\AMP(u)$ decreases or increases 
as the screening due to the cloud of virtual $\phi$ pairs is enhanced 
or reduced.

\vskip0.2cm

\noindent{\it Quantum Electrodynamics:}

The situation is very similar for QED and the behaviour of $\IM
n(u;\omega)$ and $|\AMP(u)|$ in the FRW background is qualitatively the
same as in flat spacetime, including the divergence for $u \sim u_0$.
Both polarizations have the same refractive index, since the
homogeneous plane wave in the Penrose limit is conformally flat.

If we split $\IM n(u;\omega)$ into the flat spacetime result already
found and the curvature-dependent contribution, we find
\EQ{
\IM n(u;\omega)\big|_\text{curved} = - {\alpha\over\pi} 
{1\over 2\omega} \int_0^1 d\xi\, \xi (1-\xi) ~{1\over u} 
\int_0^{1- \tfrac{u}{u_0}} {d\hat t\over \hat t^2}~\cos(z u \hat t)
\big[F(\hat t) - 1\big] \ ,
\label{ffh}
}
where
\EQ{
F(\hat t) = \Delta_{ii}(r) \sqrt{\Delta(r)} = 
{\tilde\alpha^2{\hat t}^2 (1 - \hat t)^{\tilde\alpha-1}\over
\bigl[1 - (1-\hat t)^{\tilde\alpha}\bigr]^2} \ ,
\label{ffi}
}
recalling $\hat t = 1-r$. Notice that the
subtraction $-1$ in the integrand in \eqref{ffh} is the 
flat spacetime contribution, not the contribution of the extra vacuum
polarization diagram for scalar QED. The $\hat t$ integration is
well-defined here, since the properties $F(0) = 1$, $F'(0)=0$ ensure
the absence of a singularity at $\hat t=0$.
This result for $\IM n(u;\omega)$ is plotted in Fig.~\ref{figFRWcurved}.  
The results are very similar to the scalar $A\phi^2$ theory, with an 
enhancement for $u\sim u_0$ for QED reflecting the
usual power counting $\hat t^{-2}$ factor in \eqref{ffh}.

\subsection{Black holes and the near-singularity limit}

A particularly interesting question is what happens to a dressed photon
as it approaches a spacetime singularity. This can be addressed in a
similar way, using the singular homogeneous plane wave as the Penrose
limit of a Schwarzschild black hole in the near-singularity limit. For this case, we have $\alpha_1=\frac15$ and $\alpha_2=\frac75$ and the geodesic spray defined above \eqref{pwl} is described by
\EQ{
A_{11}(u,u')&=5(uu')^{2/5}\Big(u^{1/5}-u^{\prime \,1/5}\Big)\ ,\\
A_{22}(u,u')&=\frac57(uu')^{-1/5}\Big(u^{7/5}-u^{\prime \,7/5}\Big)\ .
}
In the direction $z^1$, the geodesics focus on the point $u=0$ which is the singularity. For the other direction $z^2$, the geodesics diverge as the singularity is approached. If we define the usual Schwarzschild coordinates $(t,r,\theta,\phi)$ then by symmetry we can take the geodesics in the plane $\phi=0$. The transverse space-like coordinates $z^i$ are related to these coordinates via $z^1=r\sin\theta$ and $z^2=rdr/du$. Here, $z^1$ is orthogonal to the plane of the orbit while $z^2$ lies in the plane of the orbit \cite{Hollowood:2009qz}.

\noindent{\it $A\phi^2$ theory:}

This time, therefore, we take the initial value surface $u_0 < 0$ and 
study $\IM n(u;\omega)$ as $u$ approaches zero. In this case, the
appropriate integration variable is ${\hat t} = r-1 = -\frac tu$, and we have
\EQ{
\IM {\cal F}(u;z) =
\IM \int_{0-i\epsilon}^{\tfrac{u_0}{u} - 1} {d {\hat t}\over {\hat t}}~
e^{-i z |u| {\hat t}}~ F({\hat t}) \ ,
\label{fffa}
}
where
\EQ{
F({\hat t}) = 
\sqrt{\alpha_1 \alpha_2 {\hat t}^2(1 + {\hat t})^{-p} \over
\bigl((1+{\hat t})^{\alpha_1} - 1\bigr)
\bigl((1+{\hat t})^{\alpha_2}-1\bigr)} \ .
\label{fffb}
}
The VVM determinant function $F({\hat t})$ satisfies $F(0) = 1$,
$F'(0) = 0$ and $F({\hat t}) \simeq \sqrt{\alpha_1 \alpha_2}~ 
{\hat t}^{{p/2}}$ at large ${\hat t}$. Note that $p$ for the background geometry is defined in \eqref{fb}.

We are interested in the near-singularity behaviour of the refractive
index, so now let $u_0\rightarrow -\infty$. The integral over 
${\hat t}$ can be expressed as in \eqref{ffd}, but here, since the
upper limit of integration becomes infinity, it is possible to rotate
the contour in the complex $\hat t$ plane to run down the negative 
imaginary axis. Then, with ${\tau} = i{\hat t}$, we have the
computationally more convenient form,
\EQ{
\IM{\cal F}(u;z) = \int_0^\infty {d{\tau}\over{\tau}}~
e^{-z |u| {\tau}} G({\tau}) \ ,
\label{fffc}
} 
where $G({\tau}) = \IM F(-i{\tau})$.  
Note that the lower limit is safe from singularities since $G(0) = 0$.
At large ${\tau}$, $G({\tau}) = - 
\sqrt{\alpha_1 \alpha_2}~ \sin{\tfrac{\pi p}{4}}~
{\tau}^{{p/2}}~\bigl(1 + 
{\cal O}\bigl({\tau}^{-{\text {min}}(\alpha_1,\alpha_2)}\bigr)\bigr)$ 
and, crucially, is negative provided $p > 0$. For Schwarzschild,
$p = \frac15$.

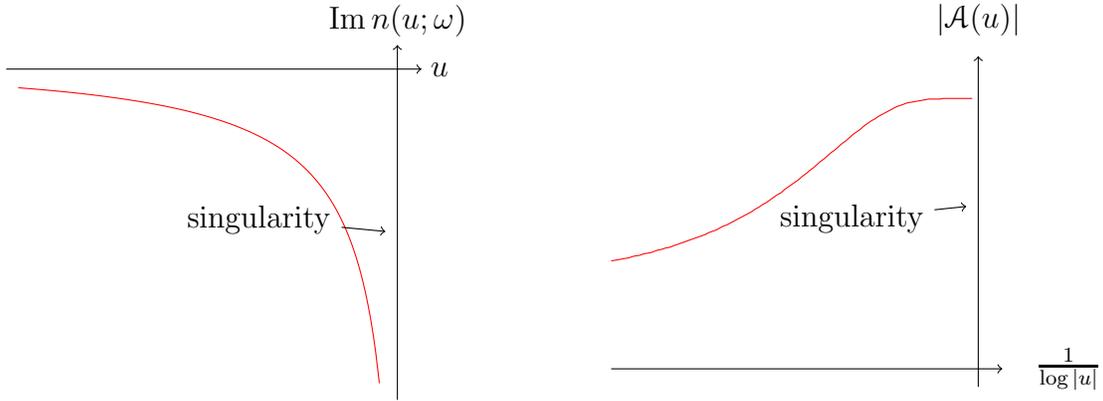
\begin{figure}[t]
  \begin{minipage}[t]{.47\textwidth}
    \begin{center}
\begin{tikzpicture}[scale=0.8]
\begin{scope}[xscale=1,yscale=40]
\draw[color=red] plot[smooth] coordinates {(0., -0.00775784) (0.06, -0.00787427) (0.12, -0.00799306) (0.18,  
-0.00811428) (0.24, -0.00823799) (0.3, -0.00836427) (0.36,  
-0.00849318) (0.42, -0.00862481) (0.48, -0.00875923) (0.54,  
-0.00889653) (0.6, -0.00903678) (0.66, -0.00918008) (0.72,  
-0.00932651) (0.78, -0.00947617) (0.84, -0.00962916) (0.9,  
-0.00978558) (0.96, -0.00994554) (1.02, -0.0101091) (1.08,  
-0.0102765) (1.14, -0.0104478) (1.2, -0.010623) (1.26,  
-0.0108024) (1.32, -0.0109861) (1.38, -0.0111742) (1.44,  
-0.0113669) (1.5, -0.0115643) (1.56, -0.0117666) (1.62,  
-0.0119739) (1.68, -0.0121864) (1.74, -0.0124044) (1.8,  
-0.012628) (1.86, -0.0128574) (1.92, -0.0130929) (1.98,  
-0.0133346) (2.04, -0.0135828) (2.1, -0.0138377) (2.16,  
-0.0140996) (2.22, -0.0143688) (2.28, -0.0146456) (2.34,  
-0.0149302) (2.4, -0.015223) (2.46, -0.0155244) (2.52,  
-0.0158346) (2.58, -0.0161541) (2.64, -0.0164832) (2.7,  
-0.0168224) (2.76, -0.0171721) (2.82, -0.0175329) (2.88,  
-0.0179051) (2.94, -0.0182894) (3., -0.0186862) (3.06,  
-0.0190963) (3.12, -0.0195202) (3.18, -0.0199586) (3.24,  
-0.0204123) (3.3, -0.020882) (3.36, -0.0213687) (3.42,  
-0.0218731) (3.48, -0.0223962) (3.54, -0.0229391) (3.6,  
-0.0235029) (3.66, -0.0240888) (3.72, -0.0246981) (3.78,  
-0.0253322) (3.84, -0.0259925) (3.9, -0.0266808) (3.96,  
-0.0273988) (4.02, -0.0281486) (4.08, -0.0289321) (4.14,  
-0.0297517) (4.2, -0.03061) (4.26, -0.0315097) (4.32, -0.0324541)  
(4.38, -0.0334463) (4.44, -0.0344903) (4.5, -0.0355902) (4.56,  
-0.0367506) (4.62, -0.0379769) (4.68, -0.0392747) (4.74,  
-0.0406509) (4.8, -0.0421128) (4.86, -0.0436689) (4.92,  
-0.0453289) (4.98, -0.0471039) (5.04, -0.049007) (5.1,  
-0.0510528) (5.16, -0.0532591) (5.22, -0.0556465) (5.28,  
-0.0582393) (5.34, -0.0610672) (5.4, -0.0641657) (5.46,  
-0.0675787) (5.52, -0.0713604) (5.58, -0.0755794) (5.64,  
-0.0803237) (5.7, -0.0857083) (5.76, -0.0918875) (5.82,  
-0.099074) (5.88, -0.107571) (5.94, -0.117832) (6., -0.13057)};
\end{scope}
\draw[->] (-0.2,0) -- (6.7,0);
\draw[<-] (6.3,0.4) -- (6.3,-5.5);
\node at (6.3,0.8) (i1) {$\IM n(u;\omega)$};
\node at (7,0) (i2) {$u$};
\node at (4,-2.5) (i3) {singularity};
\draw [->] (i3) -- (6.1,-2.7);
\end{tikzpicture}
    \end{center}
  \end{minipage}
  \hfill
  \begin{minipage}[t]{.47\textwidth}
    \begin{center}
\begin{tikzpicture}[scale=0.8]
\begin{scope}[xscale=1,yscale=700]
\draw[color=red] plot[smooth] coordinates {(0., 0.00255981)  (0.06, 0.00257613)  (0.12, 0.00259294)  (0.18, 
  0.00261026)  (0.24, 0.0026281)  (0.3, 0.00264649)  (0.36, 
  0.00266543)  (0.42, 0.00268495)  (0.48, 0.00270507)  (0.54, 
  0.0027258)  (0.6, 0.00274717)  (0.66, 0.00276918)  (0.72, 
  0.00279187)  (0.78, 0.00281526)  (0.84, 0.00283936)  (0.9, 
  0.00286419)  (0.96, 0.00288979)  (1.02, 0.00291617)  (1.08, 
  0.00294335)  (1.14, 0.00297137)  (1.2, 0.00300023)  (1.26, 
  0.00302998)  (1.32, 0.00306063)  (1.38, 0.0030922)  (1.44, 
  0.00312473)  (1.5, 0.00315823)  (1.56, 0.00319274)  (1.62, 
  0.00322828)  (1.68, 0.00326486)  (1.74, 0.00330253)  (1.8, 
  0.00334131)  (1.86, 0.00338121)  (1.92, 0.00342226)  (1.98, 
  0.00346449)  (2.04, 0.00350791)  (2.1, 0.00355256)  (2.16, 
  0.00359844)  (2.22, 0.00364558)  (2.28, 0.00369399)  (2.34, 
  0.00374368)  (2.4, 0.00379468)  (2.46, 0.00384698)  (2.52, 
  0.0039006)  (2.58, 0.00395554)  (2.64, 0.00401179)  (2.7, 
  0.00406934)  (2.76, 0.0041282)  (2.82, 0.00418834)  (2.88, 
  0.00424974)  (2.94, 0.00431236)  (3., 0.00437619)  (3.06, 
  0.00444117)  (3.12, 0.00450724)  (3.18, 0.00457436)  (3.24, 
  0.00464244)  (3.3, 0.00471142)  (3.36, 0.00478118)  (3.42, 
  0.00485164)  (3.48, 0.00492268)  (3.54, 0.00499417)  (3.6, 
  0.00506596)  (3.66, 0.0051379)  (3.72, 0.00520983)  (3.78, 
  0.00528156)  (3.84, 0.0053529)  (3.9, 0.00542363)  (3.96, 
  0.00549355)  (4.02, 0.00556242)  (4.08, 0.00563)  (4.14, 
  0.00569604)  (4.2, 0.0057603)  (4.26, 0.00582251)  (4.32, 
  0.00588243)  (4.38, 0.00593981)  (4.44, 0.00599441)  (4.5, 
  0.00604601)  (4.56, 0.00609441)  (4.62, 0.00613943)  (4.68, 
  0.00618093)  (4.74, 0.0062188)  (4.8, 0.00625298)  (4.86, 
  0.00628344)  (4.92, 0.00631022)  (4.98, 0.00633342)  (5.04, 
  0.00635317)  (5.1, 0.00636966)  (5.16, 0.00638315)  (5.22, 
  0.00639391)  (5.28, 0.00640226)  (5.34, 0.00640854)  (5.4, 
  0.00641307)  (5.46, 0.00641617)  (5.52, 0.00641814)  (5.58, 
  0.00641928)  (5.64, 0.00641989)  (5.7, 0.00642018)  (5.76, 
  0.00642031)  (5.82, 0.00642036)  (5.88, 0.00642037)  (5.94, 
  0.00642038)  (6., 0.00642038)};
\end{scope}
\draw[->] (0,0) -- (6.5,0);
\draw[->] (6.1,-0.3) --(6.1,5.2);
\node at (6.1,5.8) (i1) {$|\AMP(u)|$};
\node at (7.6,0) (i2) {$\frac1{\log|u|}$};
\node at (4,2.5) (i3) {singularity};
\draw [->] (i3) -- (5.9,2.7);
\end{tikzpicture}
   \end{center}
  \end{minipage}
  \caption{\small Plot of the imaginary part of the refractive
  index $\IM n(u;\omega)$ in $A\phi^2$ theory in a Schwarzschild
  background as $u$ approaches the singularity at $u=0$ from $u<0$.
  The right-hand diagram shows the corresponding result for the
  amplitude $|\AMP(u)|$ (plotted on an inverse log scale) showing the
  bounded behaviour following from $Q(u) \rightarrow - {\rm constant}$
  as $u\rightarrow 0$.}\label{figBHAphi2}
\end{figure}

The resulting form for $\IM n(u;\omega)$ is shown in Fig.~\ref{figBHAphi2}.
Note immediately that $\IM n(u;\omega) < 0$ and diverges for small
$u$. A key point then is whether the amplitude remains bounded, 
{\it i.e.}~whether $\IM {\cal Q}^{(1)}(u)$, given here by
\EQ{
\IM {\cal Q}^{(1)}(u) = \omega \int_{-\infty}^u du^{\prime\prime} 
\IM n(u^{\prime\prime};\omega) \ ,
\label{fffd}
}
remains finite as $u\rightarrow 0$. This is shown in the second 
plot in Fig.~\ref{figBHAphi2}, confirming that $\IM {\cal Q}^{(1)}(u) 
\rightarrow -{\rm constant}$ as $u \rightarrow 0$, ensuring that
$|\AMP(u)|$ is bounded. 

To see this explicitly, we can rescale the integral \eqref{fffc}
to get
\EQ{
\IM {\cal F}(u;z) = \int_0^\infty {d\tau\over\tau}~e^{-\tau}
G\Big(\frac{\tau}{z|u|}\Big) \ ,
\label{fffe}
}
then pick out the leading small $|u|$ behaviour from the leading term
in the expression for $G({\tau})$ for large ${\tau}$.   This
gives
\EQ{
\IM{\cal F}(u;z)~\simeq~ -\sqrt{\alpha_1 \alpha_2} ~\sin\Big(\frac{\pi p}{4}\Big)~
\Gamma\Big(\frac{p}{2}\Big) 
{1\over (z|u|)^{\tfrac{p}{2}}} \ .
\label{ffff}
}
In practice, since $\alpha_1 = \frac15$ for the Schwarzschild black hole,
the approach to the asymptotic limit is slow. However, this limit
is sufficient to show that $\IM {\cal Q}^{(1)}(u)$ is bounded and tends to a
constant as $|u|\rightarrow 0$ provided $0<\frac p2 < 1$, which is
ensured by $0<\alpha_i <1$.

The physical interpretation is as follows. With the initial value
surface set at $u_0\rightarrow -\infty$, the incoming photon is
already fully dressed as it approaches the small $u$, near-singularity
region. Here, $\IM n(u;\omega)$ is negative, corresponding to an
increasing amplitude $|\AMP(u)|$. The photon is becoming ``undressed",
i.e.~the gravitational tidal forces are stripping away its cloud
of virtual $\phi$ pairs. This reduces the level of screening and the
renormalized field amplitude increases as it reverts towards its bare
state. Since this process cannot continue indefinitely, the increase
in $|\AMP(u)|$ must be bounded, as we find. Once again, all this
is consistent with the interpretation in sections 4 and 5 of the
curved-spacetime generalisation of the optical theorem.

\noindent{\it Quantum Electrodynamics:}

Finally, we consider QED itself and look at the evolution of a dressed
photon as it approaches a Schwarzschild singularity at $u=0$.
Here, once again taking $u_0\rightarrow -\infty$, 
\EQ{
\IM n_{ij}(u;\omega) = - {\alpha\over\pi} {1\over 2\omega}
\int_0^1 d\xi\, \xi(1-\xi)~\IM {\cal F}_{ij}(u;z) \ ,
\label{fffg}
}
with
\EQ{
{\cal F}_{ij}(u;z) = \int_0^\infty {dt\over t^2} i e^{-i z t}~
\Bigl[\Delta_{ij}(u,u-t) \sqrt{\Delta(u,u-t)} -\delta_{ij} \Bigr]
\label{fffh2}
}
and with no loss-of-generality we can take ${\cal F}_{ij}(u;z) =
{\cal F}_i(u;z) \delta_{ij}$.

We use the same rescaling $\hat t = r-1 = -t/u$ as above, so that
\EQ{
{\cal F}_i(u;z) = {1\over|u|} \int_0^\infty  
{d\hat t\over {\hat t}^2}~i e^{-i z |u| {\hat t}} \bigl(F_i(\hat t) - 1\bigr)
\ ,
\label{fffg2}
}
where for the first polarization,
\EQ{
F_1(\hat t) = \sqrt{
{\alpha_1^3 \alpha_2{\hat t}^4 (1 + {\hat t})^{- q_1} \over 
\bigl((1+ {\hat t})^{\alpha_1} - 1\bigr)^3
\bigl((1 + {\hat t})^{\alpha_2} - 1\bigr)}}\ ,
\label{fffh}
}
with $q_1 = 2 - \tfrac{(3\alpha_1 + \alpha_2)}{2}$.
The equivalent result holds for the second polarization.
It is straightforward to check that at small $\hat t$,
$F_i(0) = 1, ~F'(0) = 0$ while at large $\hat t$,
$F_1(\hat t) \rightarrow \sqrt{\alpha_1^3 \alpha_2}~
{\hat t}^{{q_1}/{2}}$ and
$F_2(\hat t) \rightarrow \sqrt{\alpha_1 \alpha_2^3}~
{\hat t}^{{q_2}/{2}}$. 

These properties of $F_i(\hat t)$ ensure that the $\hat t$
integral in \eqref{fffg2} is non-singular at $\hat t = 0$, and since
the upper limit in this case is infinity (because we have taken
$u_0\rightarrow -\infty$) we can again rotate the contour just 
as in the $A\phi^2$ example and find
\EQ{
{\cal F}_i(u;z) = - {1\over|u|} \int_0^\infty {d\tau\over\tau^2}~
e^{-z |u|\tau}~\bigl(F_i(-i\tau) - 1\bigr) \ ,
\label{fffi}
}
and so 
\EQ{
\IM {\cal F}_i(u;z) = - {1\over|u|} \int_0^\infty
{d\tau\over\tau^2}~
e^{-z|u|\tau}~ G_i(\tau) \ ,
\label{fffj}
}
where we define the functions $G_i(\tau) = \IM F_i(-i\tau)$.
At small $\tau$, $G_i(\tau) = {\cal O}(\tau^2)$, while for 
Schwarzschild spacetime with $\alpha_1 = \frac15, \alpha_2 = \frac75$,
we find for large $\tau$,
\EQ{
&G_1(\tau) \rightarrow -\sqrt{\alpha_1^3 \alpha_2}~
\sin\Bigl({\pi q_1\over4}\Bigr)~\tau^{\tfrac{q_1}{2}} =
- {\sqrt{7}\over25}~\sin\Bigl({\pi\over4}\Bigr)~
\tau^{\tfrac{1}{2}} ~<~0\\
&G_2(\tau) \rightarrow -\sqrt{\alpha_1 \alpha_2^3}~
\sin\Bigl({\pi q_2\over4}\Bigr)~\tau^{\tfrac{q_2}{2}} =
{7\sqrt{7}\over25}~\sin\Bigl({\pi\over20}\Bigr)~
\tau^{-\tfrac{1}{10}} ~>~0 \ .
\label{fffk}
}

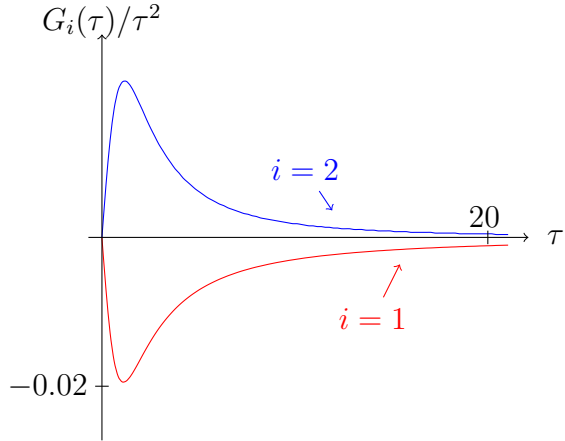
\begin{figure}[t]
    \begin{center}
\begin{tikzpicture}[scale=0.9]
\begin{scope}[xscale=1,yscale=100]
\draw[color=red] plot[smooth] coordinates {(0., -0.0000402409)  (0.06, -0.00776611)  (0.12, -0.0140686)  (0.18,  
-0.0182872)  (0.24, -0.020551)  (0.3, -0.0213588)  (0.36,  
-0.0212274)  (0.42, -0.0205519)  (0.48, -0.0195951)  (0.54,  
-0.0185185)  (0.6, -0.0174164)  (0.66, -0.0163412)  (0.72,  
-0.0153198)  (0.78, -0.0143647)  (0.84, -0.0134796)  (0.9,  
-0.0126638)  (0.96, -0.0119138)  (1.02, -0.0112252)  (1.08,  
-0.010593)  (1.14, -0.0100123)  (1.2, -0.00947832)  (1.26,  
-0.00898666)  (1.32, -0.00853328)  (1.38, -0.00811451)  (1.44,  
-0.00772707)  (1.5, -0.007368)  (1.56, -0.00703464)  (1.62,  
-0.00672462)  (1.68, -0.00643584)  (1.74, -0.0061664)  (1.8,  
-0.0059146)  (1.86, -0.00567893)  (1.92, -0.00545802)  (1.98,  
-0.00525065)  (2.04, -0.00505572)  (2.1, -0.00487224)  (2.16,  
-0.00469931)  (2.22, -0.00453612)  (2.28, -0.00438194)  (2.34,  
-0.00423609)  (2.4, -0.00409798)  (2.46, -0.00396705)  (2.52,  
-0.0038428)  (2.58, -0.00372476)  (2.64, -0.00361253)  (2.7,  
-0.00350571)  (2.76, -0.00340395)  (2.82, -0.00330693)  (2.88,  
-0.00321435)  (2.94, -0.00312592)  (3., -0.00304141)  (3.06,  
-0.00296057)  (3.12, -0.00288318)  (3.18, -0.00280905)  (3.24,  
-0.00273799)  (3.3, -0.00266983)  (3.36, -0.0026044)  (3.42,  
-0.00254156)  (3.48, -0.00248116)  (3.54, -0.00242308)  (3.6,  
-0.00236719)  (3.66, -0.00231339)  (3.72, -0.00226156)  (3.78,  
-0.00221161)  (3.84, -0.00216344)  (3.9, -0.00211697)  (3.96,  
-0.00207212)  (4.02, -0.0020288)  (4.08, -0.00198695)  (4.14,  
-0.00194649)  (4.2, -0.00190737)  (4.26, -0.00186953)  (4.32,  
-0.0018329)  (4.38, -0.00179743)  (4.44, -0.00176308)  (4.5,  
-0.00172979)  (4.56, -0.00169752)  (4.62, -0.00166623)  (4.68,  
-0.00163587)  (4.74, -0.00160641)  (4.8, -0.00157781)  (4.86,  
-0.00155004)  (4.92, -0.00152306)  (4.98, -0.00149685)  (5.04,  
-0.00147137)  (5.1, -0.00144659)  (5.16, -0.00142249)  (5.22,  
-0.00139904)  (5.28, -0.00137622)  (5.34, -0.00135401)  (5.4,  
-0.00133238)  (5.46, -0.00131131)  (5.52, -0.00129078)  (5.58,  
-0.00127078)  (5.64, -0.00125128)  (5.7, -0.00123227)  (5.76,  
-0.00121372)  (5.82, -0.00119564)  (5.88, -0.00117799)  (5.94,  
-0.00116076)  (6., -0.00114394)};
\end{scope}
\begin{scope}[xscale=1,yscale=700]
\draw[color=blue] plot[smooth] coordinates {(0, 0)  (0.06, 0.00111569)  (0.12, 0.00205083)  (0.18, 
  0.00271618)  (0.24, 0.00310834)  (0.3, 0.00327787)  (0.36, 
  0.00328929)  (0.42, 0.00319932)  (0.48, 0.0030504)  (0.54, 
  0.00287153)  (0.6, 0.00268141)  (0.66, 0.00249149)  (0.72, 
  0.00230842)  (0.78, 0.00213578)  (0.84, 0.00197521)  (0.9, 
  0.00182718)  (0.96, 0.00169146)  (1.02, 0.00156746)  (1.08, 
  0.00145436)  (1.14, 0.00135131)  (1.2, 0.00125742)  (1.26, 
  0.00117183)  (1.32, 0.00109376)  (1.38, 0.00102245)  (1.44, 
  0.000957257)  (1.5, 0.000897561)  (1.56, 0.00084282)  (1.62, 
  0.000792549)  (1.68, 0.00074631)  (1.74, 0.000703714)  (1.8, 
  0.000664415)  (1.86, 0.0006281)  (1.92, 0.000594494)  (1.98, 
  0.000563347)  (2.04, 0.000534438)  (2.1, 0.000507569)  (2.16, 
  0.00048256)  (2.22, 0.000459252)  (2.28, 0.0004375)  (2.34, 
  0.000417175)  (2.4, 0.00039816)  (2.46, 0.000380348)  (2.52, 
  0.000363645)  (2.58, 0.000347963)  (2.64, 0.000333224)  (2.7, 
  0.000319356)  (2.76, 0.000306294)  (2.82, 0.00029398)  (2.88, 
  0.000282358)  (2.94, 0.000271381)  (3., 0.000261001)  (3.06, 
  0.000251179)  (3.12, 0.000241876)  (3.18, 0.000233057)  (3.24, 
  0.00022469)  (3.3, 0.000216746)  (3.36, 0.000209197)  (3.42, 
  0.000202019)  (3.48, 0.000195188)  (3.54, 0.000188683)  (3.6, 
  0.000182484)  (3.66, 0.000176572)  (3.72, 0.000170931)  (3.78, 
  0.000165545)  (3.84, 0.000160399)  (3.9, 0.000155479)  (3.96, 
  0.000150772)  (4.02, 0.000146267)  (4.08, 0.000141953)  (4.14, 
  0.000137819)  (4.2, 0.000133856)  (4.26, 0.000130054)  (4.32, 
  0.000126406)  (4.38, 0.000122903)  (4.44, 0.000119537)  (4.5, 
  0.000116303)  (4.56, 0.000113192)  (4.62, 0.0001102)  (4.68, 
  0.00010732)  (4.74, 0.000104548)  (4.8, 0.000101877)  (4.86, 
  0.0000993027)  (4.92, 0.0000968213)  (4.98, 0.0000944282)  (5.04, 
  0.0000921192)  (5.1, 0.0000898906)  (5.16, 0.0000877387)  (5.22, 
  0.0000856603)  (5.28, 0.0000836519)  (5.34, 0.0000817107)  (5.4, 
  0.0000798337)  (5.46, 0.0000780181)  (5.52, 0.0000762614)  (5.58, 
  0.0000745611)  (5.64, 0.0000729148)  (5.7, 0.0000713204)  (5.76, 
  0.0000697758)  (5.82, 0.0000682789)  (5.88, 0.0000668278)  (5.94, 
  0.0000654208)  (6., 0.000064056)};
  \end{scope}
\draw[->] (-0.2,0) -- (6.3,0);
\draw[->] (0,-3) --(0,3);
\draw[-] (5.7,-0.1) -- (5.7,0.1);
\draw[-] (-0.1,-2.2) -- (0.1,-2.2);
\node at (5.65,0.3) (i5) {20};
\node at (-0.8,-2.2) (i9) {$-0.02$};
\node[color=blue] at (3,1) (i6) {$i=2$};
\draw[->,color=blue] (i6) -- (3.4,0.4);
\node[color=red] at (4,-1.2) (i7) {$i=1$};
\draw[->,color=red] (i7) -- (4.4,-0.4);
\node at (0,3.2) (i1) {$G_i(\tau)/\tau^2$};
\node at (6.7,0) (i2) {$\tau$};
\end{tikzpicture}
    \end{center}
  \caption{\small Plot of the functions $G_i(\tau)/\tau^2$
  derived from the VVM matrix which determine the asymptotic behaviour
  of the refractive index in QED for the two polarizations.}
\label{figBHQEDG}
\end{figure}

The small $|u|$ behaviour of the integral in \eqref{fffj} depends on
the form of the functions $G_i(\tau)/\tau^2$. These are plotted in
Fig.~\ref{figBHQEDG}.
Crucially, the asymptotic forms \eqref{fffk} show that these go to
zero at large $\tau$ fast enough, and we can define
\EQ{
&c_1=\int_0^\infty {d\tau\over\tau^2}~G_1(\tau) = -0.155\ ,\\
&c_2=\int_0^\infty {d\tau\over\tau^2}~G_2(\tau)  =~ 0.105\ .
\label{fffl}
}
This is in contrast to $A\phi^2$ theory where the corresponding
integral is divergent. As we now see, this produces quite different 
small $|u|$ behaviour for the refractive index.
Note also the difference in sign for the two polarizations. This can 
be inferred from \eqref{fffk} and is determined by the properties of
the VVM matrix, depending critically on the $\alpha_i$ parameters.
It now follows that the magnitude of the integrals in \eqref{fffj} will
increase as $|u|$ becomes smaller and the exponential decays less 
rapidly with $\tau$, until in the limit $|u|\rightarrow 0$ we find
\EQ{
\IM {\cal F}_i(u;z) \longrightarrow - {1\over|u|} \int_0^\infty
{d\tau\over\tau^2}~G_i(\tau) = - {c_i\over|u|} \ .
\label{fffm}
}

\begin{figure}[t]
    \begin{center}
\begin{tikzpicture}[scale=0.9]
\begin{scope}[xscale=1,yscale=0.02]
\draw[color=red] plot[smooth] coordinates {(0., -12.3369)  (0.06, -12.4625)  (0.12, -12.5904)  (0.18,  
-12.7208)  (0.24, -12.8537)  (0.3, -12.9892)  (0.36, -13.1273)   
(0.42, -13.2682)  (0.48, -13.4119)  (0.54, -13.5585)  (0.6,  
-13.7081)  (0.66, -13.8608)  (0.72, -14.0166)  (0.78, -14.1757)   
(0.84, -14.3382)  (0.9, -14.5042)  (0.96, -14.6738)  (1.02, -14.847)   
(1.08, -15.0242)  (1.14, -15.2052)  (1.2, -15.3904)  (1.26,  
-15.5798)  (1.32, -15.7736)  (1.38, -15.9718)  (1.44, -16.1748)   
(1.5, -16.3826)  (1.56, -16.5955)  (1.62, -16.8135)  (1.68, -17.037)   
(1.74, -17.266)  (1.8, -17.5008)  (1.86, -17.7417)  (1.92, -17.9888)   
(1.98, -18.2424)  (2.04, -18.5028)  (2.1, -18.7701)  (2.16,  
-19.0448)  (2.22, -19.3271)  (2.28, -19.6173)  (2.34, -19.9158)   
(2.4, -20.2229)  (2.46, -20.539)  (2.52, -20.8644)  (2.58, -21.1996)   
(2.64, -21.5451)  (2.7, -21.9013)  (2.76, -22.2687)  (2.82,  
-22.6478)  (2.88, -23.0392)  (2.94, -23.4435)  (3., -23.8614)  (3.06,  
-24.2935)  (3.12, -24.7405)  (3.18, -25.2033)  (3.24, -25.6827)   
(3.3, -26.1795)  (3.36, -26.6947)  (3.42, -27.2295)  (3.48,  
-27.7847)  (3.54, -28.3618)  (3.6, -28.9619)  (3.66, -29.5865)   
(3.72, -30.237)  (3.78, -30.9152)  (3.84, -31.6227)  (3.9, -32.3615)   
(3.96, -33.1337)  (4.02, -33.9417)  (4.08, -34.7878)  (4.14,  
-35.6749)  (4.2, -36.6059)  (4.26, -37.5843)  (4.32, -38.6136)   
(4.38, -39.6979)  (4.44, -40.8417)  (4.5, -42.0499)  (4.56,  
-43.3282)  (4.62, -44.6827)  (4.68, -46.1204)  (4.74, -47.6492)   
(4.8, -49.2778)  (4.86, -51.0163)  (4.92, -52.8762)  (4.98,  
-54.8704)  (5.04, -57.014)  (5.1, -59.3242)  (5.16, -61.8211)  (5.22,  
-64.5282)  (5.28, -67.4727)  (5.34, -70.6873)  (5.4, -74.2105)   
(5.46, -78.0887)  (5.52, -82.3779)  (5.58, -87.1467)  (5.64,  
-92.4796)  (5.7, -98.4823)  (5.76, -105.289)  (5.82, -113.07)  (5.88,  
-122.052)  (5.94, -132.532)  (6., -144.917)};
\end{scope}
\begin{scope}[xscale=1,yscale=0.14]
\draw[color=blue] plot[smooth] coordinates {(0., 1.39294)  (0.06, 1.40766)  (0.12, 1.42268)  (0.18, 
  1.43802)  (0.24, 1.45367)  (0.3, 1.46965)  (0.36, 1.48597)  (0.42, 
  1.50263)  (0.48, 1.51966)  (0.54, 1.53706)  (0.6, 1.55485)  (0.66, 
  1.57303)  (0.72, 1.59162)  (0.78, 1.61064)  (0.84, 1.6301)  (0.9, 
  1.65002)  (0.96, 1.6704)  (1.02, 1.69127)  (1.08, 1.71265)  (1.14, 
  1.73456)  (1.2, 1.75701)  (1.26, 1.78002)  (1.32, 1.80361)  (1.38, 
  1.82782)  (1.44, 1.85266)  (1.5, 1.87815)  (1.56, 1.90433)  (1.62, 
  1.93122)  (1.68, 1.95884)  (1.74, 1.98724)  (1.8, 2.01645)  (1.86, 
  2.04649)  (1.92, 2.07741)  (1.98, 2.10924)  (2.04, 2.14202)  (2.1, 
  2.1758)  (2.16, 2.21063)  (2.22, 2.24654)  (2.28, 2.28361)  (2.34, 
  2.32187)  (2.4, 2.36139)  (2.46, 2.40224)  (2.52, 2.44447)  (2.58, 
  2.48817)  (2.64, 2.5334)  (2.7, 2.58026)  (2.76, 2.62883)  (2.82, 
  2.6792)  (2.88, 2.73147)  (2.94, 2.78577)  (3., 2.8422)  (3.06, 
  2.90089)  (3.12, 2.96199)  (3.18, 3.02564)  (3.24, 3.09201)  (3.3, 
  3.16127)  (3.36, 3.23361)  (3.42, 3.30926)  (3.48, 3.38843)  (3.54, 
  3.47138)  (3.6, 3.55838)  (3.66, 3.64975)  (3.72, 3.74581)  (3.78, 
  3.84694)  (3.84, 3.95354)  (3.9, 4.06608)  (3.96, 4.18506)  (4.02, 
  4.31106)  (4.08, 4.4447)  (4.14, 4.58671)  (4.2, 4.7379)  (4.26, 
  4.89918)  (4.32, 5.0716)  (4.38, 5.25635)  (4.44, 5.4548)  (4.5, 
  5.66854)  (4.56, 5.89939)  (4.62, 6.14951)  (4.68, 6.42139)  (4.74, 
  6.71801)  (4.8, 7.0429)  (4.86, 7.4003)  (4.92, 7.79535)  (4.98, 
  8.23431)  (5.04, 8.72493)  (5.1, 9.2769)  (5.16, 9.90248)  (5.22, 
  10.6174)  (5.28, 11.4424)  (5.34, 12.4048)  (5.4, 13.5422)  (5.46, 
  14.9069)  (5.52, 16.5747)  (5.58, 18.6591)  (5.64, 21.3385) };
  \end{scope}
\draw[->] (-0.2,0) -- (6.4,0);
\draw[->] (6.2,-3.4) --(6.2,3.4);
\node[color=blue] at (3,2) (i6) {$i=2$};
\draw[->,color=blue] (i6) -- (3.4,0.6);
\node[color=red] at (4,-2.2) (i7) {$i=1$};
\draw[->,color=red] (i7) -- (4.4,-0.9);
\node at (6.4,3.6) (i1) {$\IM n_i(u)$};
\node at (6.7,0) (i2) {$u$};
\node at (8.2,-1.6) (i4) {singularity};
\draw[->] (i4) -- (6.4,-1.3);
\end{tikzpicture}
    \end{center}
  \caption{\small Plot of the imaginary part of the refractive
  index $\IM n(u;\omega)$ for the two polarizations in QED in a Schwarzschild
  background as $u$ approaches the singularity at $u=0$ from $u<0$.}
  \label{figBHQEDn}
\end{figure}
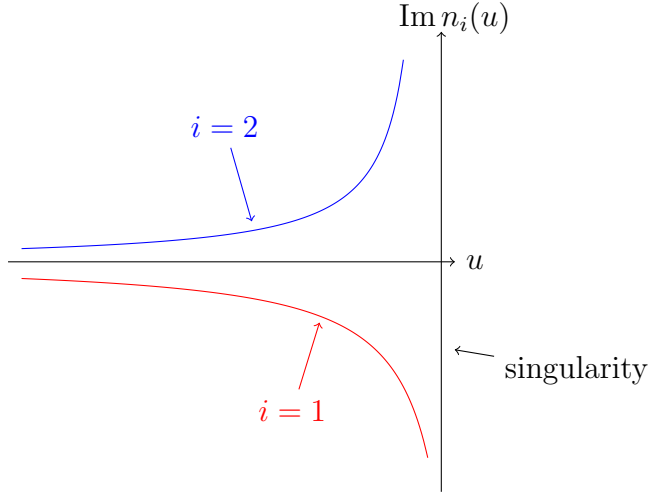

We therefore find the following results for the small $|u|$ behaviour 
of the imaginary part of the refractive index in QED 
(see Fig.~\ref{figBHQEDn}):
\EQ{
&\IM n_{1}(u;\omega) = {\alpha\over6\pi}{1\over2\omega} 
{c_i\over|u|} <0 \ ,\\
&\IM n_{2}(u;\omega) = {\alpha\over6\pi}{1\over2\omega} 
{c_2\over|u|} >0 \ .
\label{fffn}
}
$\IM n(u;\omega)$ diverges as $\frac1{|u|}$ in the near singularity
limit for both polarizations, but with opposite signs:  polarization
2 is being dressed while polarization 1 is being undressed by the
gravitational tidal forces.

The most illuminating way to see the effect on the amplitudes is to
compare $\AMP(u)$ with the amplitude $\AMP(u_1)$ for some 
fixed $u_1$  sufficiently far into the near singularity region that
the above approximations hold, then let $|u|\rightarrow 0$.
From
\EQ{
{|\AMP_i(u)|\over|\AMP_i(u_1)|} = 1 - \omega \int_{u_1}^{u}
du^{\prime\prime}~\IM n_i(u^{\prime\prime};\omega) \ ,
\label{fffo}
}
we find
\EQ{
{|\AMP_i(u)|\over|\AMP_i(u_1)|} = 1 + {\alpha\over12\pi} c_i 
\log{u\over u_1} \ ,
\label{fffp}
}
for the two polarizations. These are both logarithmically divergent,
but we can use the dynamical renormalization group to exponentiate 
to give
\EQ{
{|\AMP_i(u)|\over|\AMP_i(u_1)|} = 
\Bigl({u\over u_1}\Bigr)^{ {\alpha\over12\pi}c_i} \ .
\label{fffq}
}
Recalling \eqref{fffl} for the coefficients $c_i$, we finally find the
small $|u|$ behaviour:
\EQ{
&|\AMP_1(u)| ~\sim~ |u|^{-0.155 {\alpha\over12\pi}}\ , \\
&|\AMP_2(u)| ~\sim~ |u|^{0.105 {\alpha\over12\pi}}\ ,
\label{fffr}
}
so in fact only $|\AMP_1(u)|$ diverges, while for the second polarization
$|\AMP_2(u)| \rightarrow 0$ as $|u|\rightarrow 0$. The diverging case is for the polarization along the direction $z^1$ for which the geodesics focus. 

We see therefore that the evolution of the field amplitude as the
singularity is approached is determined by the properties of the VVM
matrix, which is encoded in the signs of the coefficients $c_i$ and 
differs for the two polarizations. For polarization 2, which correspond to the direction $z^2$ along which the geodesics diverge and for which 
$\IM n(u;\omega)$ is positive, the photon becomes increasingly 
dressed -- the screening increases and the amplitude falls to zero.
For polarization 1, however, for which the geodesics focus, the gravitational tidal forces near the
singularity strip away the vacuum polarization cloud of virtual 
$\phi$ pairs and the photon becomes undressed, the screening is
reduced and the amplitude increases. In the purely scalar $A\phi^2$
theory, this process is bounded and the amplitude rises to a limiting
value. In QED, however, we see the consequence of the theory requiring
wave function renormalization. As the singularity is approached, the
renormalized, dressed photon is being restored in real time to its 
bare state, but since for QED the ratio of bare to renormalized fields
is UV divergent, in this case the amplitude rises without bound as 
$|u|\rightarrow 0$ for one of the polarization states.

\section{Conclusions}

This work was motivated by the apparent paradox that, when vacuum
polarization is taken into account, photons propagating in curved
spacetime can experience a refractive index with a negative imaginary
part, corresponding to an amplification of the amplitude. This is in
direct contradiction to the conventional flat-spacetime optical
theorem, which relates $\IM n(\omega)$ to a decay rate and is
necessarily positive or zero.

In this paper, we have addressed this issue from the point of view of
an initial value problem, tracing the evolution of a renormalized
(dressed) quantum field as it propagates through curved
spacetime. Different initial conditions were used, corresponding to an
initially bare or dressed field, and renormalization issues carefully
addressed. The simplifications following from being able to replace   
the background spacetime by its Penrose plane-wave limit were fully
exloited, allowing in many cases a very detailed analysis of the field
evolution using a Laplace transform solution of the one-loop equation
of motion.

The physical picture that emerges is that in a curved spacetime
background, gravitational tidal forces act on the virtual 
$e^+e^-$ cloud which dresses a renormalized photon field in such a way
as to increase or decrease the degree of dressing. In particular,
gravity can ``undress" a renormalized photon, returning the field
towards its bare state. As the degree of dressing is reduced, so is
the level of screening and in consequence the amplitude increases. 
This effect can therefore produce a negative $\IM n(u;\omega)$,
though the effect is constrained to be local, and bounded
(with the exceptions above of theories with UV-divergent wave function
renormalization taken at the limit of the intial-value surface or
a spacetime singularity).
Indeed, the propagation of a renormalized photon through
a time-dependent curved background resembles in many ways the
initial transient phase of propagation in flat spacetime when an
interaction is abruptly turned on.

In addition, in studying scalar $A\phi^2$ theories, as well as QED,
we found many examples of below-threshold decays which would be 
kinematically forbidden in flat spacetime. These decay rates are 
non-perturbative in the curvature. In particular, the analytic
structure and nature of the thresholds was studied in detail in the 
class of symmetric plane waves, where a rich pattern of
curvature-dependent thresholds emerged. It is probable that these can
be interpreted in terms of group representations, since this class of
homogeneous plane waves admits an enhanced symmetry described 
by a Heisenberg algebra.

The central result of this paper, which provides the formal framework
for all of these phenomena, is the formulation of a generalised
optical theorem valid in curved spacetime. In a plane-wave spacetime,
this has the form:
\EQ{
P_{A\to\phi\phi}(u) = 
4\w \int_{u_0}^u du^{\prime\prime}
\IM n(u^{\prime\prime};\w)  =
 -{2\over\w} \int_{u_0}^u du^{\prime\prime}\, 
\int_{u_0}^{u^{\prime\prime}} du' ~ 
\IM \tilde \Pi_\text{SK}(u^{\prime\prime},u';\w,p) \ ,
\label{ga}
}
and relates the total $A\rightarrow\phi\phi$ decay probability 
to the integral of the imaginary part of the vacuum
polarization. The key point is that while this allows 
$\IM n(u;\omega) < 0$ {\it locally}, when integrated over the 
entire trajectory from the surface $u=u_0$ at which the interaction is
turned on,  $\int_{u_0}^u du^{\prime\prime}
\IM n(u^{\prime\prime};\w)$ must be positive to ensure consistency
with unitarity.

This confirms the theme of our earlier work that, despite the many
novel and unexpected effects due to vacuum polarization, quantum field
theories in curved spacetime indeed respect the fundamental principles
of causality and unitarity.  However, in so doing, many of the basic and
widely-assumed theorems of QFT in flat spacetime, notably dispersion
relations, the analytic structure of Green functions and scattering
amplitudes, as well as the optical theorem, must be reformulated in curved
spacetime.

\vspace{0.5cm}
\begin{center}
{\tiny**************}
\end{center}
\vspace{0.5cm}

This research was supported in part by the STFC grant ST/G000506/1.
We would like to thank the TH Division, CERN for hospitality
while much of this work was carried out.

\startappendix

\Appendix{Weak Curvature Expansions}\label{ap2}

In this appendix, we quote the results for the weak coupling expansions of the refractive index for scalar and spinor QED generalizing the case of massless scalar fields \eqref{bbp}
\EQ{
n(u;\w)=1-\frac{e^2}{(4\pi)^2}\frac1{360m^4}R_{uu}(u)-
\frac{e^2}{(4\pi)^2}\frac{i \omega}{840m^6}\dot R_{uu}(u)+\cdots\ ,
}
The case for scalar QED follows from \eqref{fl}:
\EQ{
n_{ij}(u;\w)=\delta_{ij}-&\frac{\alpha}{360\pi m^2}\big(R_{uu}(u)\delta_{ij}+
2R_{iuju}\big)\\ &-\frac{i\alpha \omega}{1680\pi m^4}\big(\dot R_{uu}(u)\delta_{ij}+
2\dot R_{iuju}(u)\big)+\cdots\ ,
}
Here, $R_{uu}=R_{1u1u}+R_{2u2u}$ is a component of the Ricci tensor.
The result for (spinor) QED can be extracted from \cite{Hollowood:2009qz}
\EQ{
n_{ij}(u;\w)=\delta_{ij}-&\frac{\alpha}{180\pi m^2}\big(13R_{uu}(u)\delta_{ij}-4
R_{iuju}(u)\big)\\ &-\frac{i\alpha \omega}{1260\pi m^4}\big(25\dot R_{uu}(u)\delta_{ij}-6
\dot R_{iuju}(u)\big)+\cdots\ ,
}
The first correction here is the original Drummond-Hathrell result \cite{Drummond:1979pp}. The higher corrects are then non-linear in the curvature.

These expansions allow us to extract the leading order behaviour of the amplitude generalizing the expression for the scalar field in \eqref{bbq}. For scalar QED
\EQ{
\AMP_i(u_1)=\AMP_j(u_2)\exp\left[\frac{\alpha \omega^2}{1680\pi m^4}\big(\dot R_{uu}(u_1)\delta_{ij}+2\dot R_{iuju}(u_1)-\dot R_{uu}(u_2)\delta_{ij}-2\dot R_{iuju}(u_2)\big)\right]
}
and (spinor) QED
\EQ{
\AMP_i(u_1)=\AMP_j(u_2)\exp\left[\frac{\alpha \omega^2}{1260\pi m^4}\big(25\dot R_{uu}(u_1)\delta_{ij}-6
\dot R_{iuju}(u_1)-25\dot R_{uu}(u_2)\delta_{ij}+6
\dot R_{iuju}(u_2)\big)\right]\ .
}

\vfill\eject

\Appendix{Penrose limit and de Sitter space}\label{a1}

A key element of our analysis of field propagation in curved
spacetimes has been the simplification brought about by the Penrose
limit. It is remarkable that a highly non-trivial validation of this
method can be found using the symmetric plane waves of the third type
considered in section 6.

To see this, consider the spacetime $dS_3\times\mathbb R$, with metric 
\EQ{
ds^2=-dt^2+\cosh^2(\alpha
t)\,\big(d\theta^2+\sin^2\theta\,d\phi^2\big)+dz^2\ ,
\label{Aa}
}
The three-dimensional de Sitter space metric is given in global
coordinates where $(\theta,\phi)$ are the coordinates on $S^2$.
A congruence of null geodesics consists of $t=z$ with 
$\theta=\theta_0$ and $\phi=\phi_0$ fixed. 
The associated null coordinates are
\EQ{
u=\frac1{\sqrt2}(z+t)\ ,\qquad v=\frac1{\sqrt2}(z-t)\ .
\label{Ab}
}
The Penrose limit is now trivial to take: we identify
$x^1=\theta-\theta_0$ and $x^2=\sin^{-2}(\theta_0)(\phi-\phi_0)$
and expand in powers
of $v$ and $x^a$ keeping terms of order 2 with $v$ having weight 2 and
$x^a$ weight 1:
\EQ{
ds^2\Big|_\text{Penrose limit}=2du\,dv+\cosh^2(\sigma u)dx^a\,dx^a\ .
\label{Ac}
}
with $\sigma=(\sqrt2\alpha)^{-1}$. In this case, the Penrose limit
is the symmetric plane wave of the ``wrong sign' kind with 
$\sigma_1=\sigma_2=i\sigma$. This geometry 
violates the null energy condition since in Brinkmann coordinates 
$R_{uu}=-2\sigma^2<0$, but we can nevertheless consider this, 
as in section 6, as a valid fixed background.

The remarkable feature that allows the test of the Penrose limit
method is that the Green functions and spectral density are known
exactly in this case for the original spacetime $dS_3\times\mathbb R$,
as well as its symmetric plane wave Penrose limit, due to the high
degree of symmetry of the de Sitter space.  In particular, the Green
functions admit a K\"all\'en-Lehmann representations from which the 
spectral density $\rho(M)$ can be extracted. In the limit $M\to 0$,
we will show how this exact result reproduces the spectral density
\eqref{ezg} already found in the plane wave background.

Using the notation of \cite{Bros:2008sq,Bros:2009bz}, the
K\"all\'en-Lehmann representation is
\EQ{
\tilde 
G_+(\nu;\tilde x,\tilde x')\tilde 
G_+(\mu;\tilde x,\tilde x')=\alpha^{-1}\int_0^\infty d\kappa^2\,
\rho_{\nu,\mu}(\kappa)
\tilde G_+(\kappa;\tilde x,\tilde x')\ .
\label{Ad}
}
Here, $\tilde G_+(\nu;\tilde x,\tilde x')$ is the Wightman function and
for a (minimally-coupled) field of mass $m$
\EQ{
\nu^2=\frac{m^2}{2\sigma^2}-1\ ,
\label{Ae}
}
where $\tilde x$ are coordinates on $dS_3$. The K\"all\'en-Lehmann 
weight-function for three-dimensional de Sitter space is
explcitly\footnote{This follows directly from equations (4) and (38) of
  \cite{Bros:2009bz}.} 
\EQ{
\rho_{\nu,\mu}(\kappa)=\frac{\sinh^2(\pi\kappa)}{2^5
\pi\kappa
\cosh\frac{\pi(\kappa+\nu+\mu)}2\cosh\frac{\pi(\kappa-\nu+\mu)}
2\cosh\frac{\pi(\kappa+\nu-\mu)}2\cosh\frac{\pi(\kappa-\nu-\mu)}2}\ .
\label{Af}
}

The Green functions for a field of mass $m$ on 
the product space $dS_3\times\mathbb R$ follows in a simply way by noticing
that momentum along the $z$ direction acts as an effective
contribution to the three-dimensional de Sitter mass. Splitting the
coordinates as $x=(z,\tilde x)$, we have
\EQ{
G_+^{(m)}(x,x')=\int\frac{dp}{2\pi}\,e^{ip(z-z')}
\tilde G_+\big(\sqrt{\tfrac{m^2+p^2}{2\sigma^2}-1};\tilde x,\tilde x'\big)\ .
\label{Ag}
}
Taking the Fourier transform along the $z$ direction, we have
\EQ{
G_+^{(m)}(p;\tilde x,\tilde x')=\tilde
G_+\big(\sqrt{\tfrac{m^2+p^2}{2\sigma^2}-1};\tilde x,\tilde x'\big)\ .
\label{Ah}
} 

The vacuum polarization for the scalar $A\phi^2$ theory is
\EQ{
\Pi(x,x')=i e^2 G_+^{(m)}(x,x')^2\ .
\label{Ai}
}
Using 
the K\"all\'en-Lehmann representation for $dS_3$ in
\eqref{Ad}, we therefore find
\SP{
\int dz\, e^{-ip(z-z')}\Sigma(x,x')&=e^2
\int\frac{dq}{2\pi}G_+^{(m)}(q;\tilde
x,\tilde x')G_+^{(m)}(p-q;\tilde x,\tilde x')\\ &
=\int_0^\infty dM^2\,\rho(M)G_+^{(M)}(p;\tilde x,\tilde x')\ ,
\label{Aj}
}
where the weight function is 
\EQ{
\rho(M)=\frac{e^2}{\sqrt2\sigma}\int\frac{dq}{2\pi}
\rho_{\nu,\mu}(\kappa)\ ,
\label{Ak}
}
with
\EQ{
\nu^2=\frac{m^2+q^2}{2\sigma^2}-1\ ,\qquad\mu^2=
\frac{m^2+(p-q)^2}{2\sigma^2}-1\ ,\qquad
\kappa^2=\frac{M^2+p^2}{2\sigma^2}-1\ .
\label{Al}
}
In particular, $\rho(M)$ is precisely the
spectral density for the vacuum polarization on 
$dS_3\times\mathbb R$ and the decay rate
for $A\to\phi\phi$, for massless $A$, is given by 
$\frac\pi\omega\rho(0)$.

Now we carefully take the ``geometric optics" and ``weak curvature" 
limits described in section 2.1, which led to the use of the Penrose limit.
First of all, we identify the external
momentum as $p=\tfrac\omega{\sqrt2}$, where $\omega$ is the light-cone
momentum. In the geometric optics limit, due to the
form of the function $\tilde\rho_{\nu,\mu}(\kappa)$, the
integrand in \eqref{Ak} only has support in the neighbourhood of
$q=\tfrac p2$, {\it i.e.}~where the external momentum is shared
equally by the two particles in the final state.
Consequently the arguments of all the hyperbolic functions in
\eqref{Af}, except the one involving 
$\kappa-\nu-\mu$, are large and therefore 
may be replaced by exponentials. This means
we can approximate
\EQ{
\rho_{\nu,\mu}(\kappa)\simeq\frac{1}{2^3
\pi\kappa(1+e^{\pi(\nu+\mu-\kappa)})}\ .
\label{Am}
}
In particular, we can parameterize
$q=\tfrac{\omega\xi}{\sqrt2}$ and restrict the $\xi$ integral to the
interval $[0,1]$ and then expand for large $\omega$
\SP{
\nu+\mu-\kappa&=\frac{\omega}{2\sigma}\big(\xi+(1-\xi)-1\big)
+\frac{m^2}{2\omega\sigma\xi(1-\xi)}+\cdots\\ &
=\frac{m^2}{2\omega\sigma\xi(1-\xi)}+\cdots\ .
\label{An}
}
It is important for the validity of these expansions 
that the support of the integrand lies away from the
points $\xi=0,1$.
Finally, putting all this together gives the spectral density at $M=0$:
\EQ{
\rho(0)=
\frac{e^2}{(4\pi)^2}\int_0^1d\xi\,\frac1{
1+e^{\frac{\pi m^2}{2\omega\sigma\xi(1-\xi)}}}\ .
\label{Ao}
}

This is precisely equal to \eqref{ezg} for $M=0$. This confirms that
the results we have obtained directly from the Penrose limit
spacetimes are indeed the correct approximations to those of the full
background spacetime when we impose the physically-motivated
conditions described in section 2.

\end{document}